\documentclass[prb,twocolumn,aps,showpacs,superscriptaddress,epsfig]{revtex4-2}
\usepackage{amssymb}
\usepackage{amsbsy}
\usepackage{amsmath}
\usepackage{epsfig}
\usepackage{graphicx}
\usepackage{subfigure}
\usepackage{ulem}
\usepackage[toc,page]{appendix}
\usepackage{soul}
\usepackage{float}
\usepackage{hyperref}
\hypersetup{
	colorlinks=true,
	linkcolor=blue,
	filecolor=cyan,      
	urlcolor=magenta,
	citecolor=red,
}

\begin{document}
\title{Anomalous pumping in the non-Hermitian Rice-Mele model}
\author{Abhishek Kumar}
\email{akumar0@umass.edu}
\affiliation{National Institute of Science Education and Research, Jatni, Odisha 752050, India}
\affiliation{Department of Physics, University of Massachusettes, Amherst, MA 01003, USA}

\author{Sarbajit Mazumdar}
\email{sarbajit.mazumdar@niser.ac.in}
\affiliation{National Institute of Science Education and Research, Jatni, Odisha 752050, India}
\affiliation{Homi Bhabha National Institute, Training School Complex, Anushakti Nagar, Mumbai 400094, India }

\author{S D Mahanti}
\affiliation{Department of Physics and Astronomy,
Michigan State University, MI 48824, USA}

\author{Kush Saha}
\email{kush.saha@niser.ac.in}
\affiliation{National Institute of Science Education and Research, Jatni, Odisha 752050, India}

\affiliation{Homi Bhabha National Institute, Training School Complex, Anushakti Nagar, Mumbai 400094, India }


\begin{abstract}
We study topological charge pumping (TCP) in the Rice-Mele (RM) model with irreciprocal hopping. The non-Hermiticity gives rise to interesting pumping physics, owing to the presence of skin effect and exceptional points. In the static 1D RM model, we observe two independent tuning knobs that drive the topological transition, viz., non-Hermitian parameter $\gamma$ and system size $N$. To elucidate the system-size dependency, we made use of the finite-size generalized Brillouin zone (GBZ) scheme. This scheme captures the state pumping of topological edge modes in the static 1D RM model and provides further insight into engineering novel gapless exceptional edge modes with the help of adiabatic drive. Finally, we apply three types of adiabatic protocols to study TCP in the 1+1D RM model. We further explain the number of pumped charges (in each period) using a non-Bloch topological invariant. This exactly explains the presence of different pumping phases in the non-Hermitian RM model as we tune the non-Hermitian parameter $\gamma$. We observe that in a non-Hermitian system, even a trivial adiabatic protocol can lead to pumping that has no Hermitian counterpart.
\end{abstract}
\maketitle

\section{Introduction}
Topological charge pump (TCP) proposed by Thouless\cite{PhysRevB.27.6083} allows charge transportation in an insulator without any voltage bias. Instead, charges are pumped with time-periodic and adiabatic variations of the Hamiltonian. The number of pumped charges in a full cycle turns out to be a topological invariant. Although the idea of quantized charge transport was introduced almost three decades back and subsequently verified in nanoscale devices such as acoustic waves\cite{PhysRevB.56.15180}, quantum dots\cite{Bumenthal_2007,PhysRevB.77.153301}, it has only been recently realized in a many-body set-up. 
Specifically, using a highly controllable optical superlattices of ultracold atoms, pumping of atomic clouds per cycle and topological aspects of this phenomena have been revealed \cite{Nakajima_2016,Lohse_2015}. This has led to a considerable amount of recent theoretical studies on TCP in interacting systems  \cite{Kuno_2017,PhysRevB.98.245148,PhysRevA.101.053630,PhysRevResearch.2.042024}. 

The study of topology in non-Hermitian systems has also been a subject of extensive research interest in recent times, both theoretically and experimentally \cite{PhysRevB.84.205128,PhysRevA.87.012118,PhysRevLett.116.133903,PhysRevX.8.031079,PhysRevLett.118.045701,PhysRevLett.118.040401,PhysRevLett.120.146402,PhysRevB.99.081302,PhysRevB.97.121401,PhysRevB.99.201103,Ghatak_2019,PhysRevB.101.045415,PhysRevX.9.041015,PhysRevLett.123.016805,doppler2016dynamically,xu2016topological,PhysRevLett.124.250402,PhysRevB.104.195102,li2021quantized,PhysRevLett.126.215302,Li_2022,PhysRevResearch.5.L022050}. This is because the non-Hermiticity has dramatic consequences on the topological phenomena of Hermitian models. For example, the notion of standard bulk-boundary correspondence (BBC) is no longer valid    \cite{PhysRevLett.121.026808,PhysRevLett.121.136802,PhysRevLett.125.226402} due to the non-Hermitian skin effect (NHSE) \cite{PhysRevLett.123.066404,PhysRevLett.124.056802,PhysRevLett.124.086801,PhysRevLett.121.136802, PhysRevB.105.075128} where all bulk states in a finite system become localized near the boundary. It has also been shown that the BBC can be restored in some specific non-Hermitian models using a topological invariant quantity defined over both GBZ and biorthogonal space ~\cite{PhysRevLett.121.086803,PhysRevLett.123.066404,PhysRevX.9.041015,PhysRevLett.121.026808}. In addition, non-Hermitian systems can host exceptional points (EPs) \cite{PhysRevB.99.041202,PhysRevLett.118.045701,PhysRevLett.123.066405} which are analogous to degenerate points in Hermitian systems. These EPs can have inherent topology and can annihilate with each other to give rise to degenerate points\cite{PhysRevB.101.045415}.

Although there has been a vast literature on non-Hermitian systems, focusing on bulk-boundary correspondence and characterising boundary modes, only a handful number of studies addressed TCP. For example, topological charge pumping \cite{PhysRevB.95.184306} has been addressed in a non-Hermitian network model. Subsequently, dynamical topological pumping has been discussed in a pseudo-Hermitian model with real spectrum\cite{doppler2016dynamically}. Also, spontaneous charge pumping stemming from state conversion of EPs \cite{PhysRevLett.118.093002,Zhang_2019} has been discussed in the {\it static} non-Hermitian SSH model \cite{PhysRevA.99.032109}. A non-adiabatic charge pumping in Floquet chains has also been studied \cite{Fedorova_2020,PhysRevResearch.2.023235}. Similar to our case, Ref.~\cite{PhysRevA.98.042120} studies TCP using irreciprocal hopping but in the absence of NHSE. In addition, a few recent computational demonstrations have revealed that the interplay between non-Hermitian skin modes and topological boundary modes, leading to the delocalization of topological modes in the SSH model, may assist adiabatic state pumping \cite{Longhi_2018,PhysRevB.103.195414}. However, the main focus of all of these studies lies in characterizing topological boundary modes and/or establishing Bulk-boundary correspondences (BBC).   Our aim here is to address the following set of questions: \textit{a) Can we improve on the understanding of state pumping phenomena that result as an interplay between skin and topological modes in a 1D non-Hermitian model? b) If yes, then can this lead to some exotic pumping phenomena under some adiabatic drive? c) Can these phases be described by a non-Bloch topological invariant? d)What are the interesting tuning knobs to explore these phases? e) Can we use point and line gaps to characterize the nature of pumping?} We then further extend the insights acquired from the \textit{static} 1D RM model to the adiabatically driven 1+1D RM model. 

To answer these questions, we introduce the non-Hermiticity term in the static RM model such that the model exhibits both skin modes and topological boundary modes.  a) We found that the non-Hermiticity can induce pumping features in the static RM model analogous to the dynamic RM model. To analytically elaborate the nature of pumping, we use the generalized Brillouin zone (GBZ) scheme \cite{PhysRevLett.127.116801}, which is found to contain a signature of topological boundary modes. It also allows us to show that the non-Hermiticity-induced skin modes and topological boundary modes survive together for moderately large values of non-Hermitian parameter $\gamma$.  This is in contrast to the earlier continuum GBZ schemes\cite{PhysRevLett.121.086803, PhysRevLett.123.066404} where the GBZ spectra do not capture information about topological boundary modes and their evolution as a function of non-Hermitian parameters. We observed that as we tune the non-Hermitian parameter $\gamma$, the left localized edge state moves to the right boundary in the system. This pumping nature can also be reversed by changing the sign of the non-Hermitian parameter $\gamma$, in the system. b) In the dynamic RM model, we demonstrate non-Hermiticity-driven richer pumping phenomena, exploring three distinct driving protocols namely, clockwise nontrivial protocol, anti-clockwise nontrivial protocol, and trivial protocol. For non-trivial clockwise and anti-clockwise protocols,  we find that the topological pumping due to the adiabatic and periodic variation of the model parameters retains for small non-Hermitian parameter $\gamma$, however, for moderate $\gamma$, the in-gap states no longer survive, leading to the breakdown of conventional pumping. Surprisingly, for sufficiently large $\gamma$, the in-gap states reemerge and they are localized over a single unit cell at the right boundary as opposed to the small $\gamma$ features. We dub this as ``anomalous pumping". For trivial pumping protocol, no pumping feature is obtained as expected for low and moderate values of $\gamma$. Remarkably, for stronger $\gamma$, we find in-gap states even in trivial pumping protocol. c) To further understand the pumping phenomena and to establish a non-Bloch BBC, we use the concept of the non-Bloch Chern number \cite{PhysRevB.105.075128}, which can fully characterize the possible pumping phases in the 1+1D RM model. 
d) As previously described in the case of the static RM model, the non-hermitian parameter $\gamma$ is shown to be capable of tuning the pumping within the system. We have now observed that the same can be achieved by varying the system size ($N$). This phenomenon gives rise to a system size-dependent topological phase transition in the static case, subsequently leading to a novel size-dependent topological pumping transition in the dynamic case. Moreover, for an odd number of unit cells, edge modes are found to be glued by the exceptional points as opposed to the even number of unit cells. This is typically known as ``exceptional edge modes", and was first obtained in a two-layered tight-binding model\cite{Sone_2020}. e) Finally, we characterise these features by studying the type of gaps in the spectrum: point and line gap. For pumping in the low $\gamma$ regime, we find the system exhibiting a line gap. In contrast, for anomalous pumping at higher $\gamma$ values, a transition from the line gap to the point gap or vice versa occurs over a full pumping period. For trivial pumping protocol, however, the system maintains point gap topology as opposed to the non-trivial protocols.

The rest of the paper is organized as follows: In Sec.~\ref{sec:Hamiltonian}, we introduce the model Hamiltonian in the presence of the non-Hermitian term $\gamma$. This is followed by the analysis of the spectrum of the static 1D RM model and the investigation of the pumping by tuning the non-Hermitian parameter. 
After that, we provide analytical results for both skin and topological boundary modes localized at the boundaries. Following this, we analyzed the complex energy spectra for the non-Hermitian RM model utilizing the finite-size GBZ scheme. Moving on, we explain the emergence of second-order EPs in the static 1D model and characterise the gaps in the complex energy spectra. We, then, review Hermitian pumping under trivial and non-trivial pumping protocols in Sec.~\ref{sec:dynamicalRM}. This is followed by the conventional and anomalous pumping for the different values of the non-Hermitian term.  Subsequently, we use the concept of non-Bloch Chern number to analyze the pumping phases in the dynamic case. We also provide an analytical description of how the variation in system size gives rise to exceptional edge modes in the dynamic case. Thereafter, we demonstrate the connections between various pumping protocols and the presence of gaps in complex energy spectra i.e. point gaps and line gaps. Finally, we conclude with a detailed summary of our findings and discuss possible future prospects regarding non-Hermitian TCP.

\section{\label{sec:Hamiltonian} Model Hamiltonian}
We begin with the RM model Hamiltonian with two atoms per unit cell: \cite{PhysRevLett.49.1455}
\begin{align}
H(t) =& \sum_i w(t)(a^{\dagger}_{i+1}b_{i}+\text h.c)+\sum_i(v_{l}(t)a^{\dagger}_{i}b_{i} + v_{r}(t)b^{\dagger}_{i}a_{i} )\nonumber\\&~~~~~~~~~~~~~~~~~~~~~~+ \sum_{i}\Delta(t)(a^{\dag}_{i}a_{i}-b^{\dag}_{i}b_{i}),
\label{eqn:ham0}
\end{align} 
where $i$ is the lattice index representing unit cell of two atoms, $a^{\dagger}_{i}$ ($b_{i}$) fermionic creation (annihilation) operator for atom $A$ ($B$) in the $i$th unit cell. Equation (\ref{eqn:ham0}) is a time-dependent model where $v_{l/r}(t)$ denotes intracell left ($l$) and right ($r$) hopping,  $w(t)$ denotes intercell hopping and $\Delta(t)$ is a staggered onsite potential. Here, $w(t)$, $v(t)$ and $\Delta(t)$ are in general periodic function of time $t$ with a period $T$. Note that, for $\Delta(t)=0$, $w(t)= \text{constant}$, and $v_r(t)=v_{l} (t)=\text{constant}$, we recover SSH model. 

To introduce non-Hermiticity, we consider left and right {\it intracell} hopping to be asymmetric, i.e., $v_{l}(t) = v(t)-\gamma$ and $v_{r}(t) = v(t)+\gamma$, where $\gamma$ is a measure of non-Hermiticity in this model, and we assume it to be time-independent. 
We note that there are several proposals for realizing asymmetric hopping in experiments with ultracold systems using reservoir engineering\cite{PhysRevX.8.031079,PhysRevB.102.235151,ren2021chiral}, and with imaginary gauge fields \cite{Longhi_2015,PhysRevLett.77.570,PhysRevB.56.8651,PhysRevB.92.094204}.

Assuming periodic boundary condition and performing a Fourier transform to  momentum space, Eq.~(\ref{eqn:ham0}) can be written as 
\begin{align}
H(k)=& (v(t)+w(t)\cos k)\sigma_{x}+(w(t)\sin k - i\gamma)\sigma_{y}\nonumber\\&~~~~~~~~~~~~~~~~~~~~~~~~~~~~~~~~~~~~~~~~~~ +\Delta\,\sigma_{z},
\label{eqn:hamiltonian_momentum_space}
\end{align}
where $\sigma$'s are the Pauli matrices in sublattice space.  Since $H$ is not Hermitian, the {\it right} and {\it left} eigenstates ($|\phi_{n}^{k}\rangle$ and $|\chi_{n}^{k}\rangle$) satisfy the eigenvalue equation
\begin{equation}
H(k)|\phi_{n}^{k}\rangle = E_{n}(k)|\phi_{n}^{k}\rangle,~H^{\dag}(k)|\chi_{n}^{k}\rangle = E^{*}_{n}(k)|\chi_{n}^{k}\rangle,
\label{eqn:left_right_eigenstates}
\end{equation}
where $n$ is the band index.

\begin{figure}
	\centering
	\subfigure[]{
	\includegraphics[width=0.48\textwidth]{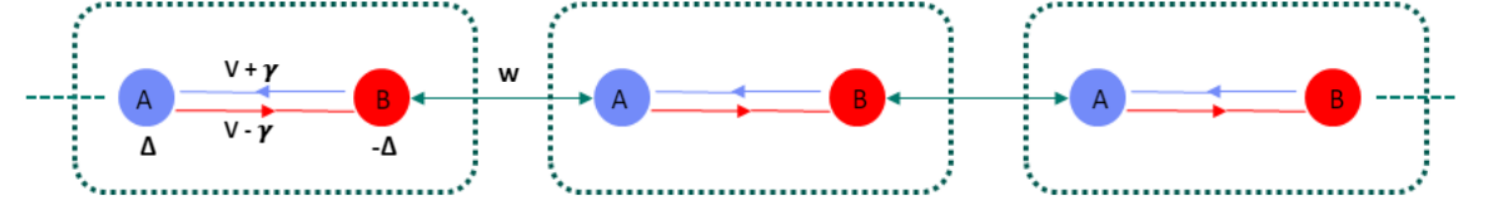}}\\
\vspace{-1\baselineskip}
\subfigure[]{
	\includegraphics[width=0.155\textwidth, height=3.0cm]{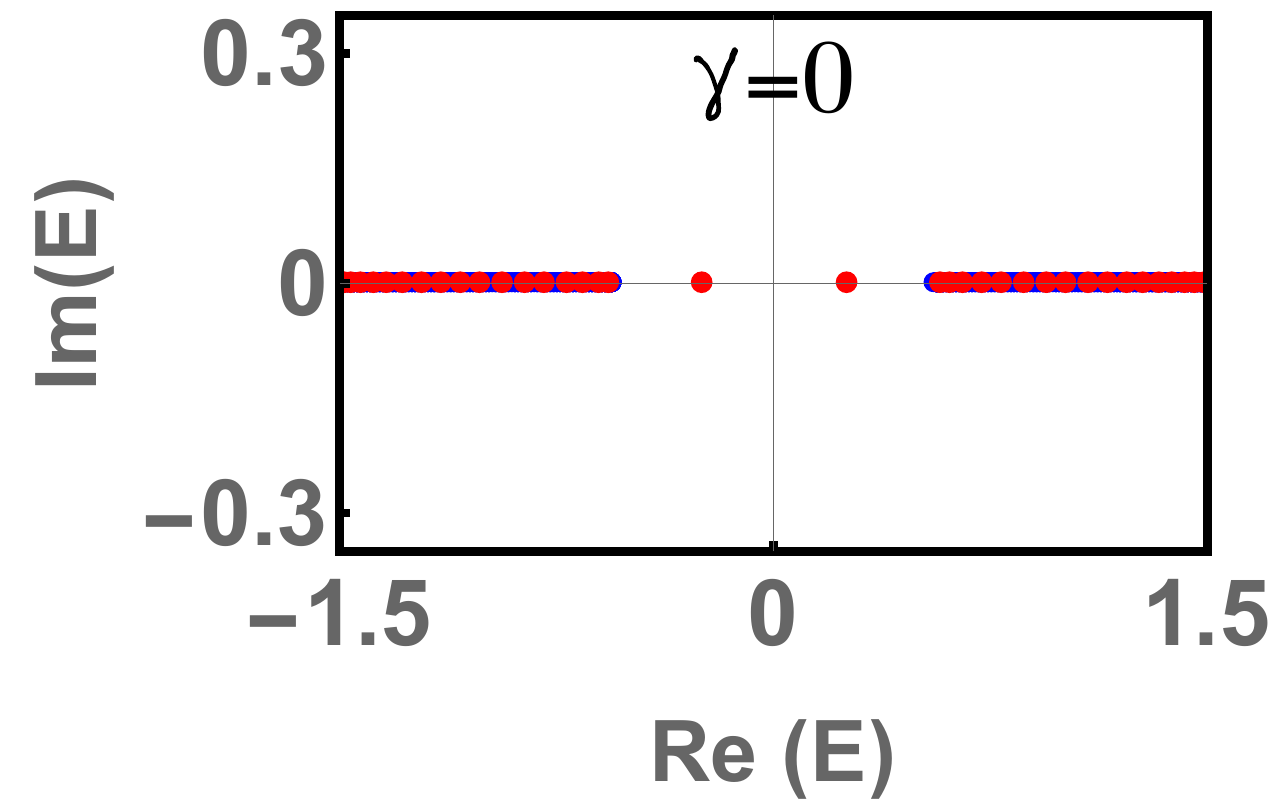}}
\subfigure[]{
	\includegraphics[width=0.145\textwidth,height=3.0cm]{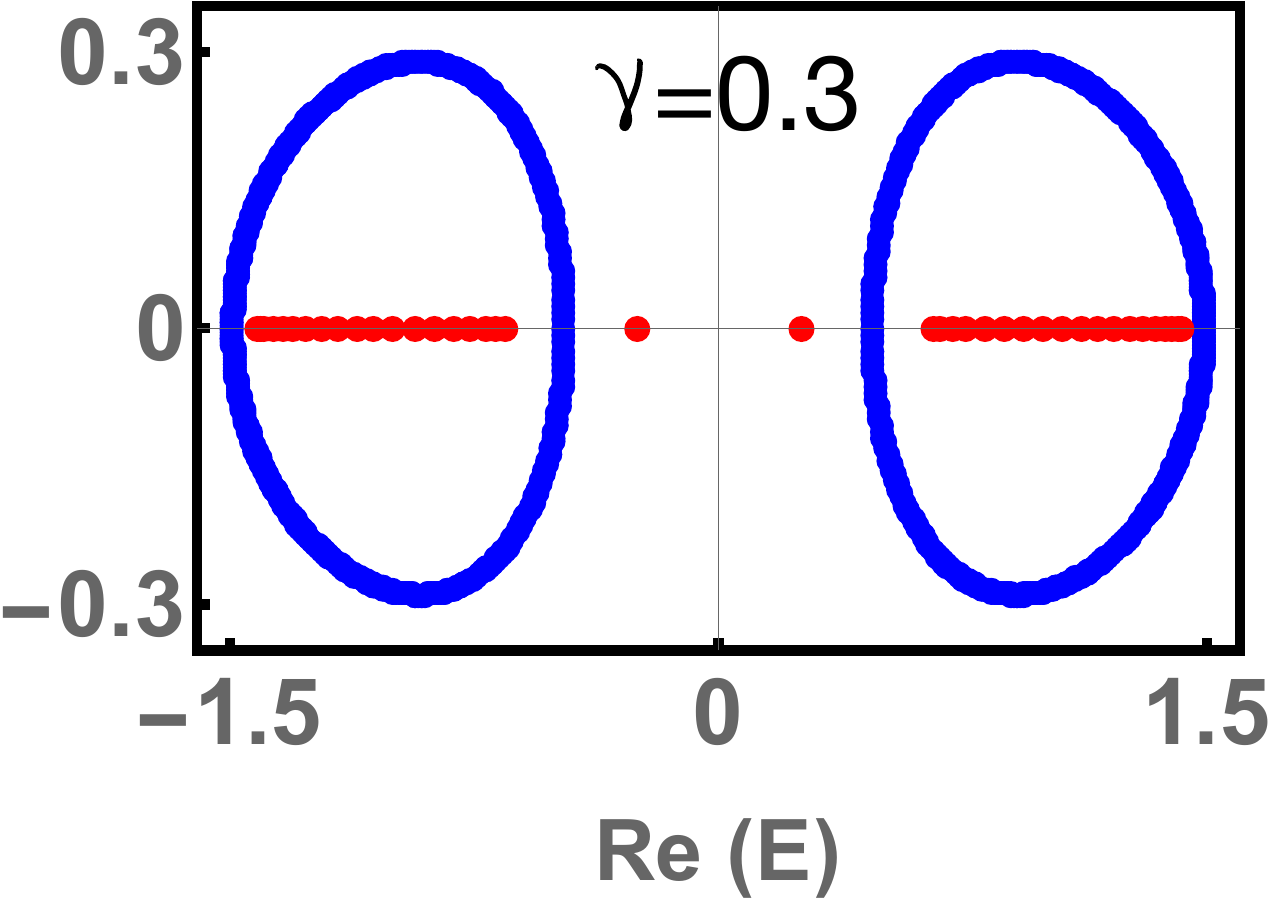}}
\subfigure[]{
	\includegraphics[width=0.145\textwidth,height=3.0cm]{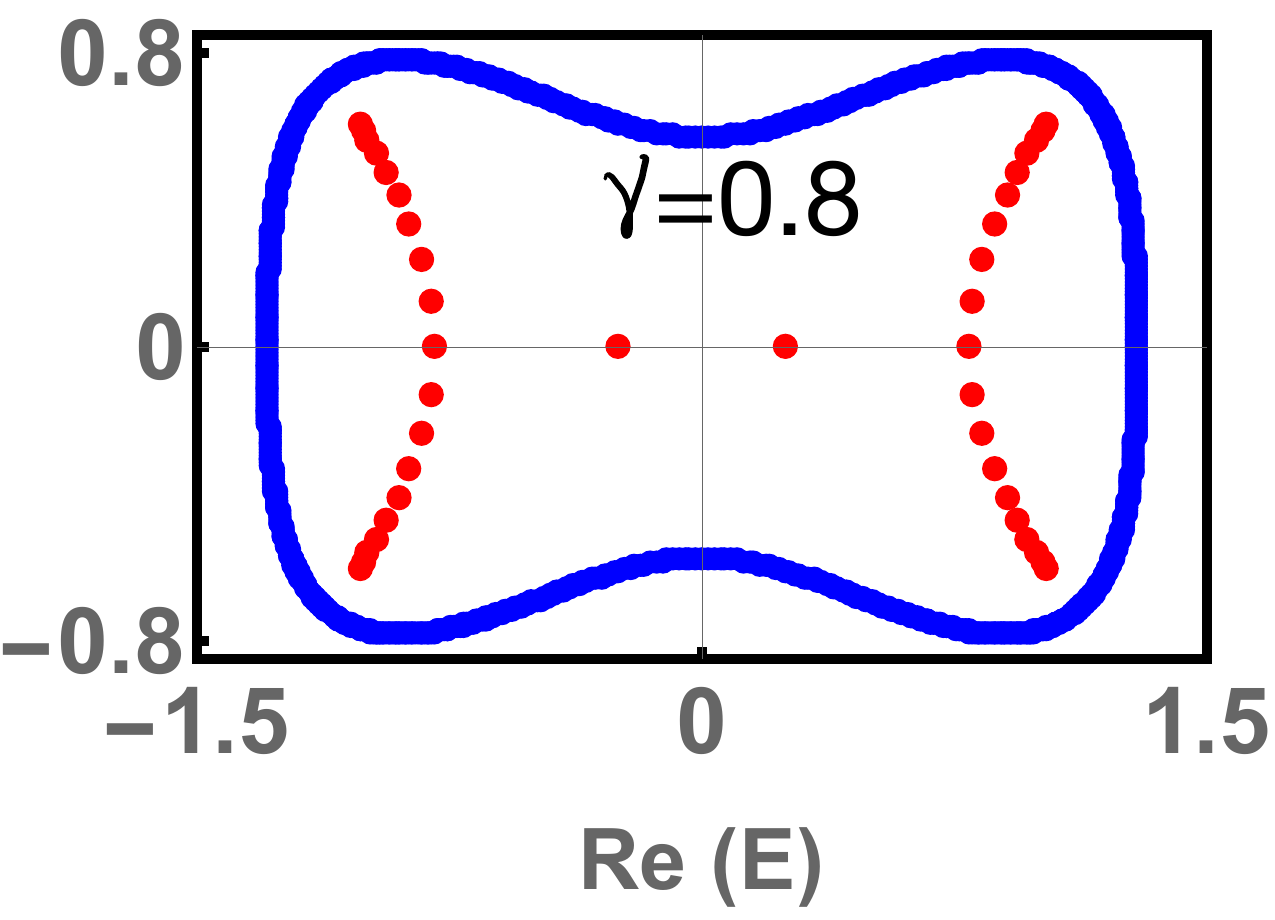}}\\
\vspace{-1\baselineskip}
\subfigure[]{
	\includegraphics[width=0.155\textwidth,height=3cm]{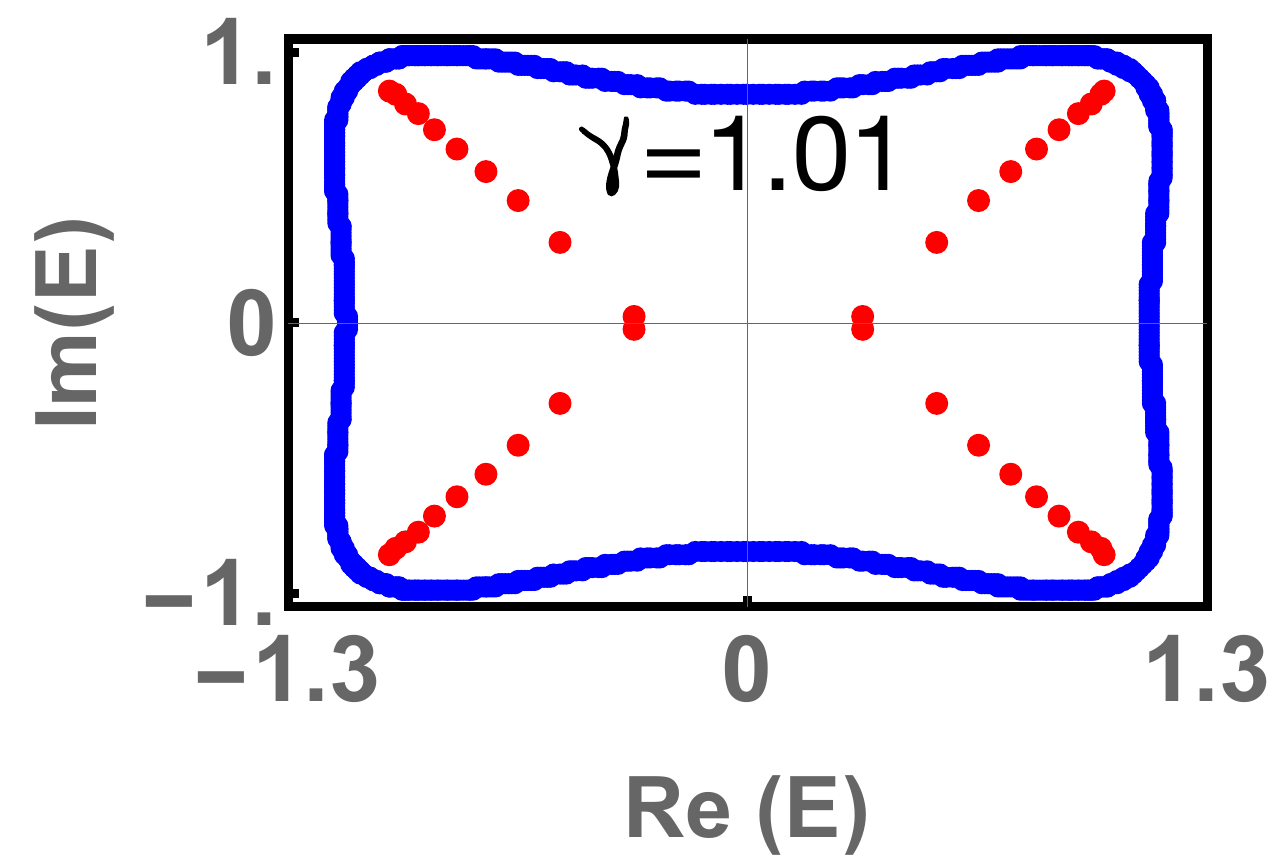}}
\subfigure[]{
	\includegraphics[width=0.145\textwidth,height=3cm]{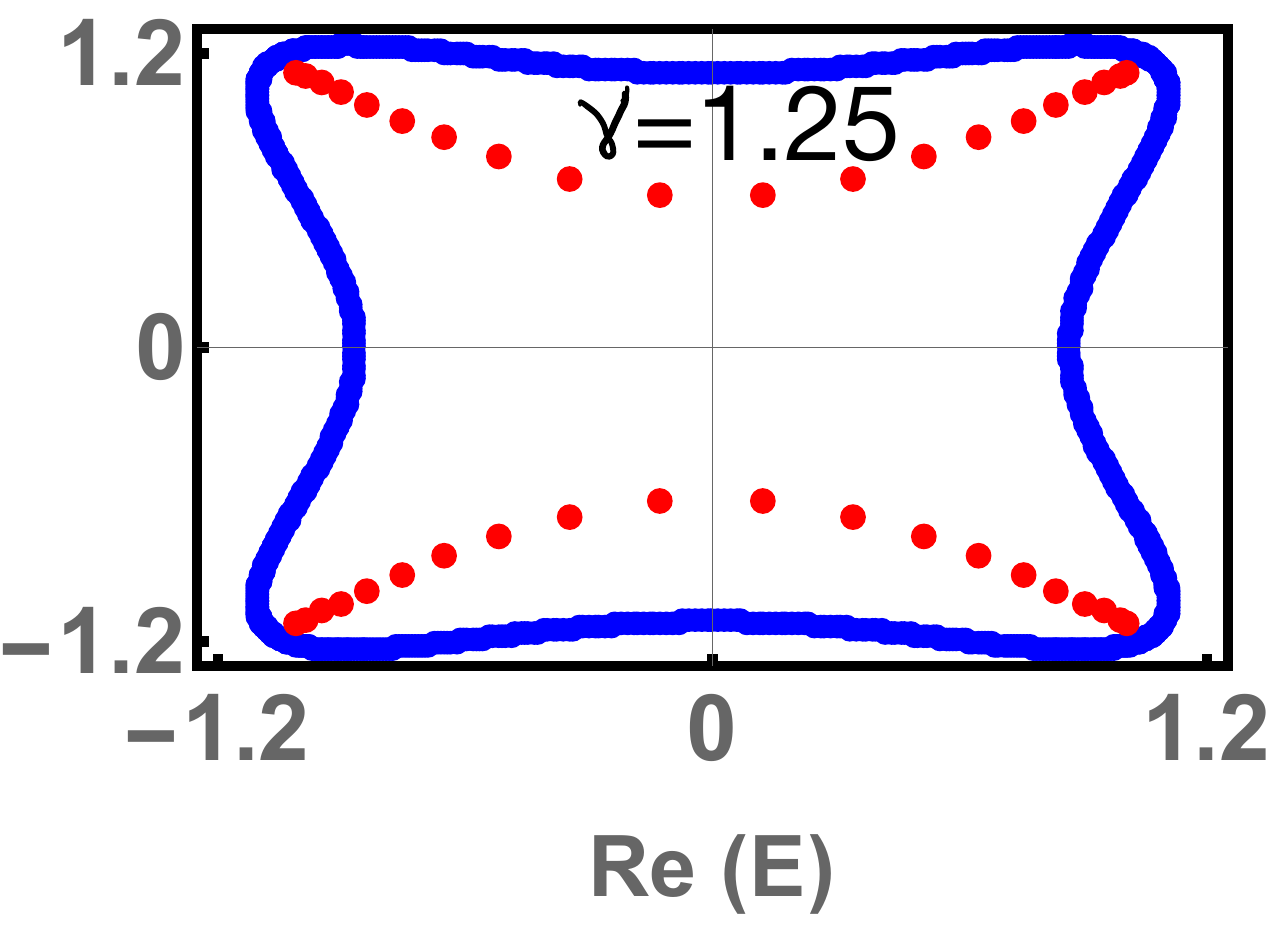} }
\subfigure[]{
	\includegraphics[width=0.145\textwidth,height=3cm]{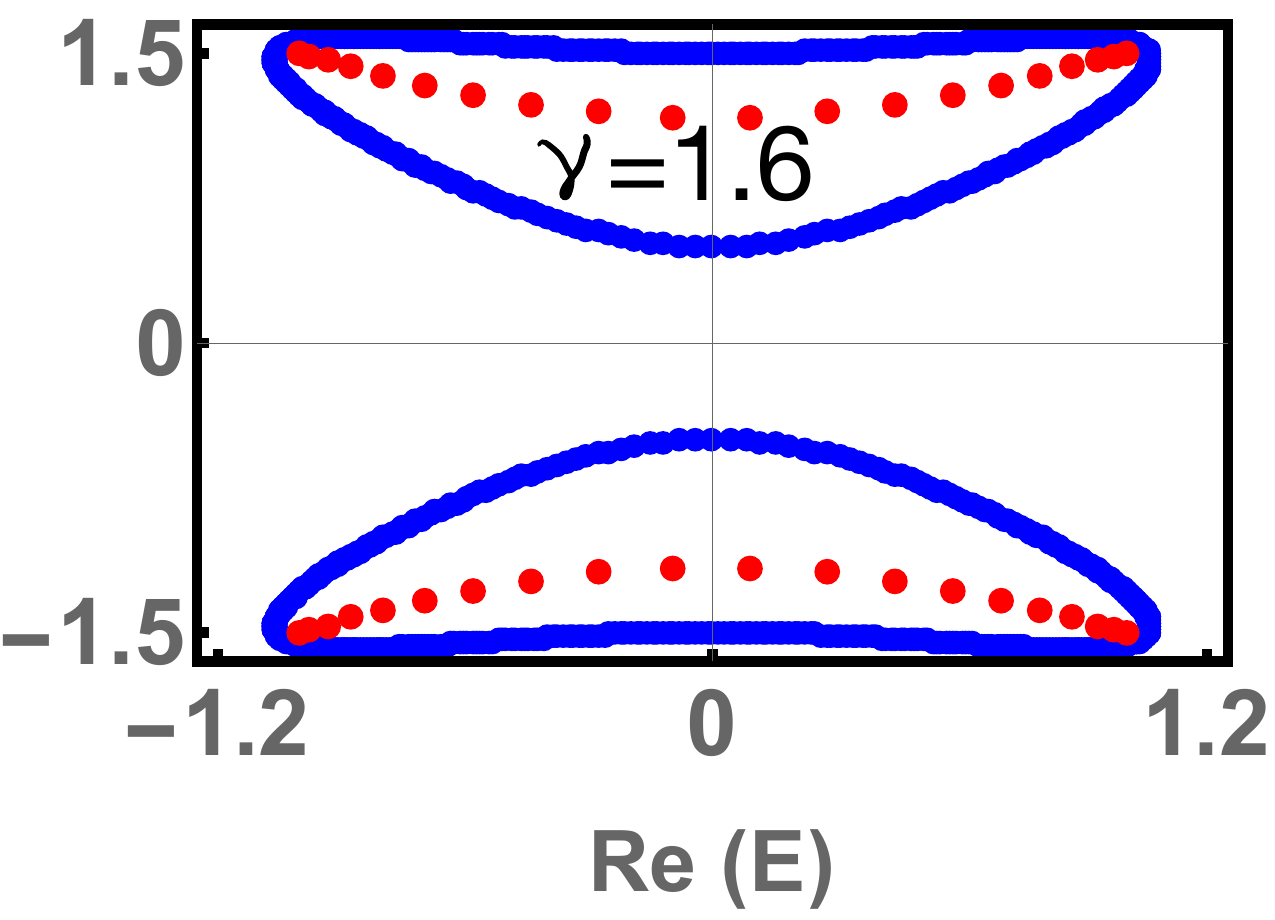}}
    \vspace{-1\baselineskip}
    \caption{\textbf{Spectral analysis with varying $\mathbf{\gamma}$ for non-Hermitian RM model: }(a) A schematic of asymmetric hopping-induced non-Hermitian RM model. The parameter $\gamma$ is used to introduce non-Hermiticity. (b) Plot of the complex spectrum of the non-Hermitian RM model. The OBC (red dots) and PBC (blue dots) spectra are plotted for values of 
	non-Hermitian parameter $\gamma =0.3,\, 0.8,\, 1.01,\, 1.25,\,$ and $1.6$. If we consider the base Energy to be $E_B=0$, then we observe that spectra in (d), (e) and (f) exhibit a Point gap, whereas in (c) and (g)  there is a Line gap in the PBC spectrum. By varying $\gamma$, we see that the PBC spectrum undergoes a transition from the Point gap to the Line gap. The OBC spectrum is plotted for system size $N = 20$ and the PBC spectrum is plotted in the thermodynamic limit.~The other parameters of the Hamiltonian take the value $v = 0.5$,~$w = 1$, and $\Delta = 0.25$.}
	\label{fig:complex_spectrum_size_40}
\end{figure}

\subsection{Static non-Hermitian case and anomalous pumping}\label{sub:unconpumping}
In this subsection, we analyze spectrum and anomalous pumping of boundary modes of the RM Hamiltonian as shown in Fig.~\ref{fig:complex_spectrum_size_40}a (cf. Eq.~(\ref{eqn:ham0})), considering all parameters $v,\,w,\,\Delta$ to be independent of time. For $\gamma\ne0$, the eigenspectrum is obtained to be
\begin{align}
E(k) = \pm\sqrt{v^{2}+w^{2}+\Delta^{2}-\gamma^{2}+2w\,(v\,\cos k -i\gamma\sin k)}
\end{align}
Evidently, the energy eigenvalue is complex for $0< k< 2\pi$. Fig.~\ref{fig:complex_spectrum_size_40} (b-g) demonstrates real and imaginary part of the energy spectrum using periodic boundary condition (PBC) and open boundary condition (OBC). Both the OBC (red dotted curve) and PBC (blue dotted curve) lie on the real axis for $\gamma=0$. For non-zero and moderate $\gamma$, the OBC spectrum remains purely real while the PBC spectrum becomes complex. This indicates that the non-Hermitian RM model considered here exhibits skin effect as opposed to the non-Hermitian RM model discussed in Ref.~\onlinecite{PhysRevA.98.042120}. We note that the OBC spectrum remains on the real axis as long as $v\le\gamma$. At $v=\gamma$, two EPs are obtained (not shown here). As $\gamma$ increases, the OBC spectrum becomes complex (cf. Fig ~\ref{fig:complex_spectrum_size_40}c). Subsequently, the disjoint PBC spectrum merges into a single curve. For reasonably large $\gamma$, the PBC spectrum again splits into two distinct curves, whereas the OBC spectrum changes its pattern as shown in Fig.~\ref{fig:complex_spectrum_size_40}d-f. Note that the spectra resemble the non-Hermitian SSH model as shown in Ref.~\onlinecite{PhysRevB.99.201103} except for the presence of edge modes at non-degenerate energies. 

\begin{figure}
	\centering
	\subfigure []
	{\includegraphics[width=0.15\textwidth,height=2.5cm]{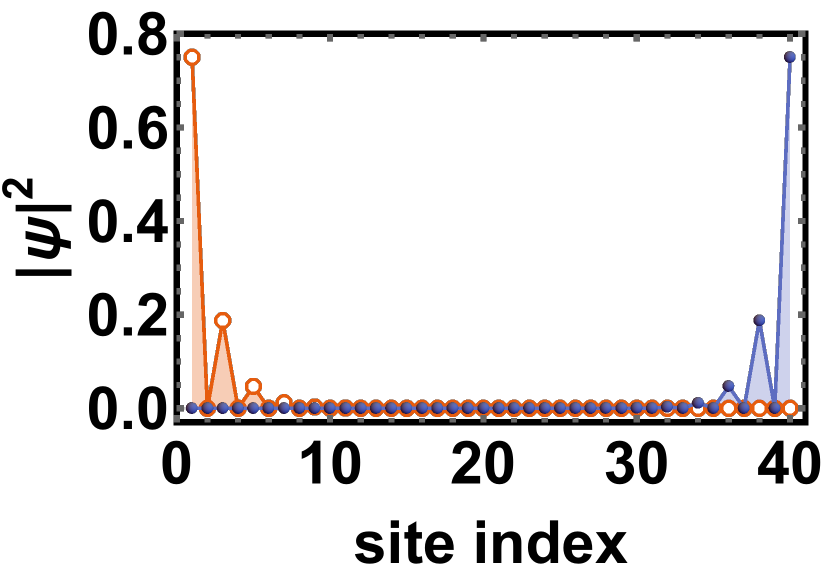}\label{fig:wavefunction_size_40_0}}
	\subfigure []
	{\includegraphics[width=0.15\textwidth,height=2.5cm]{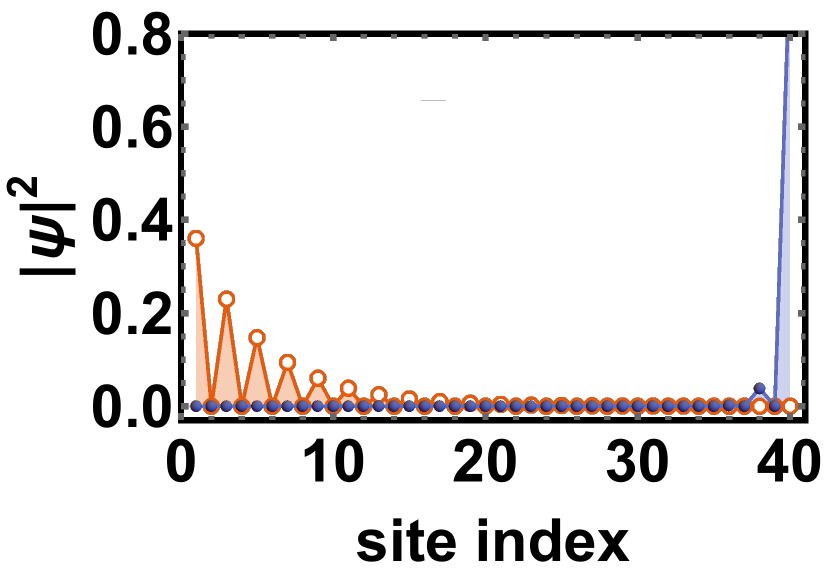}\label{fig:wavefunction_size_40_0.3}}
	\subfigure []
	{\includegraphics[width=0.15\textwidth,height=2.5cm]{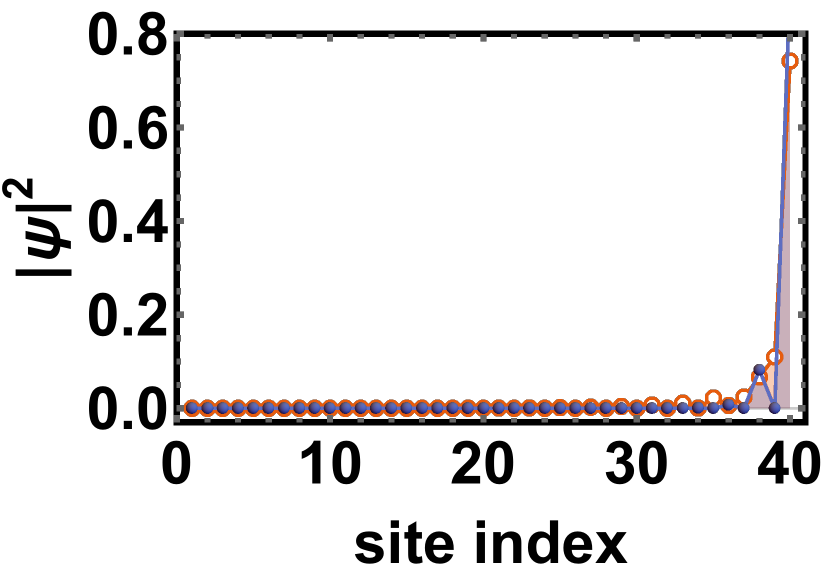}\label{fig:wavefunction_size_40_0.8}}
	\vspace{-1\baselineskip}
	\caption{\textbf{Anomalous edge state pumping: } Plot of the right wave function of the topological boundary states with $E =- \Delta$ (blue dots) and $E = \Delta$ (orange circle) for (a) $\gamma = 0$,  (b) $\gamma=0.3$ and (c) $\gamma=0.8$ with system size $N = 20$. The left localized edge state moves to the right boundary with the increase in the non-Hermitian parameter $\gamma$, leading to anomalous edge state pumping. All other parameters are the same as in Fig.~\ref{fig:complex_spectrum_size_40}.}
	\label{fig:wavefunction_size_40}
\end{figure}

In Fig.~\ref{fig:complex_spectrum_size_40}b, the red dots at $E = \pm\Delta$ on the real axis for $\gamma=0$ represent topological edge modes localized at the left and right boundary, respectively. The corresponding edge modes are shown in Fig.~\ref{fig:wavefunction_size_40}a. For $\gamma\ne 0$, both the skin modes and topological edge modes coexist and can be distinguished using GBZ as will be evident shortly. For finite $\gamma$, the degree of localization of the topological modes changes as compared to the case with $\gamma=0$. This is evident from the Fig.~\ref{fig:wavefunction_size_40}b. Additionally, the edge mode localized at the left boundary starts moving to the right with increasing $\gamma$, and it finally merges with the skin modes on the right near $\gamma\sim 1$. This gives rise to the topological state pumping due to the variation of $\gamma$, in contrast to the typical edge mode pumping in the dynamic RM model. Fig.~\ref{fig:wavefunction_size_40}b-\ref{fig:wavefunction_size_40}c illustrates the movement of the left edge mode from to the right. We note that the topological pumping discussed above follows a preferred direction since $v_r>v_l$. By choosing $v_r<v_l$,  one can change the direction of pumping. Interestingly, the evolution of the OBC spectrum in Fig.~\ref{fig:complex_spectrum_size_40} can also be obtained by tuning the system size $N$ while fixing the non-Hermitian parameter $\gamma$ (Appendix \ref{appen A}). This unique characteristic enables us to explore system-size-dependent topological phase transitions and the occurrence of anomalous pumping phenomena, both in static and dynamic scenarios, solely by altering the system size.

\subsection{Topological edge states and skin modes}\label{sub:analytics}

We next turn to find the exact analytical solutions for both the skin and topological modes using a finite-size GBZ scheme. This too
further explains the topological transition under
OBC, as shown in Fig.~\ref{fig:complex_spectrum_size_40}. We note that the generalized Brillouin zone (GBZ) is a useful tool for understanding topological physics in non-Hermitian systems. The GBZ spectrum and edge states analysis seems to provide sufficient information about a non-Hermitian system and to characterize topological aspects
\cite{PhysRevLett.121.086803,PhysRevLett.124.086801,PhysRevLett.125.226402,PhysRevB.105.045422}. In addition, there are proposals on experimentally realizable non-Hermitian models with asymmetric hopping for both interacting fermions and bosons\cite{PhysRevB.102.235151,PhysRevB.102.035153}. Interestingly, these studies are based on only spectrum analysis and edge modes physics. In terms of the experiment, finding the left and right eigenvectors is routine nowadays as pointed out in Ref.~\cite{RevModPhys.93.015005}. Thus we
believe that the complex energy spectrum together
with GBZ will suffice to extract essential physics in the non-Hermitian version.

To justify the different nature of the topological boundary and skin modes, we now provide an analytical solution of these modes following Ref.~\onlinecite{PhysRevLett.127.116801}. For simplicity, we restrict to modes associated with the real eigenvalues obtained from the OBC. We find the wavefunctions by solving the eigenvalue equation $H|\Psi\rangle = E|\Psi\rangle$ for an open chain using an ansatz $|\Psi\rangle \equiv (\phi_{1A},\phi_{1B},\phi_{2A},\phi_{2B},\dots,\phi_{NA},\phi_{NB})^{T}$ where $\phi_{n(A/B)} = z^{n}u_{(A/B)}$ with $n = 1,2,\dots,N$. This gives the two bulk recursion relations
\begin{eqnarray}
v_{r}\phi_{pA}-(\Delta + E)\phi_{pB} + w\phi_{(p+1) A} &=& 0,\label{eqn:bulk_recursion1}\\
w\phi_{pB}+(\Delta - E)\phi_{(p+1) A} + v_{l}\phi_{(p+1) B} &=& 0,
\label{eqn:bulk_recursion2}
\end{eqnarray}
where $p = 1,2,\dots,N-1$ and boundary relations
\begin{eqnarray}
(\Delta - E)\phi_{1A}+v_{l}\phi_{1B} &=& 0,\label{eqn:boundary_relation1}\\
v_{r}\phi_{NA}-(\Delta + E)\phi_{NB} &=& 0.
\label{eqn:boundary_relation2}
\end{eqnarray}
Solving the bulk equations together with the ansatz $|\Psi\rangle$, we obtain a quadratic equation in $z$ as
\begin{eqnarray}
v_{l}wz^{2} + z(v_{l}v_{r}+\Delta^{2}+w^{2}-E^{2}) + v_{r}w = 0.
\label{eqn:quadratic_z}
\end{eqnarray}
For real values of $\it{E}$,~Eq.~(\ref{eqn:quadratic_z}) leads to two roots $z_{1}$ and $z_{2}$ which are complex conjugate to each other when the coefficients of the quadratic equation are real. Thus assuming real quadratic coefficients,  we can write $z_{1} = |z|e^{i\Theta}$ and $z_{2} = |z|e^{-i\Theta}$. This gives energy eigenvalues
\begin{equation}
E = \pm\sqrt{v_{l}v_{r}+w^{2}+\Delta^{2}+2w\sqrt{v_{r}v_{l}}\cos \Theta}
\label{eqn:energy_eigenvalue}
\end{equation}
and $|z| = \sqrt{v_{r}/v_{l}}$. Further, the general solution of  $|\Psi\rangle$ is given as $|\Psi\rangle = s_{1}|\Psi_{1}\rangle + s_{2}|\Psi_{2}\rangle$ where $|\Psi_{(1,2)}\rangle = (\phi_{1A}^{(1,2)},\phi_{1B}^{(1,2)},\phi_{2A}^{(1,2)},\phi_{2B}^{(1,2)},\dots,\phi_{NA}^{(1,2)},\phi_{NB}^{(1,2)})^{T}$ with $\phi_{n(A/B)}^{(1,2)} = z_{(1,2)}^{n}u^{(1,2)}_{(A/B)}$ and $n = 1,2,\dots,N$. The values of $\Theta$ are obtained from the boundary relations in Eqs.~(\ref{eqn:boundary_relation1}-\ref{eqn:boundary_relation2}) and introducing the definition of $\phi$ in terms of $z$.~This lead to a transcendental equation

\begin{figure}
	\centering
	\subfigure []
	{\includegraphics[width=0.23\textwidth,height=3.5cm]{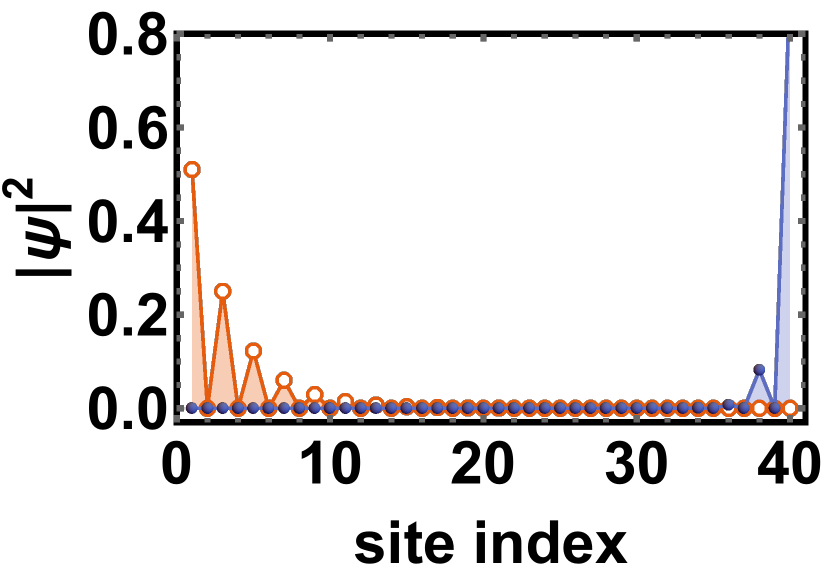}\label{fig:wavefunction_edge_16_0.2}}
	\subfigure []
	{\includegraphics[width=0.23\textwidth,height=3.5cm]{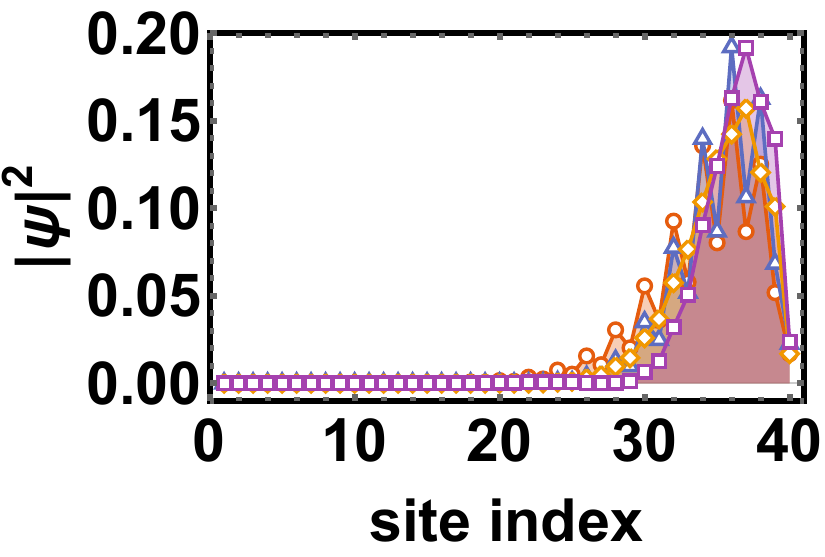}\label{fig:wavefunction_skin_16_0.2}}
	\vspace{-1\baselineskip}
	\caption{\textbf{Edge modes and skin modes localization:} Plot of the right wavefunction of the (a) topological boundary states and (b) skin modes only for a few energies $\gamma = 0.2$ and system size $N = 20$. The localized edge states are present only at one sublattice site either at A or B. In contrast, the skin modes are present at both the sublattice sites. All other parameters are the same as in Fig.~\ref{fig:complex_spectrum_size_40}}.
	\label{fig:wavefunction_edge_skin_size_16}
\end{figure}
\begin{equation}
\sin (N\Theta)= -\chi\sin ((N+1)\Theta),
\label{eqn:transcendental_equation_skin}
\end{equation}
where $\chi = \sqrt{v_{r}v_{l}}/w$. All real $\Theta$ solutions lie between $0$ and $\pi$. The endpoints are not included as we get $z_{1} = z_{2}$. For $\chi>\chi_c$, we obtain one complex and $(N-1)$ real solutions for $\Theta$, while for $\chi<\chi_c$ all solutions are real. The {\it complex} solution $\Theta=\pi+i\theta$, in turn, leads to the solutions for the topological edge modes, where $\theta$ satisfies the transcendental equation
\begin{equation}
N\theta = \frac{1}{2}\log \left(\frac{\chi e^{-\theta} - 1}{\chi e^{\theta}-1}\right).
\label{eqn:transcendental_equation_edge}
\end{equation}
In the thermodynamic limit, $\theta \to -\ln (\chi)$.~With this,~Eq.~(\ref{eqn:energy_eigenvalue}) gives $E \to \pm \Delta$, representing the edge  modes. The corresponding wavefunctions are obtained to be 
\begin{eqnarray}
\phi_{nA} &=& (-1)^{n+1}\frac{2s_{1} u^{(1)}_{A}}{\chi-e^{\theta}}\left[|z|^{n}(\chi\sinh (n\theta)-\sinh ((n-1)\theta))\right]\label{eqn:edge_eigenstate1}\nonumber\\
\phi_{nB} &=& (-1)^{n+1}\frac{2s_{1}\chi(E-\Delta) u^{(1)}_{A}}{v_{l}(\chi - e^{\theta})}\left[|z|^{n}\sinh (n\theta)\right].
\label{eqn:edge_eigenstate2}
\end{eqnarray} 

with $n= 1,2,\dots,N$. Evidently, the edge state probabilities vanish either at sublattices A or sublattices B. 

In contrast to the topological boundary modes, the energy eigenvalues of skin modes are obtained by substituting the {\it real} $\Theta$ solution in Eq.~(\ref{eqn:energy_eigenvalue}). The corresponding eigenstates are   
\begin{eqnarray}
\phi_{nA} &=& \frac{2is_{1} u^{(1)}_{A}}{\chi+e^{-i\Theta}}\left[|z|^{n}(\chi\sin (n\Theta)+\sin ((n-1)\Theta))\right]\label{eqn:skin_eigenstate1}\nonumber\\
\phi_{nB} &=& \frac{2is_{1}\chi(E-\Delta) u^{(1)}_{A}}{v_{l}(\chi + e^{-i\Theta})}\left[|z|^{n}\sin (n\Theta)\right]
\label{eqn:skin_eigenstate2}
\end{eqnarray}
with $n= 1,2,\dots,N$. In our case $|z| > 1$ (since $v_{r} > v_{l}$) which causes the eigenstates of all skin modes to localize at the right boundary. These skin modes are present on both the sites (A and B) as shown in Fig.~\ref{fig:wavefunction_edge_skin_size_16}, corroborating the distinct nature of the topological boundary and skin modes. Note that the wavefunctions obtained here denote the {\it right} wavefunctions introduced earlier.  

A similar analysis for the {\it left} wavefunction gives left boundary-localized skin modes instead of right boundary. This is easy to see since the left wave function is obtained using an eigenvalue equation with $H^{\dag}$ instead of $H$. It is equivalent to performing an interchange of $v_{l}$ and $v_{r}$ in the above analysis which in turn gives $|z| < 1$, and hence we obtain results for left boundary modes.

\subsection{GBZ scheme and complex energy spectrum}
\label{GBZ_complex}

In this subsection, we provide solutions for {\it complex} energy spectrum and study the evolution of topological boundary modes in the GBZ spectrum. We note that the prescription of finding $\Theta$ solutions and related quantities discussed above and also presented in Ref.\onlinecite{PhysRevLett.127.116801} are limited to real or purely imaginary spectrum only. Since the quadratic coefficients of Eq.~(\ref{eqn:quadratic_z}) are complex, we need a more general approach, as given below, for complex spectrum. The complex eigenvalue solutions can be obtained using  

\begin{figure}
	\centering
	\subfigure []
	{\includegraphics[width=0.155\textwidth,height=3cm]{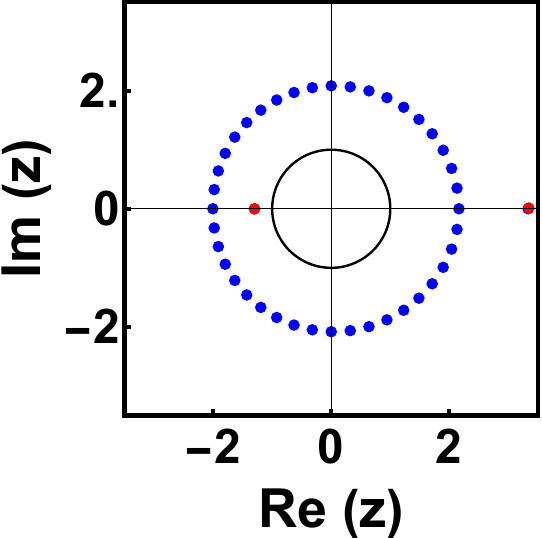}\label{fig:gbz_0.8_20}}
	\subfigure []
	{\includegraphics[width=0.155\textwidth,height=3cm]{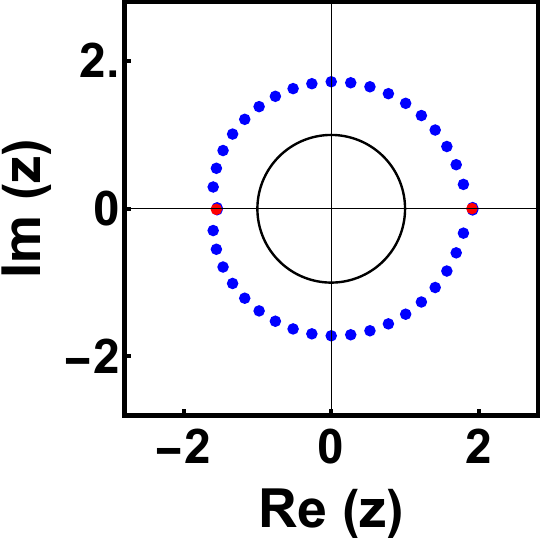}\label{fig:gbz_1.01_20}}
	\subfigure []
	{\includegraphics[width=0.155\textwidth,height=3cm]{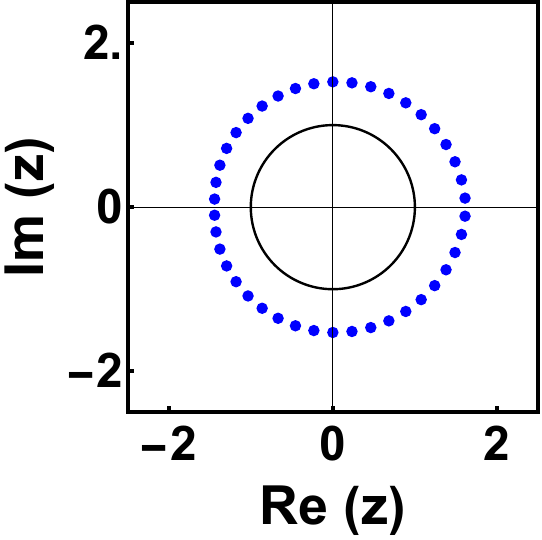}\label{fig:gbz_1.01_50}}\\
	\vspace{-1\baselineskip}
	\subfigure []
	{\includegraphics[width=0.155\textwidth,height=3cm]{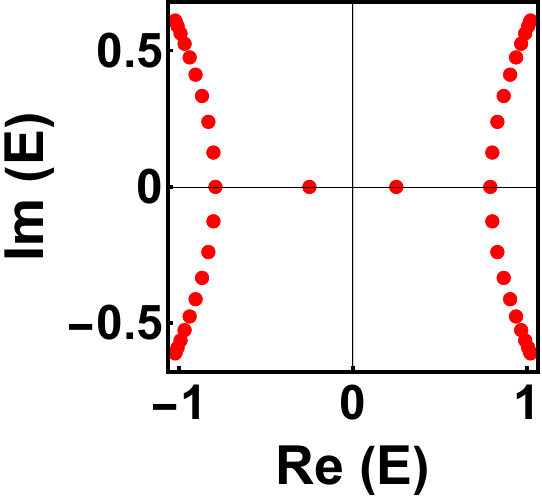}\label{fig:gbz_spectrum_0.8_20}}
	\subfigure []
	{\includegraphics[width=0.155\textwidth,height=3cm]{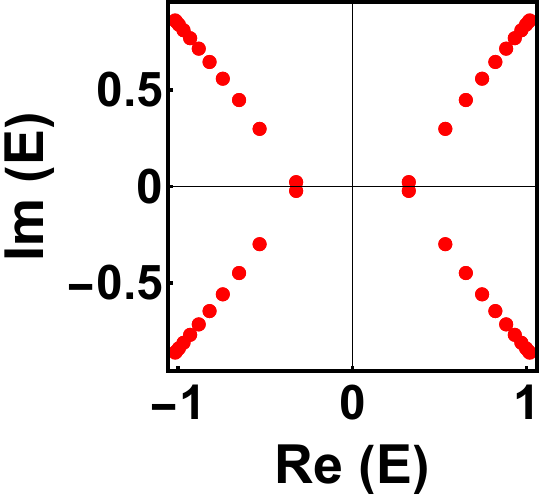}\label{fig:gbz_spectrum_1.01_20}}
	\subfigure []
	{\includegraphics[width=0.155\textwidth,height=3cm]{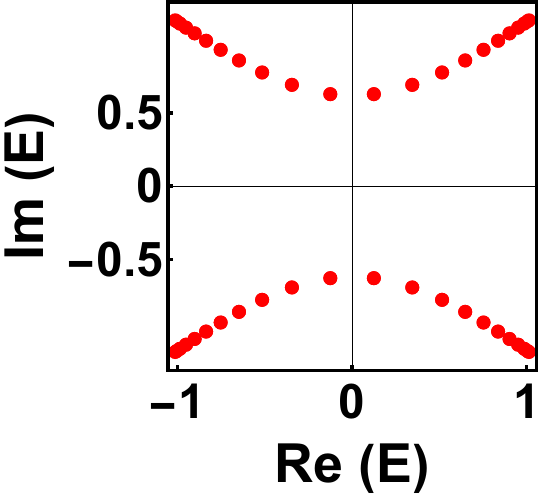}\label{fig:gbz_spectrum_1.25_20}}
    \vspace{-1\baselineskip}
    \caption{\textbf{GBZ and complex spectrum of the non-Hermitian RM model:} The GBZ solutions (a, b, c) and corresponding OBC spectrum (d, e, f) are plotted for values of non-Hermitian parameter   (a, d)~ $\gamma=0.8$, (b, e)~ $\gamma=1.01$, and (c, f)~$\gamma=1.25$. The plots are at system size $N = 20$. The black curves represent the GBZ solutions for the parent Hermitian system in the thermodynamic limit.  All other parameters are the same as in Fig.~\ref{fig:complex_spectrum_size_40}.}
	\label{fig:gbz_complex_spectrum_size_40}
\end{figure}
\vspace{-1cm}

\begin{equation}
E = \pm\sqrt{v_{l}v_{r}+w^{2}+\Delta^{2}+wv_{l}(z_{1} +z_{2})},
\label{eqn:complex_energy_eigenvalue}
\end{equation}
where $z_{1}$ and $z_{2}$ are solutions of equations
\begin{eqnarray}
z_{1}z_{2} &=& \frac{v_{r}}{v_{l}}\label{eqn:gbz0},\\
v_{l}(z_{2}^{N+1}-z_{1}^{N+1}) &=& -w(z_{2}^{N}-z_{1}^{N}).
\label{eqn:gbz}
\end{eqnarray}
Note that, the sum and product of roots of quadratic Eq.~(\ref{eqn:quadratic_z}) lead to Eq.~(\ref{eqn:complex_energy_eigenvalue}) and~(\ref{eqn:gbz0}), respectively. Also, Eq.~(\ref{eqn:gbz}) results from the boundary relations in Eqs.~(\ref{eqn:boundary_relation1}-\ref{eqn:boundary_relation2}) together with the definition of $\phi$ in terms of $z$.
Then the values of $z_{1}$ and $z_{2}$ obtained from Eq.~(\ref{eqn:gbz0}-\ref{eqn:gbz}) represent the GBZ solutions. Clearly, the solution set for both $z_{1}$ and $z_{2}$ are the same since they satisfy the same polynomial equations of order 2(N+1). We restrict to only one set of solutions~(say $z_{1}$). They are plotted with their corresponding complex spectra (cf. Eq.~(\ref{eqn:complex_energy_eigenvalue})) in Fig.~\ref{fig:gbz_complex_spectrum_size_40}. Notice that the 
solutions pointed out by red dots represent edge modes. They are well separated and asymmetrically placed {\it{ w.r.t}} the zone center in the extended Brillouin zone, indicating the different degrees of localization of the boundary modes. This can be understood by mapping $z$ with $e^{i k}$ where $k\rightarrow k+i\kappa$. Accordingly, the localization of these topological boundary modes differs from the skin modes. The result obtained in Fig.~\ref{fig:wavefunction_size_40} also corroborates this. As we increase $\gamma$, the solutions for edge modes merge with the skin modes on the real energy axis to form an EP. It also results in cusp and kinks in the GBZ spectrum. For reasonably large values of $\gamma$, we recover the GBZ spectrum without any distortion, indicating the absence of edge modes due to strong non-Hermiticity. 

Moreover, we also differentiate between the approach followed to obtain purely real or purely imaginary spectrum and complex spectrum with both real and imaginary parts being non-zero. In short, we justify that the $z_{1} = |z|e^{i\theta}$ and $z_{2} = |z|e^{-i\theta}$ forms a pair of complex conjugate roots of quadratic Eq.(\ref{eqn:quadratic_z}), only if the energy eigenvalues are purely real or purely imaginary since this will lead to a quadratic equation with real coefficients. However, if we have complex energy (with both real and imaginary parts non-zero) then the coefficients of the quadratic equation are no longer real and hence the roots may not occur as complex conjugate pairs which in turn leads to Eq.~(\ref{eqn:complex_energy_eigenvalue}). The approach discussed in Ref.~\onlinecite{PhysRevLett.127.116801} has missed this crucial point and thus they do not observe any topological transition associated with complex spectrum (see Fig.(S5) of Ref.~\onlinecite{PhysRevLett.127.116801}) as shown in Fig.~\ref{fig:gbz_complex_spectrum_size_40} of our work.

\begin{figure}
	\centering
	\subfigure []
	{\includegraphics[width=0.155\textwidth,height=3cm]{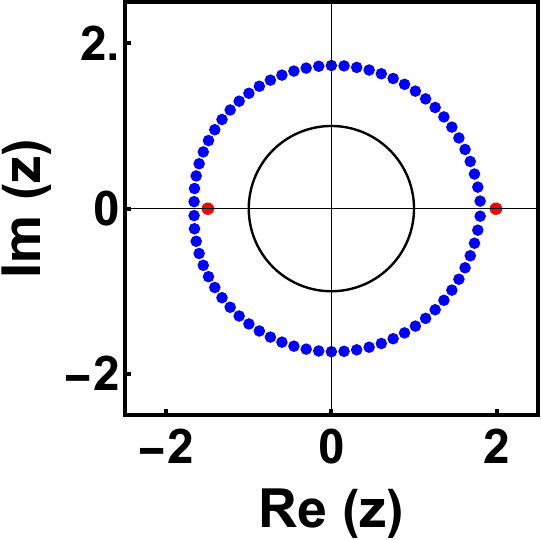}\label{fig:gbz_0.8_2011}}
	\subfigure []
	{\includegraphics[width=0.155\textwidth,height=3cm]{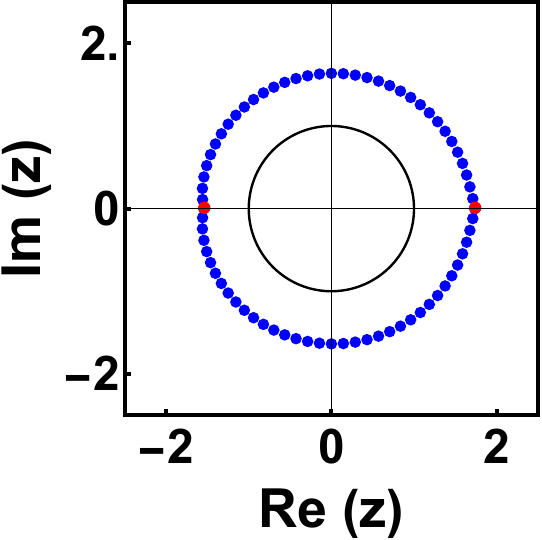}\label{fig:gbz_1.01_2011}}
	\subfigure []
	{\includegraphics[width=0.155\textwidth,height=3cm]{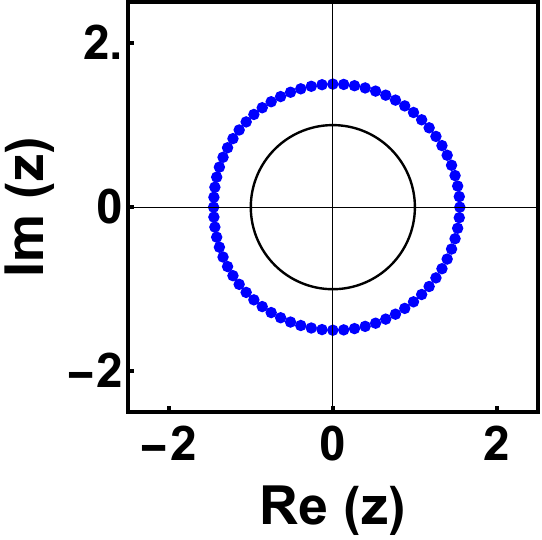}\label{fig:gbz_1.01_5011}}\\
	\vspace{-1\baselineskip}
	\subfigure []

	{\includegraphics[width=0.15\textwidth,height=3cm]{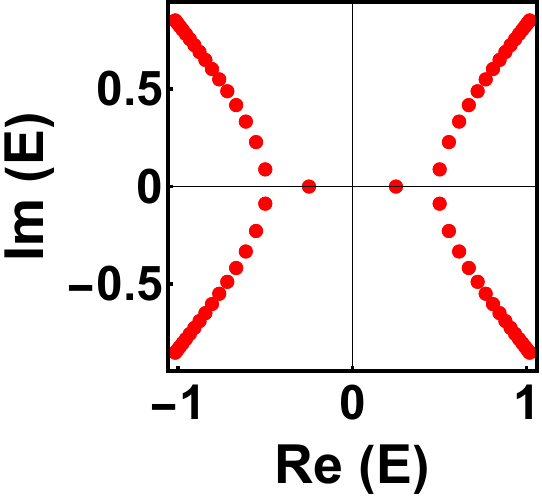}\label{gamma11}}
	\subfigure []
	{\includegraphics[width=0.145\textwidth,height=3cm]{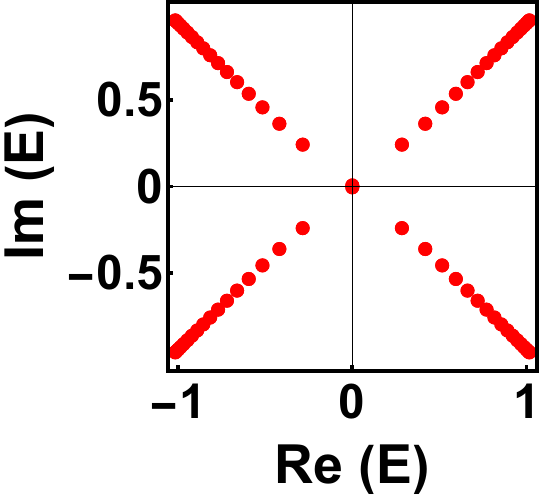}\label{gamma22}}
	\subfigure []
	{\includegraphics[width=0.15\textwidth,height=3cm]{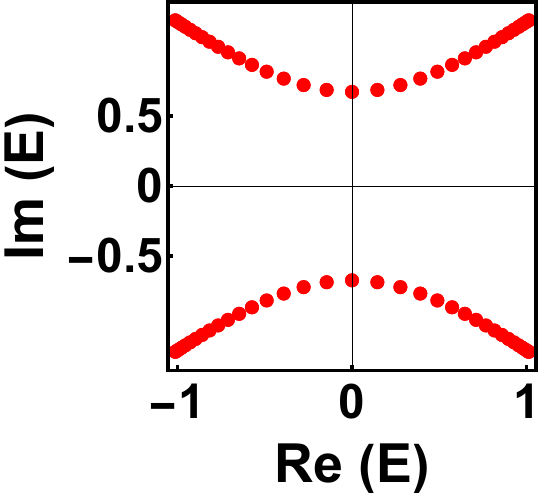}\label{gamma33}}
	\vspace{-1\baselineskip}
	\caption{ \textbf{GBZ spectrum and Topological boundary modes for odd system size: } The GBZ solutions (a, b, c) for odd system size are plotted. The black curves represent the GBZ solutions for the parent Hermitian system in the thermodynamic limit. The corresponding evolution of topological boundary modes for odd  $N$ in the complex spectrum, due to variation of $\gamma$. For $\gamma=\gamma_{c}$  a second order EP forms at $E=0$ (b). The spectra (d, e, f) are plotted for values of non-Hermitian parameter $\gamma = 1$, $\gamma_{c} = 1.095$ and $\gamma=1.3$ respectively. The system size is taken to be $N=35$ and all other parameters are the same as in Fig.~\ref{fig:complex_spectrum_size_40}.}
	\label{fig:wavefunction_size_odd}
\end{figure}
\subsection{Second order exceptional points for odd system-size}

\label{second order EP, system size}

It is well known that the bulk properties of the non-Hermitian Hamiltonian are significantly affected by a small change in the boundary\cite{PhysRevA.99.022127,PhysRevB.99.155431}. Very recently, it has been demonstrated that topological phase transition can be obtained by varying the system size in a coupled non-Hermitian system \cite{li2020critical} exhibiting NHSE and nonreciprocal chains with impurity \cite{PhysRevResearch.2.043167}. In view of this, it is imperative to assess the fate of the obtained results discussed above with the variation of system size. While the numerical approach for addressing boundary and system size-sensitive issues can produce unreliable results due to numerical errors \cite{PhysRevLett.127.116801}, we investigate the same using the analytical approach discussed in the preceding sections.  Interestingly, for our model, we find that when the system size ($N$) is odd, an EP of second-order (merging of two distinct edge modes) emerges precisely at $E=0$ for a specific critical value of $\gamma=\gamma_c$. Note that the nature of this EP is completely different from the case when $N$ is even as evident from Fig.~(\ref{fig:complex_spectrum_size_40}d-f)). The critical value $\gamma_c$ for a fixed configuration of the parameters $(v, w, \Delta)$ can be obtained using  (\ref{eqn:quadratic_z}) and (\ref{eqn:gbz0})-(\ref{eqn:gbz}):
\begin{eqnarray}
v_{l}wz_{1}^{2} + z_{1}(v_{l}v_{r}+\Delta^{2}+w^{2}) + v_{r}w = 0\label{eqn:EP1},\\
v_{l}z_{1}^{2N+2}+wz_{1}^{2N+1}-wr^{N}z_{1}-v_{l}r^{N+1}=0
\label{eqn:EP2}
\end{eqnarray}
Here, $r=\dfrac{v_r}{v_l}$. Solving Eqs. (\ref{eqn:EP1}) and (\ref{eqn:EP2}), we obtain the exact value of $\gamma_{c}$ for which the EP emerges at $E=0$. It is worth noting that the solutions of these two equations yield two real values for $\gamma$ ($-\gamma_{c}$, $\gamma_{c}$) when $N$ is odd, however for even $N$, no real solutions exist. This particular characteristic implies that the EP originating at zero energy is a distinctive feature of odd system size. In Fig.~\ref{fig:wavefunction_size_odd}, the complex energy spectrum shows the evolution of the boundary modes and  
the EP emerges at $E=0$ for $\gamma_{c}=1.095$. These findings highlight the critical roles played by both the system's boundaries and its system size in inducing boundary-sensitive effects.

\begin{figure}
    \centering
    \includegraphics[width =0.4\textwidth]{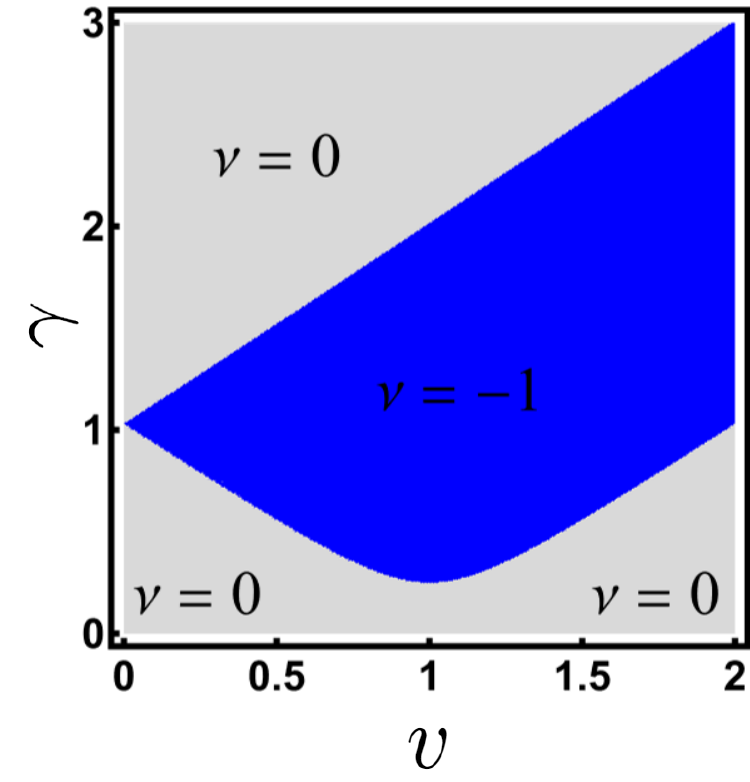}
    \caption{\textbf{Point gap and Line gap phases: }Phase diagram in the parameter $(v-\gamma)$ space representing Point gap and Line gap. In the Blue region $\nu=-1$, indicating the presence of a Point gap in the system. The other parameters are $w=1$ and $\Delta=0.25$.}
    \label{fig:Phase diagram}
\end{figure}

\subsection{Line gap and point gap}
\label{lp topology}
In this subsection, we provide a precise characterization of the gaps in the complex energy spectrum, shown in, Fig.~\ref{fig:complex_spectrum_size_40}. Unlike Hermitian systems, the complex energy eigenvalues of non-Hermitian Hamiltonians do not exhibit a natural ordering from the lowest to the highest values. Thus defining an energy gap in such systems becomes non-trivial. It is well known that two distinct types of complex energy gaps exist in non-Hermitian spectra, namely the point gap and the line gap. A point gap can be characterized as a specific point within the complex plane with no energy eigenstates. Any non-Hermitian Hamiltonian possessing a point gap can be continuously deformed into a unitary matrix while preserving both the point gap and its associated symmetries \cite{PhysRevX.9.041015, PhysRevLett.124.086801}. In contrast, a line gap is defined as an infinite line within the complex plane without any crossing in eigenenergies. In addition, it is known that a non-Hermitian Hamiltonian featuring a line gap can be converted to a Hermitian Hamiltonian through a unitary transformation provided that the line gap and its corresponding symmetries are preserved. Therefore, systems exhibiting a point gap do not have an equivalent Hermitian counterpart and for such a system, the \textit{winding invariant} can be expressed as (\cite{PhysRevB.103.045420,PhysRevLett.123.206404})

\begin{equation}
     \nu=\frac{1}{2\pi i} \sum_{n=1}^2\int_{0}^{2\pi} ~dk ~\partial_{k} \log[E_{n}(k)-E_B],
\end{equation}
where $\nu$ represents the winding number of $E_{n}(k)$ around the base point $E_B$, $n$ denotes the band index, and the summation is over all the bands. 
For systems exhibiting a point gap, the winding invariant will yield $ \nu=\pm 1$, whereas for a spectrum with a line gap results in $\nu=0$. The sign of the winding invariant depends upon the direction of the winding. In Fig.~\ref{fig:complex_spectrum_size_40}(d-f) we see that there is a clear point gap in the spectrum, whereas in Fig. \ref{fig:complex_spectrum_size_40} (c, g) there is a line gap separating the spectral loops, which suggests that the transition from the line gap to the point gap can be achieved by tuning $\gamma$. Interestingly, the same can also be achieved by varying the other parameters of the system. Fig.~\ref{fig:Phase diagram}, evidences that the transition from a point gap to a line gap can also be achieved by varying the parameter $v$.

\begin{figure}
	\centering
	\subfigure []
	{\includegraphics[width=0.15\textwidth,height=2.5cm]{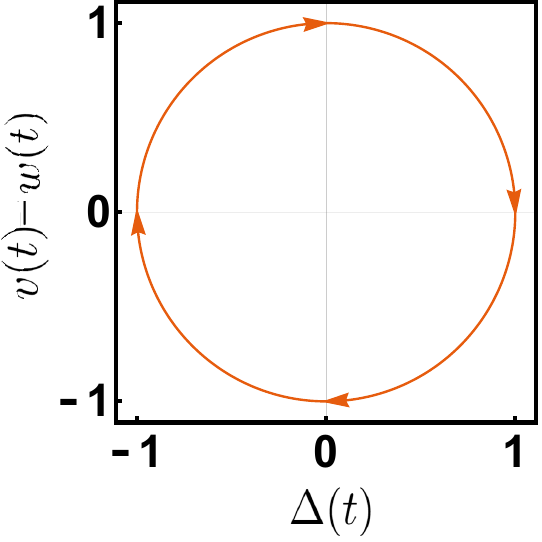}\label{fig:wavefunction_size_40_0_1}}
	\subfigure []
	{\includegraphics[width=0.15\textwidth,height=2.5cm]{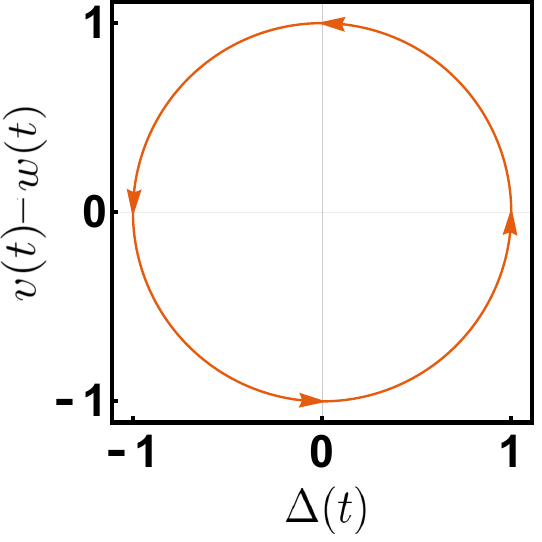}\label{fig:wavefunction_size_40_0.3_1}}
	\subfigure []
	{\includegraphics[width=0.15\textwidth,height=2.5cm]{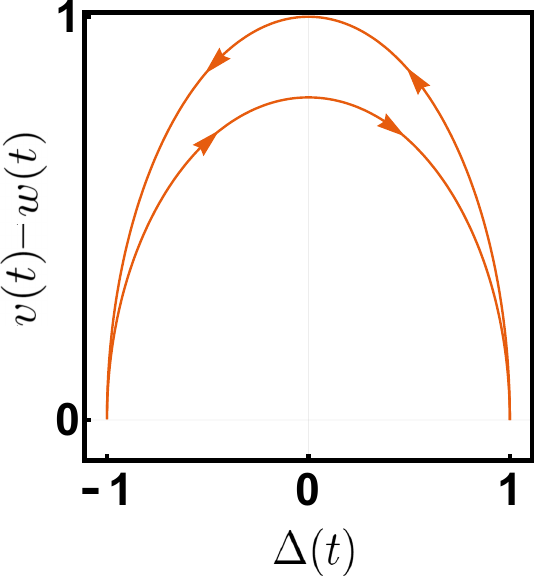}\label{fig:wavefunction_size_40_0.8_1}}
	\vspace{-1\baselineskip}
	\caption{\textbf{Driving protocols in parameter space: } Schematics of (a) clockwise non-trivial protocol (b) anticlockwise non-trivial protocol and (c) trivial protocol. }
	\label{fig:protocols}
\end{figure}

\begin{figure}	
\subfigure []	
{	\includegraphics[width=0.15\textwidth,height=3.0cm]{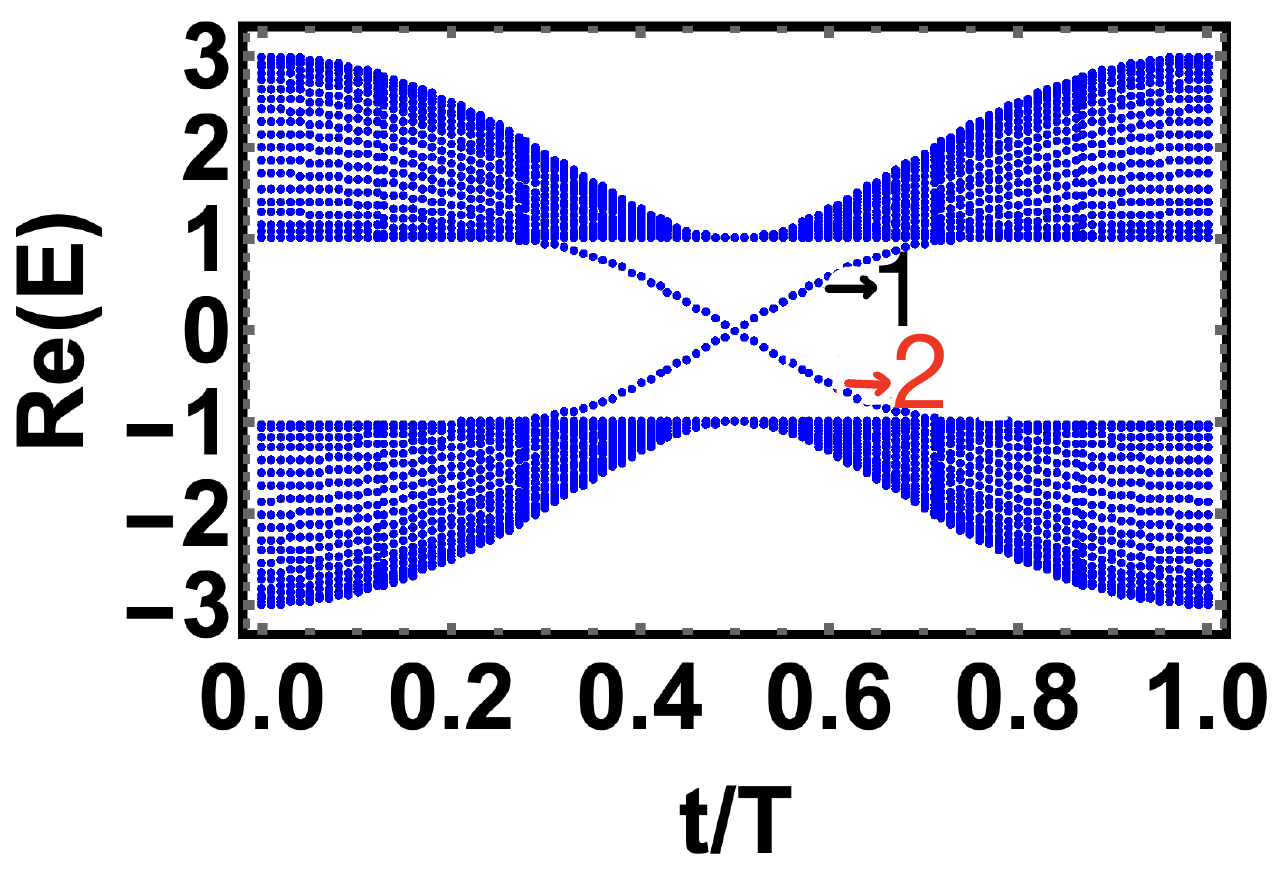}}
\subfigure []
	{\includegraphics[width=0.15\textwidth,height=3.0cm]{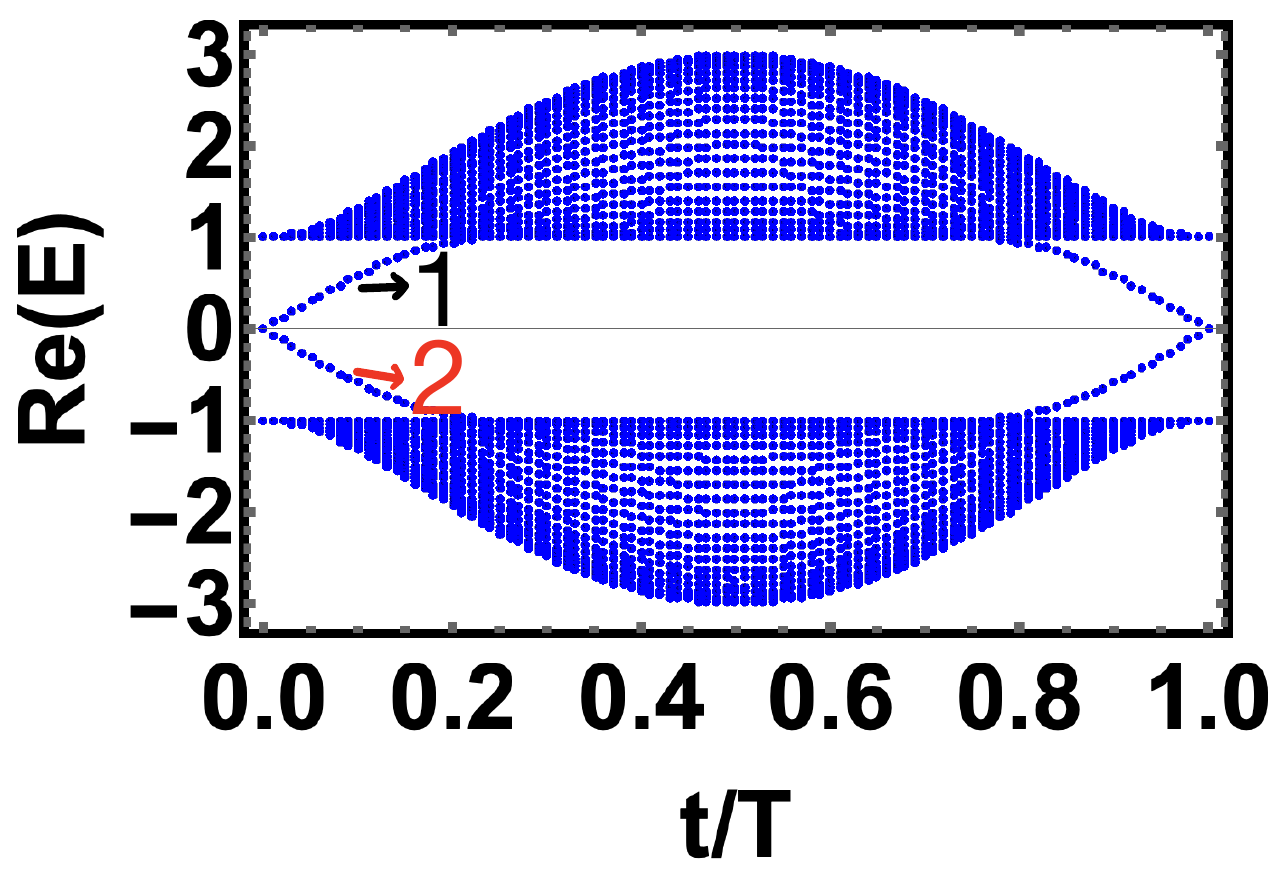}}
	\subfigure []{
	\includegraphics[width=0.15\textwidth,height=3.0cm]{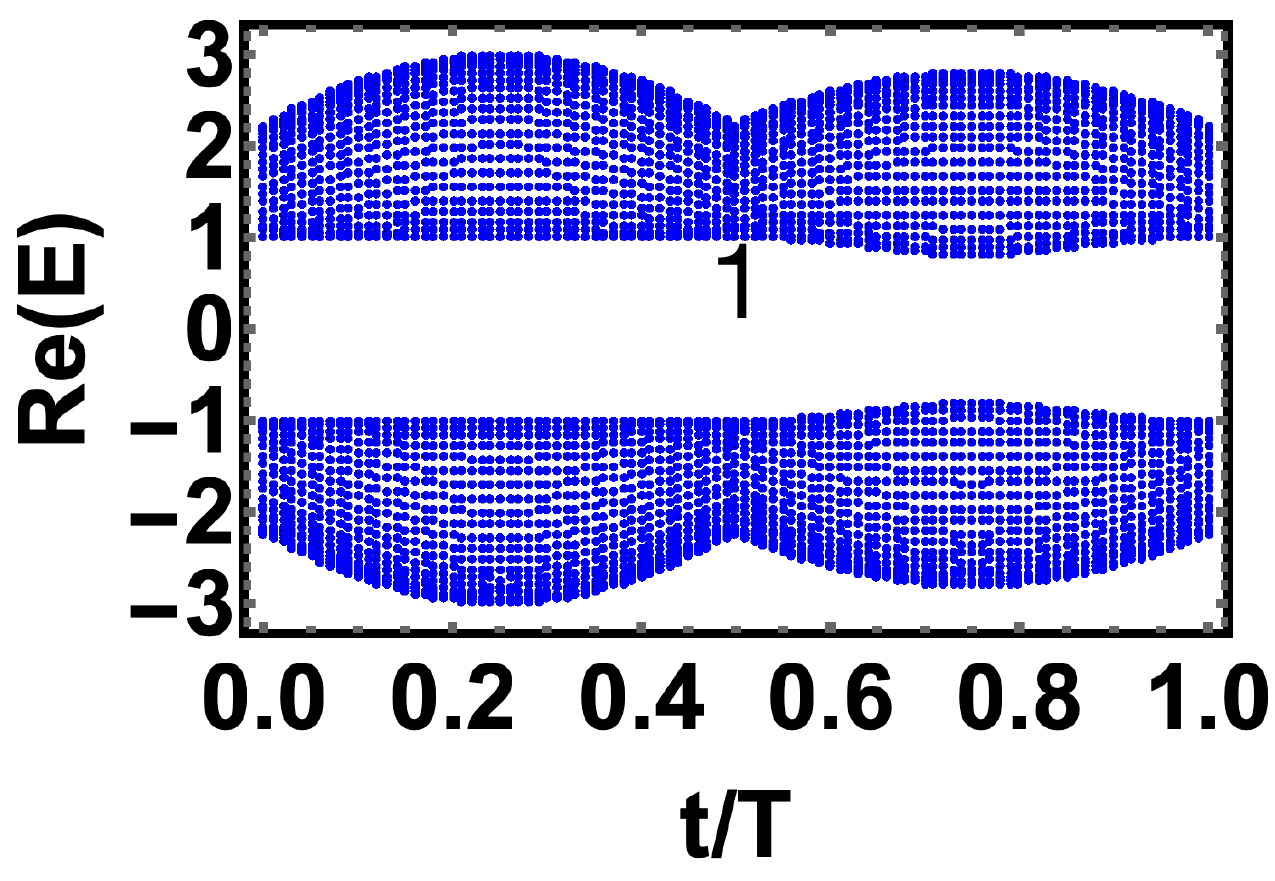}}\\
\vspace{-1\baselineskip}
\subfigure []{
	\includegraphics[width=0.15\textwidth,height=3.0cm]{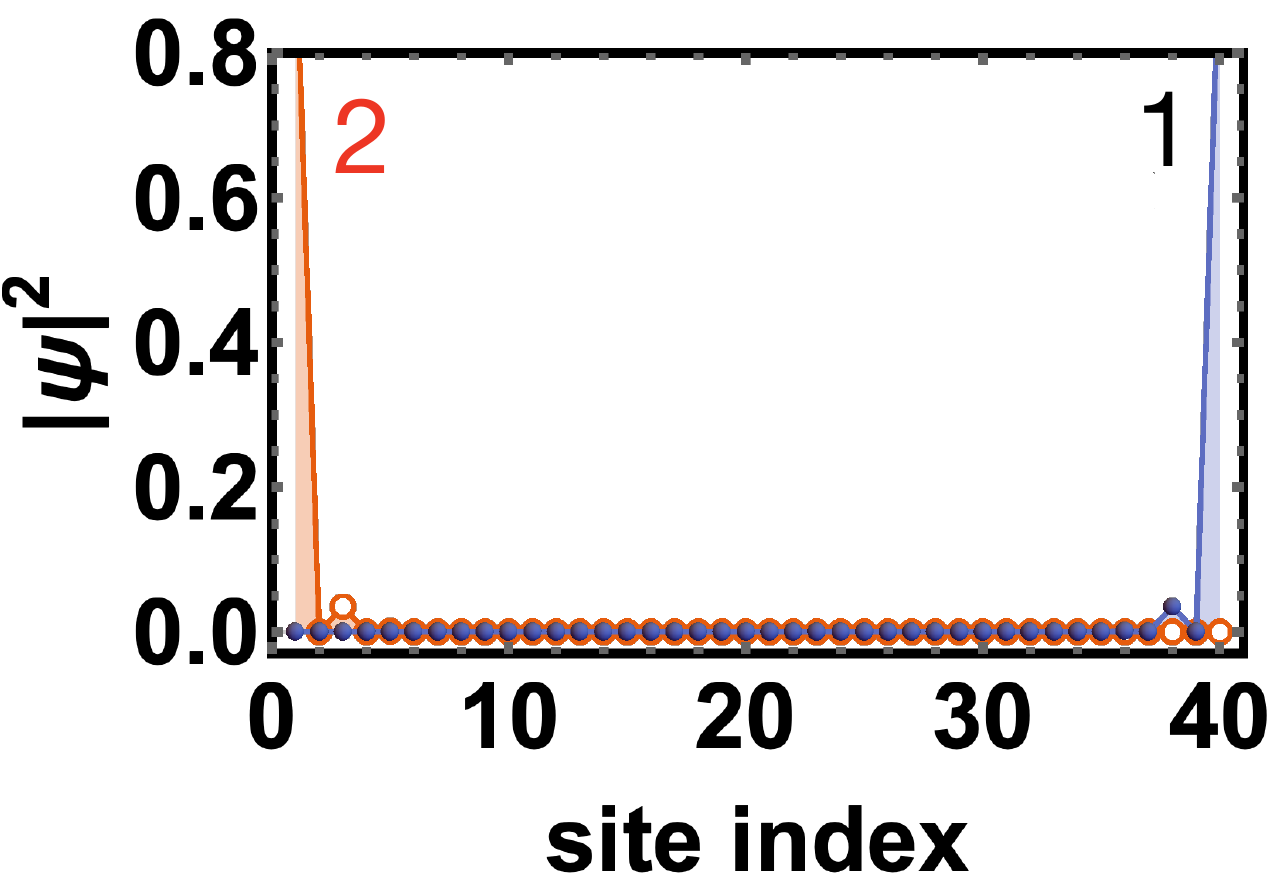}}
\subfigure []{
	\includegraphics[width=0.15\textwidth,height=3.0cm]{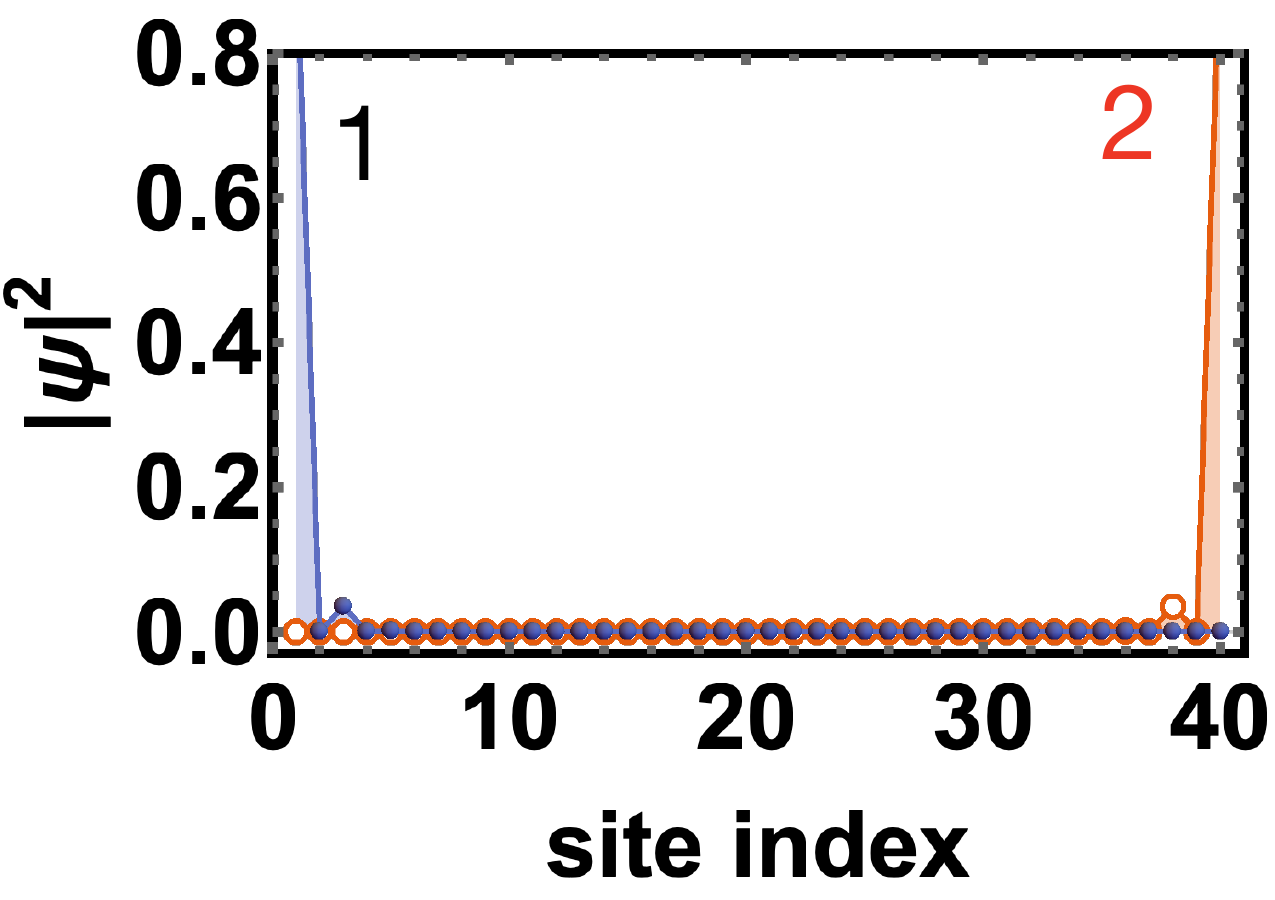}}
\subfigure []{
	\includegraphics[width=0.15\textwidth,height=3.0cm]{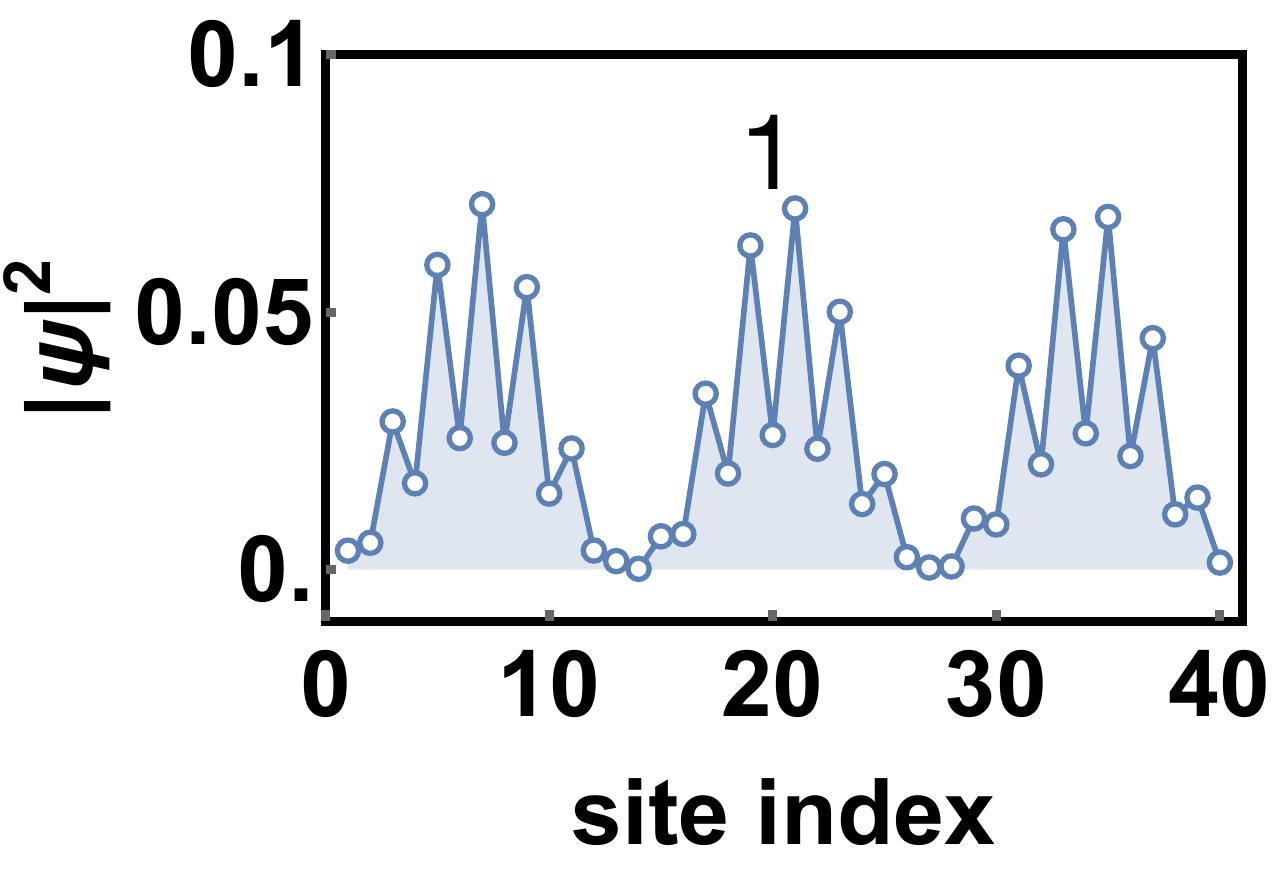}}
\vspace{-0.8\baselineskip}
	\caption{\textbf{Energy spectrum and smooth pumping in Hermitian RM model: }Instantaneous energy (a-c) are plotted for the Hermitian RM model with three different pumping protocols (a) clockwise, (b) anti-clockwise, and (c) trivial, for system size $N = 20$. (a) and (b) are non-trivial protocols with winding numbers 1 and -1 while protocol (c) is trivial and has winding number 0. The pumping is present in the case of non-trivial protocols but not in trivial one. The net particle current direction of (b) is opposite to that of (a). The wavefunctions of two topological boundary states are plotted for clockwise (d) and anticlockwise (e) non-trivial protocols at $\gamma = 0$ for two instants of times $t/T=0.6$ and $t/T=0.1$, respectively. For the trivial protocol, the wavefunction is plotted for bulk energy at $t/T=0.5$ (f).}\label{fig:Hermitian_RM}\end{figure}

\section{Dynamic RM model and charge pumping}\label{sec:dynamicalRM}
Having discussed the static case, we first review the dynamic Hermitian RM model, allowing periodic and adiabatic modulation of the model parameters. It turns out that the charge pumping in the dynamic RM model depends on the driving protocols. We shall thus consider three driving protocols to study topological charge pumping in the non-Hermitian RM model.

\subsection{Hermitian RM model and different driving protocols}\label{subsec:review}
We consider three driving protocols  (a) clockwise non-trivial protocol: $v(t) = 1 + \cos (\omega\, t)$, $w(t) = 1$, and $\Delta(t) = \sin (\omega\, t)$; (b) anticlockwise non-trivial protocol: $v(t) = 1 - \cos (\omega t)$, $w(t) = 1$, and $\Delta(t) = \sin (\omega\, t)$; (c) trivial protocol: $v(t) = 1 + \sin (\omega\, t)$, $w(t) = 1$, and $\Delta(t) = \cos (\omega\, t)$ for $t\in [0,T/2)$ and $v(t) = 1 - 0.8\sin (\omega\, t)$, $w(t) = 1$, and $\Delta(t) = \cos (\omega\, t)$ for $t\in [T/2,T)$. Here $T$ is the pumping period. The reason for choosing these different pumping protocols is attributed to the fact that they are easily accessible in experiment \cite{Lohse_2015}, and they allow to investigate of all possible topology which include several distinct pumping characteristics. Fig.~\ref{fig:protocols} represents the three different driving protocols in the parameter space.

\begin{figure}
	\subfigure []{
	\includegraphics[width=0.15\textwidth,height=2.5cm]{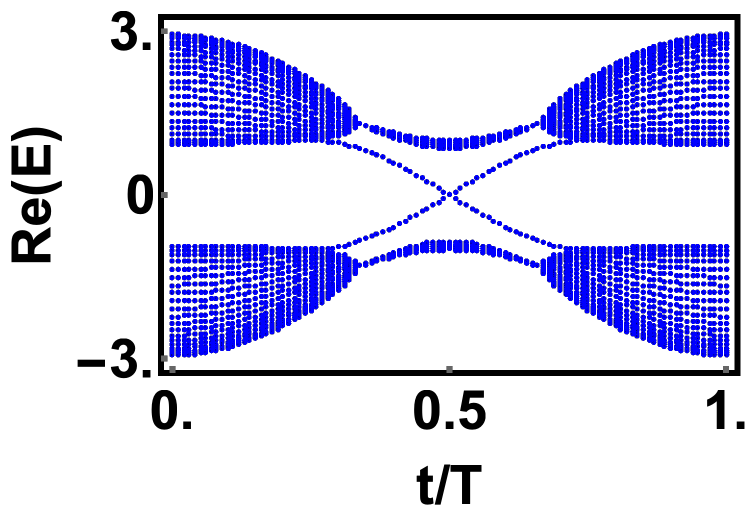}}
\subfigure []{
	\includegraphics[width=0.15\textwidth,height=2.5cm]{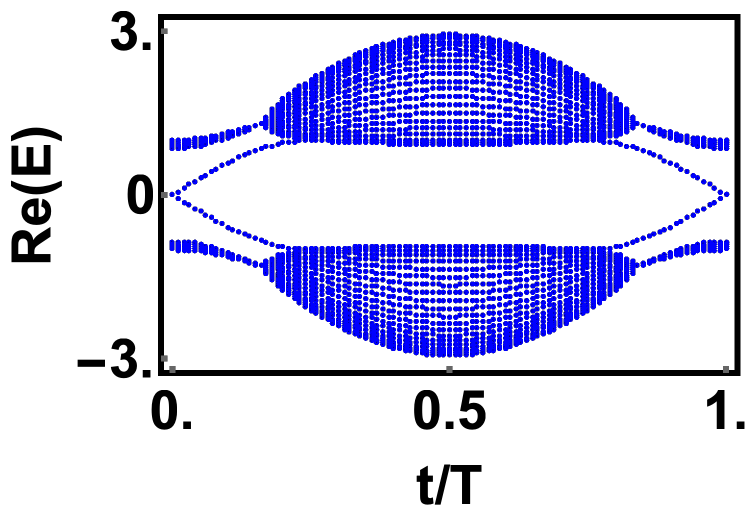}}
\subfigure []{
	\includegraphics[width=0.15\textwidth,height=2.5cm]{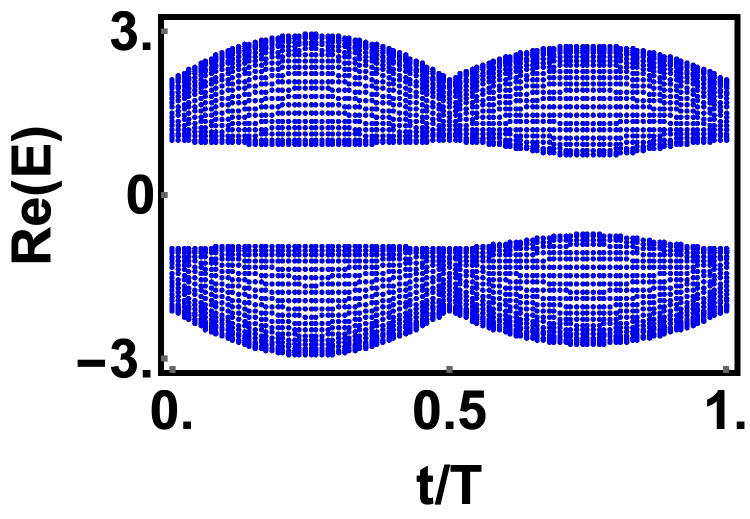}}
\\\vspace{-1\baselineskip}
\subfigure []{
	\includegraphics[width=0.15\textwidth,height=2.5cm]{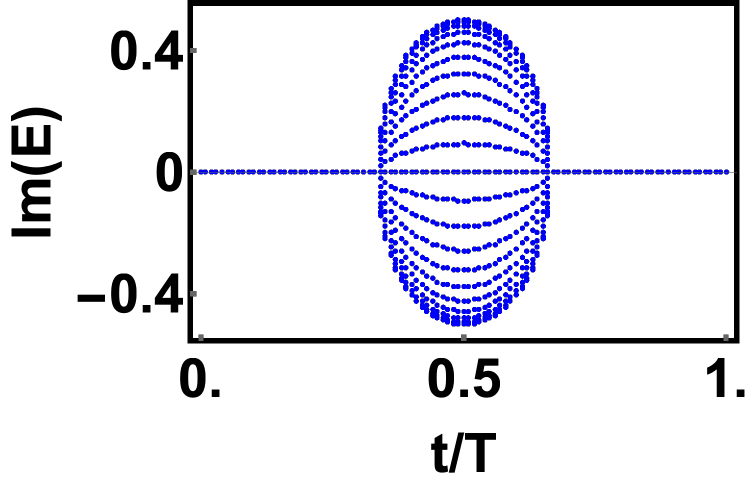}}
\subfigure []{
	\includegraphics[width=0.15\textwidth,height=2.5cm]{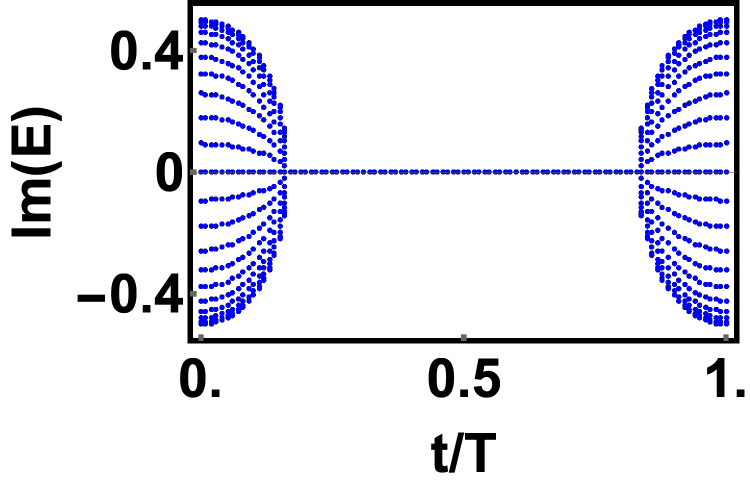}}
\subfigure []{
	\includegraphics[width=0.15\textwidth,height=2.5cm]{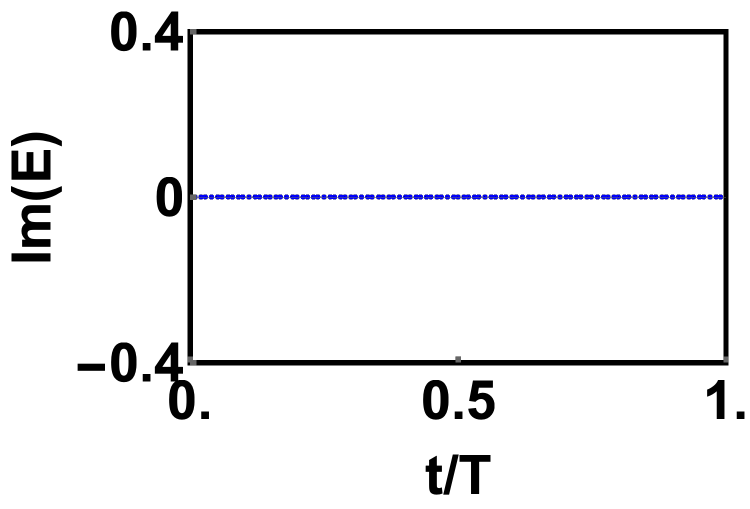}}\\
\vspace{-1\baselineskip}
\subfigure []{
	\includegraphics[width=0.15\textwidth,height=2.5cm]{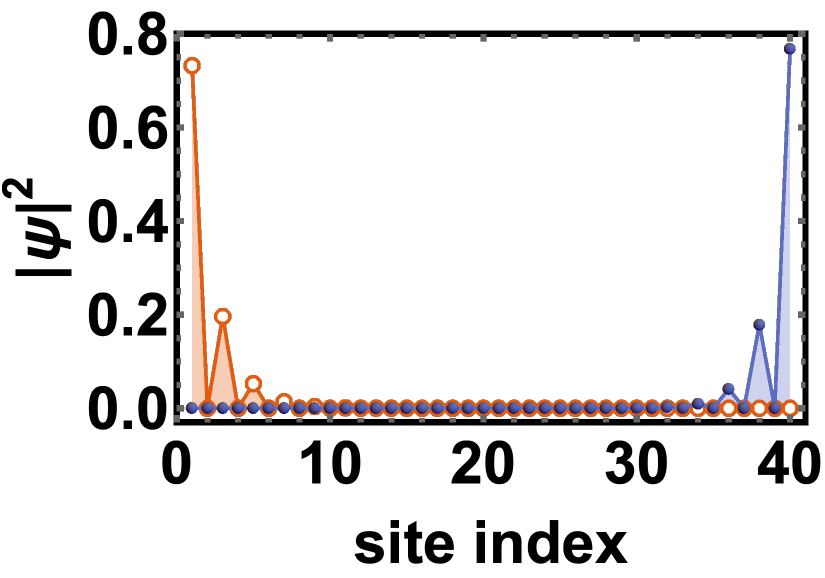}}
\subfigure []{
	\includegraphics[width=0.15\textwidth,height=2.5cm]{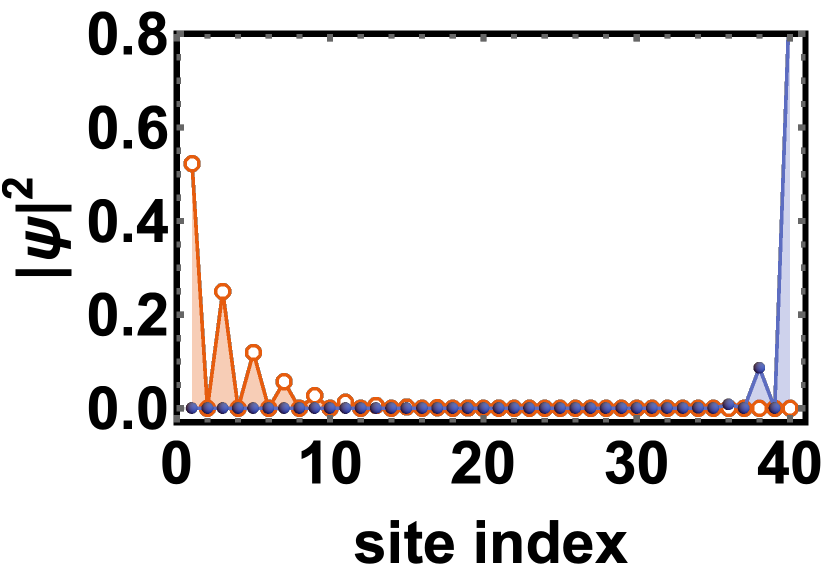}}
\subfigure []{
	\includegraphics[width=0.15\textwidth,height=2.5cm]{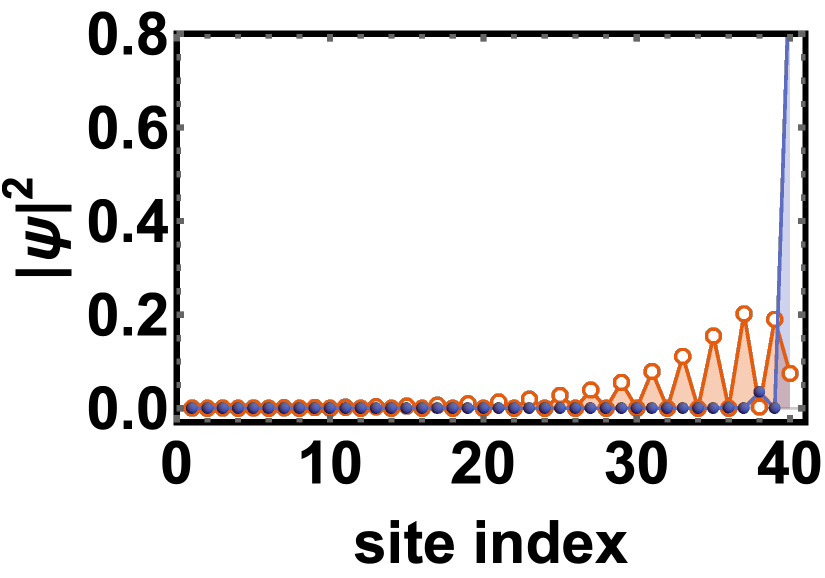}}
\vspace{-1\baselineskip}
	\caption{\textbf{Pumping in non-Hermitian RM model: }Plot of the real and imaginary part of the complex instantaneous spectrum for clockwise non-trivial ((a), (d)), anticlockwise non-trivial ((b), (e)), and trivial ((c), (f)) protocols at $\gamma = 0.5$ and $N =20$. The eigenvalues corresponding to in-gap edge states are purely real. The plot of the right wavefunction for the in-gap edge states of clockwise non-trivial protocol at an instant (g) $t/T = 0.53$, (h) $t/T = 0.6$, (i) $t/T = 0.7$. The orange circle and blue dots represent negative and positive energy edge states, respectively. Clearly, the left localized state moves towards the right edge with increasing time and it finally merges with the right localized skin modes. The system size takes value $N = 20$.}
	\label{fig:complex_instantaneous_spectrum_gamma_0.5_cat}
\end{figure}

Fig.~\ref{fig:Hermitian_RM} (a-c) illustrates the instantaneous spectrum corresponding to three different pumping protocols for a Hermitian RM model. Evidently, we see in-gap states in the bulk for the non-trivial clockwise and anticlockwise driving protocols. For the non-trivial clockwise sequence, one in-gap state with $E<0$ moves to the upper band with $ E>0$ as a function of time. Throughout this evolution, this state remains localized at the right boundary and finally moves to the bulk. Similarly, the other in-gap state moving from $E>0$ to $E<0$ is localized on the left boundary. The localization of these in-gap states is shown in Fig.~\ref{fig:Hermitian_RM} (d-f). The edge modes are not confined within a single unit cell, but exponentially decay towards the bulk. For the anticlockwise sequence, the in-gap state localized at the right boundary moves from $E>0$ to $E<0$, and the other one being localized at the left boundary moves from $E<0$  to $E>0$. This gives a notion of edge states being pumped in energy at the boundaries and the number of pumped states can be considered as a topological invariant.  We note that the pumping mechanism is similar to a conveyor belt in the energy-position space \cite{2015arXiv150902295A,shen2012topological} i.e., with each pumping cycle the Wannier centers of valence and conduction band are transported unidirectionally toward opposite edges of the chain where already existing edge state moves into high or low energy band to avoid state overlap. In contrast to clockwise and anticlockwise sequences, the pumping is absent for the trivial protocol. However, the presence of non-Hermitian term can give rise to in-gap states even in the trivial case as will be evident shortly. 
\begin{figure}
		\subfigure []{
	\includegraphics[width=0.15\textwidth,height=2.5cm]{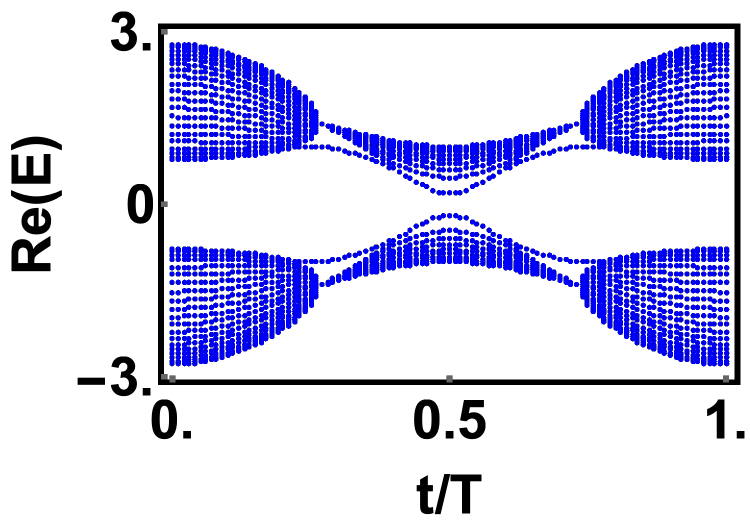}}
	\subfigure []{
	\includegraphics[width=0.15\textwidth,height=2.5cm]{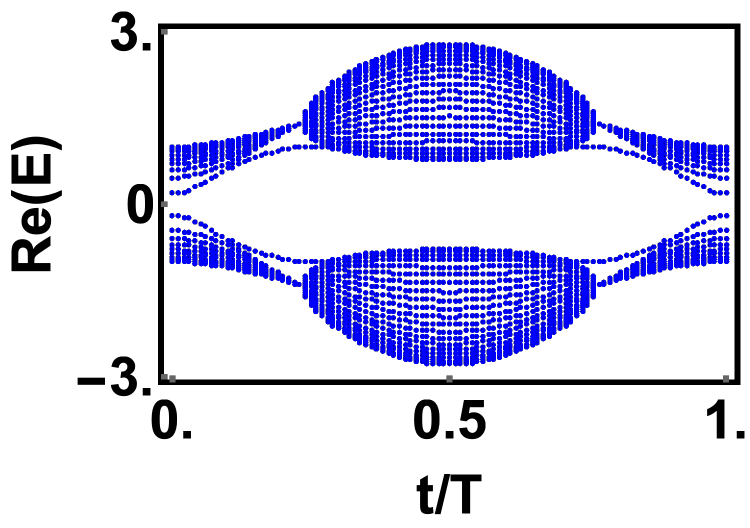}}
	\subfigure []{
	\includegraphics[width=0.15\textwidth,height=2.5cm]{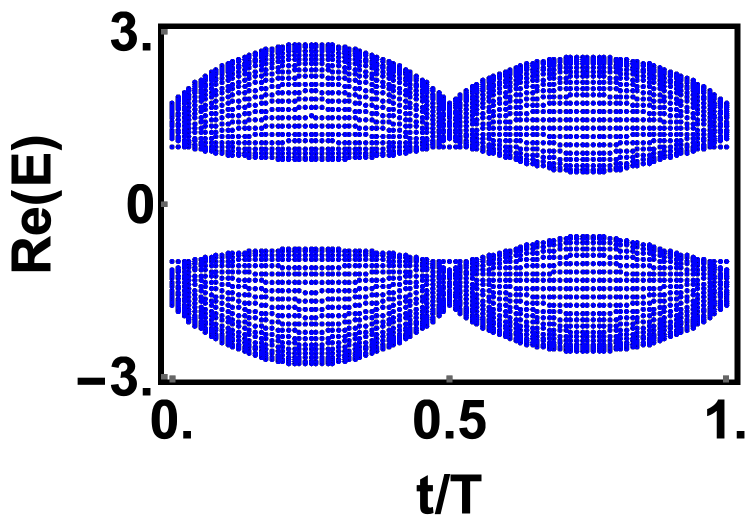}}\\
\vspace{-1\baselineskip}
	\subfigure []{
	\includegraphics[width=0.15\textwidth,height=2.5cm]{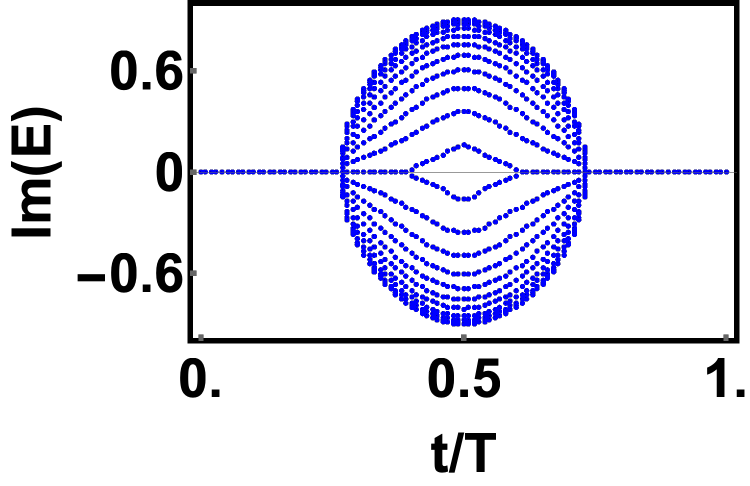}}
	\subfigure []{
	\includegraphics[width=0.15\textwidth,height=2.5cm]{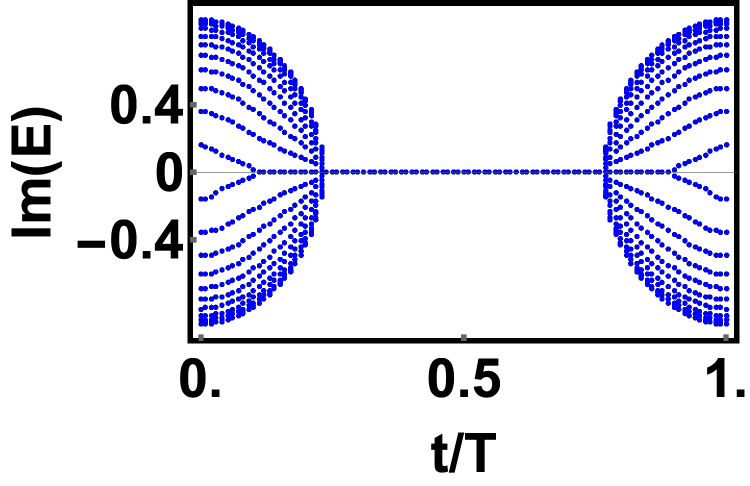}}
	\subfigure []{
	\includegraphics[width=0.15\textwidth,height=2.5cm]{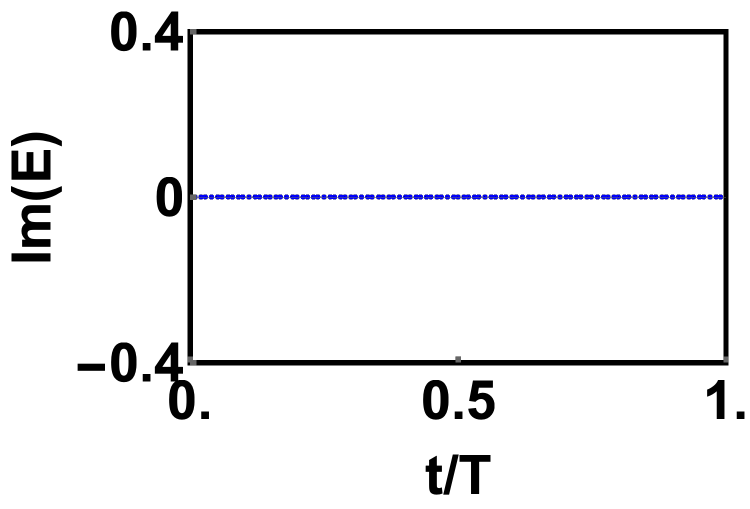}}
\vspace{-1\baselineskip}
	\caption{\textbf{Breakdown of pumping in non-Hermitian RM model: }Plot of the real and imaginary part of the complex instantaneous spectrum for clockwise non-trivial (a), anti-clockwise non-trivial (b) and trivial (c) protocols at $\gamma = 0.9$ for system size $N =20$. The eigenvalues pick up some imaginary part and the real part opens up a gap, indicating the breaking of pumping.}
	\label{fig:complex_instantaneous_spectrum_gamma_0.5_clockwise_size_40}
\end{figure}

\subsection{Non-Hermitian RM model and topological pumping}\label{subsec:non-hermitianpumping}
To investigate the effect of non-Hermitian term, we take $v_l(t)=v(t)-\gamma$ and $v_r(t)=v(t)+\gamma$ as before. Following the finite-size GBZ scheme for dynamical parameters and using Eq.~(\ref{eqn:complex_energy_eigenvalue}), the complex instantaneous spectrum for the non-Hermitian RM model is plotted in Fig.~\ref{fig:complex_instantaneous_spectrum_gamma_0.5_cat} (a-f) for all three protocols mentioned before. The in-gap states survive even in the presence of small and moderate $\gamma$ for both non-trivial sequences. Additionally, the complex instantaneous spectra contain special points ( analogous to degenerate points in Hermitian system) with energies $E=\pm\sqrt{w^2+\Delta^2}$ in both the valence and conduction bands at two instants of time $t/T=(1/3, 2/3)$. These special points turn out to be exceptional points~(EPs) as the eigenstates corresponding to the eigenvalues coalesce at the same instant of time $t$. The appearance of these EPs can be attributed to the fact that the left hopping $v_l(t)$ vanishes for the given $\gamma$ at $t/T=(1/3, 2/3)$. Consequently, no solutions of $z$ exist for $v_l=0$ as evident from Eq.~(\ref{eqn:gbz}).~This is in conjunction with the EPs obtained in the static case with $v_l=\gamma$. Note that the energy has an imaginary part only in between these two EPs. Also, the energies corresponding to the topological edge modes as shown in Fig.~\ref{fig:complex_instantaneous_spectrum_gamma_0.5_cat} are always real. In contrast, the weak and moderate non-Hermiticity does not lead to any in-gap states in the trivial sequence. Moreover, the spectrum remains real even in the presence of weak and moderate values of $\gamma$.   

\begin{figure}
		\subfigure []{
	\includegraphics[width=0.15\textwidth,height=2.5cm]{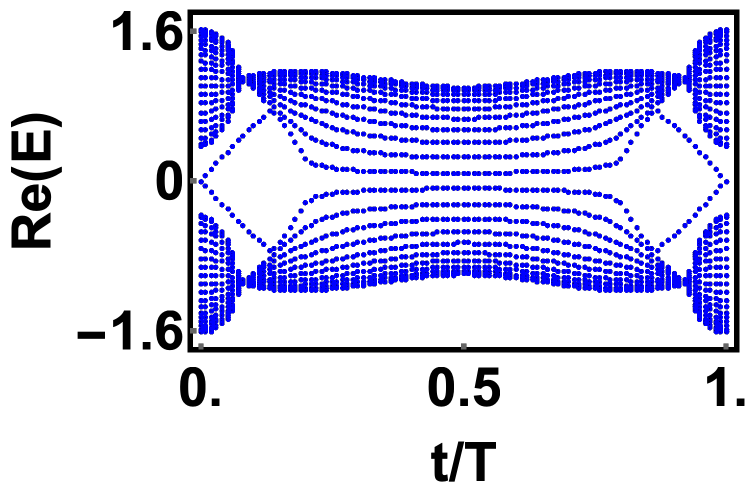}}
	\subfigure []{
	\includegraphics[width=0.15\textwidth,height=2.5cm]{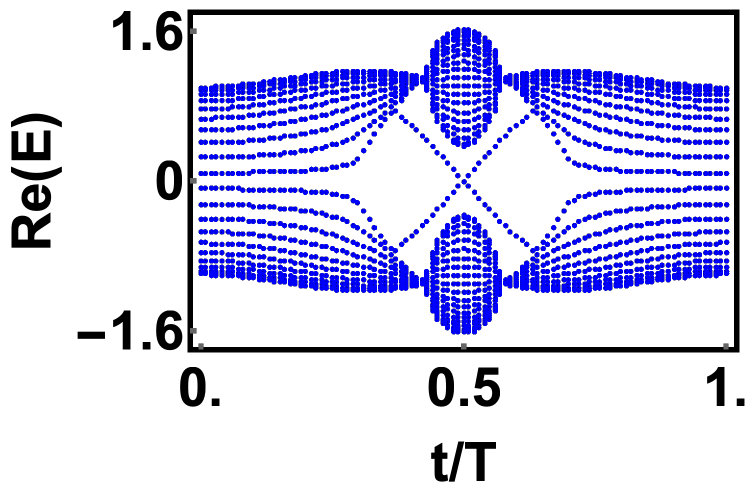}}
	\subfigure []{
	\includegraphics[width=0.15\textwidth,height=2.5cm]{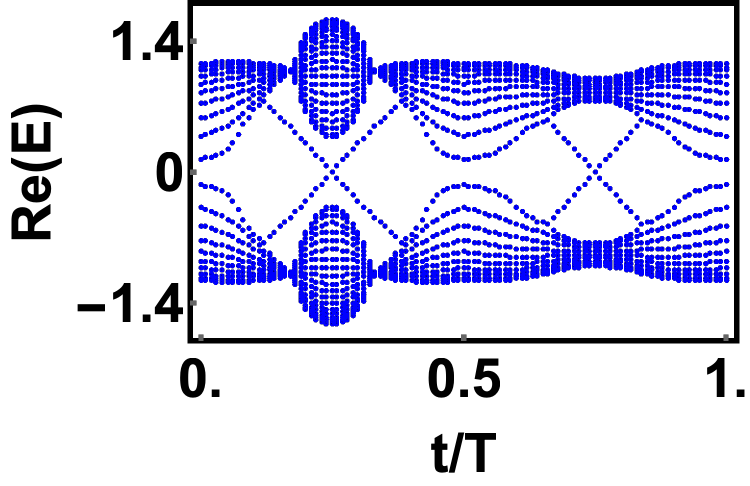}}
\\
\vspace{-1\baselineskip}
	\subfigure []{
	\includegraphics[width=0.15\textwidth,height=2.5cm]{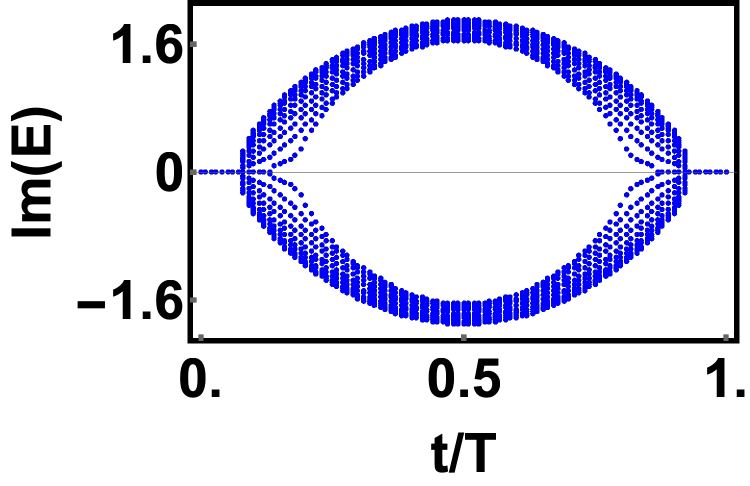}}
	\subfigure []{
	\includegraphics[width=0.15\textwidth,height=2.5cm]{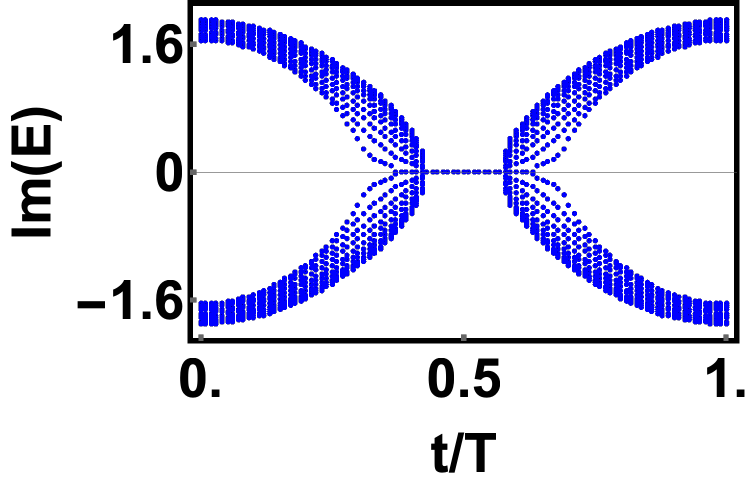}}
	\subfigure []{
	\includegraphics[width=0.15\textwidth,height=2.5cm]{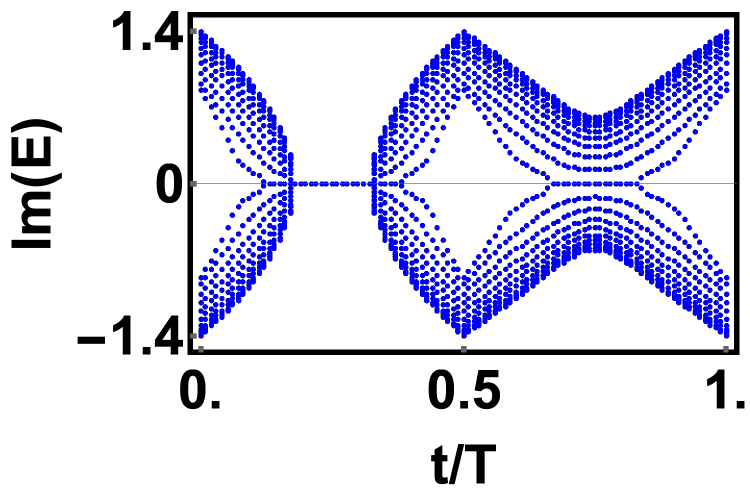}}
\\
\vspace{-1\baselineskip}
	\subfigure []{
	\includegraphics[width=0.15\textwidth,height=2.5cm]{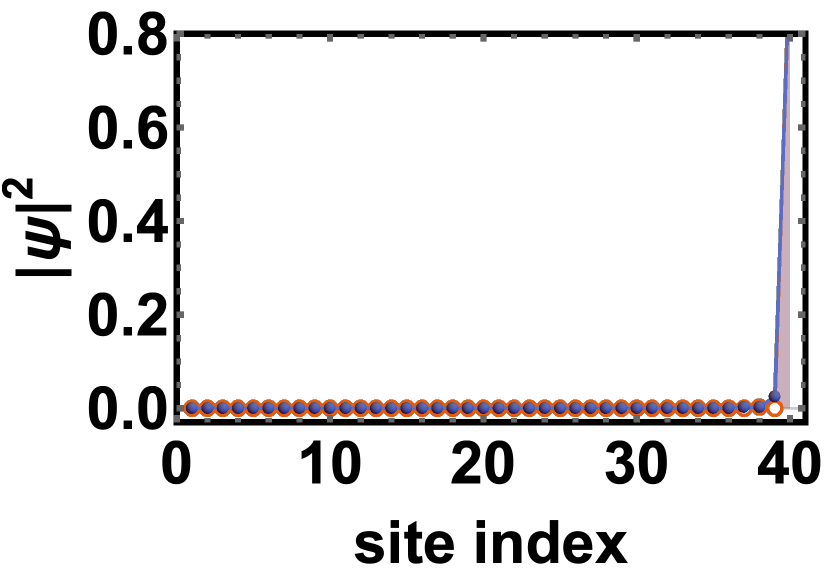}}
	\subfigure []{
	\includegraphics[width=0.15\textwidth,height=2.5cm]{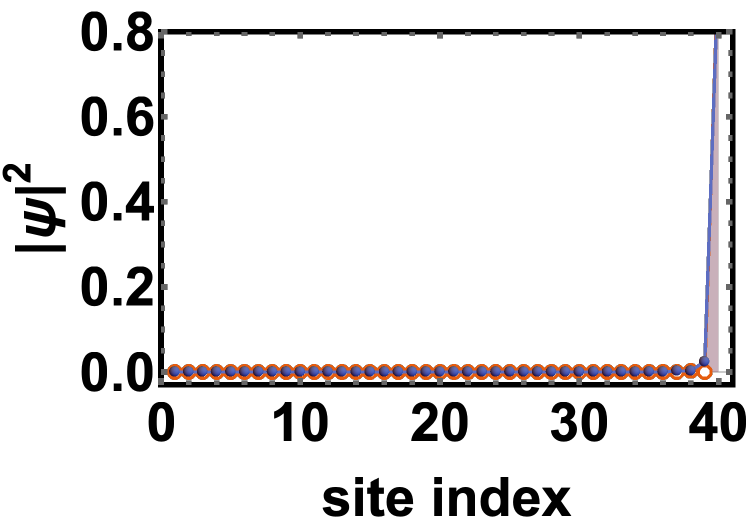}}
	\subfigure []{
	\includegraphics[width=0.15\textwidth,height=2.5cm]{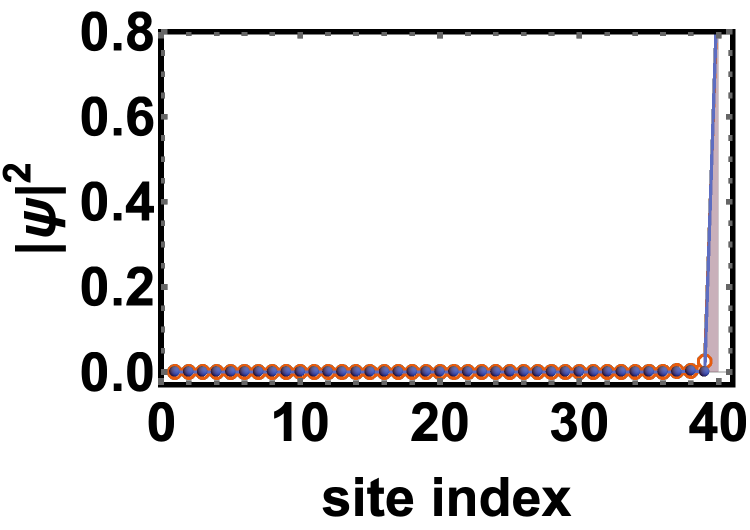}}
\vspace{-1\baselineskip}
	\caption{\textbf{Emergence of anomalous pumping in non-Hermitian RM model: }Plot of real and imaginary part of complex instantaneous spectrum for clockwise non-trivial ((a), (d)), anticlockwise non-trivial ((b), (e)), and trivial ((c), (f)) protocols at $\gamma = 1.9$ and $N =20$. The eigenvalues corresponding to the in-gap edge states are purely real. Right wavefunctions are plotted for (a) clockwise non-trivial, (b) anticlockwise non-trivial, and (c) trivial protocols at instants $t/T = 0.05$, $0.55$, and $0.3$ respectively. The orange circle and blue dots represent negative and positive energy edge states respectively. Here both edge states are always localized at the right boundary.}
	\label{fig:complex_instantaneous_spectrum_gamma_1.9_cat}
 \end{figure}

In the instantaneous spectrum of non-Hermitian Hamiltonian, the bands consist of localised skin modes instead of extended bulk modes. Since $v_{r} > v_{l}$, the right wavefunctions of the skin modes are localised at the right edge whereas the left wavefunctions are localised at the left boundary. Fig.~\ref{fig:complex_instantaneous_spectrum_gamma_0.5_cat} (g-i) demonstrates the evolution of the topological edge state for the clockwise non-trivial cycle. The edge mode crossing $E>0$ to $E<0$ remains localized at the left boundary near the half period of a full cycle and starts moving towards the right boundary with increasing time and finally merges with skin modes. Note that this is in contrast to the conventional pumping in the Hermitian case where the edge mode localized at the boundary moves to the bulk rather than to the other edge. Thus the notion of topological pumping is different in the non-Hermitian RM model, exhibiting skin modes. However, it does not affect the topological invariant as it gets measured in the bulk gap near $E = 0$. This proves the validity of bulk-boundary correspondence even in the non-Hermitian RM model. On increasing non-Hermiticity, gap opens in the real instantaneous spectrum as evident from the Fig.~\ref{fig:complex_instantaneous_spectrum_gamma_0.5_clockwise_size_40}. Consequently, the imaginary energies of the in-gap states corresponding to the topological boundary modes now have finite non-zero values. This reflects the destruction of charge pumping, giving rise to topological transition in the RM model due to stronger non-Hermiticity.

On further increasing the parameter $\gamma$, we observe an anomalous pumping phenomenon which is unique to non-Hermitian system and does not have any Hermitian analogue. In this case, the in-gap states are found for both trivial and non-trivial protocols as shown in Fig.~\ref{fig:complex_instantaneous_spectrum_gamma_1.9_cat}. Similar to the conventional cases, the energies of in-gap states are real. However,  these states are localized over a single unit cell at the right boundary for all the protocols (see Fig.~\ref{fig:complex_instantaneous_spectrum_gamma_1.9_cat} (g-i)). This is in contrast to the earlier non-Hermitian cases where edge modes move from one boundary to another either with the non-Hermitian parameter $\gamma$ or with time ($t/T$).  We will soon topologically characterize these in-gap states using the non-Bloch chern number.  Additionally, the large $\gamma$ also leads to shifting of in-gap states either from the zone center (clockwise) or from the zone boundary  (anticlockwise) of the full cycle to the zone boundary (clockwise) or to the zone center (anticlockwise) of the complete cycle.  This is attributed to the change in the parameters,  $v_l$ and $v_r$ due to $\gamma$. In the dynamic scenario, the pumping phenomena can also be achieved by altering the system size $N$ while keeping $\gamma$ fixed (Appendix \ref{appen A2}). This distinct feature portrays the system-size-dependent topological transitions in the 1+1D RM model similar to the static case.   

.

\begin{figure}
    \centering
    \includegraphics[width =0.4\textwidth]{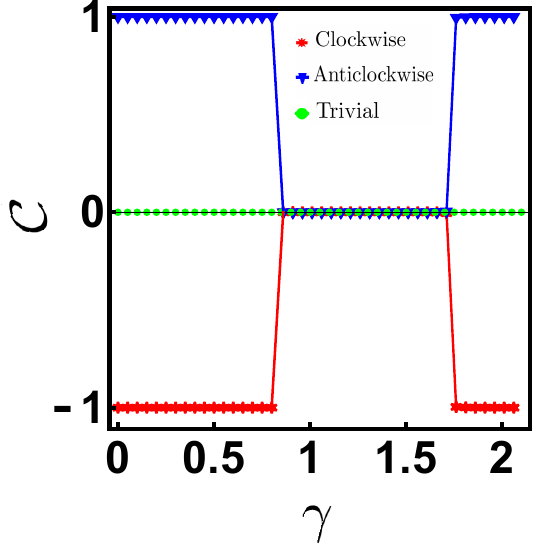}
    	\caption{\textbf{Non-Bloch Chern number in non-Hermitian RM model: } Plot of cylinder non-Bloch Chern number ($\mathbf{\mathcal{C}}$)  for  three pumping protocol. The cylinder height is taken to be $N=20$ along the open boundary and $100$ points are taken for the time $(t)$ grid.  }
	\label{fig:non-bloch chern}
\end{figure}

\subsection{Non-Bloch Chern number}
\label{non-bloch}

To find the topological nature of the pumping under different protocols, we use the non-Bloch topological invariant discussed in several recent articles \cite{PhysRevLett.121.136802, PhysRevB.105.075128, PhysRevB.100.035102,PhysRevLett.127.270602}. The conventional Bloch waves, characterized by a pure phase factor, $e^{ik}$, can be represented by $\beta$. This  implies that the usual Bloch phase factor $e^{ik}$ is replaced by $\beta \equiv |\beta|e^{ik}$ in the OBC, and beyond the phase factor $\beta$ possesses a non-unit modulus: $|\beta|=\sqrt{|{v_r}/{v_l}|}$ for our model as derived in Sec.~\ref{sub:analytics}. This suggests that the wave vector acquires an imaginary part: $k \rightarrow k - i \operatorname{ln} |\beta|$. Thus for the driven non-Hermitian Hamiltonian, a non-Bloch cylinder Hamiltonian $H(\tilde{k},t)$ can be constructed via $\tilde{k}\rightarrow k - i \operatorname{ln} |\beta|$. 
Accordingly, the cylinder Chern number can be defined as the integral over the generalized Brillouin zone (GBZ)\cite{PhysRevLett.121.136802, PhysRevB.105.075128},
\begin{equation}
    \mathbf{\mathcal{C}}_{n}=\dfrac{1}{2\pi}\int_{\text{GBZ}}\epsilon_{ij} \mathbf{\Omega}^{n}_{ij}~d^{2}\mathbf{\tilde{k}}
    \label{cylin chern}
\end{equation}
where  $\tilde{\mathbf{k}}\equiv (\tilde{k},t)$, $n$ is the band index, $i,j=1,2$, $\epsilon_{12}=-\epsilon_{21}=1$,  $\mathbf{\Omega}^{n}_{ij}=i\langle \partial_{i} \chi_{n}^{\tilde{\mathbf{k}}}|\partial_{j}\phi_{n}^{\tilde{\mathbf{k}}}\rangle$,  $\partial_{1}\equiv\partial_{t}$ , $\partial_{2}\equiv\partial_{k}$, the {\it right} and {\it left} eigenstates ($|\phi_{n}^{\tilde{\mathbf{k}}}\rangle$ and $|\chi_{n}^{\tilde{\mathbf{k}}}\rangle$), satisfying the biorthonormal condition $\langle \phi_{n}^{\tilde{\mathbf{k}}}| \chi_{n}^{\tilde{\mathbf{k}}} \rangle=1$. We note that the non-Bloch Chern number can also be realized as the change of the Berry phase along circular sections of the GBZ in the direction of the open boundary as $\mathbf{\mathcal{C}}=\displaystyle{\dfrac{1}{T} \int_{0}^{T} \frac{d W(t)}{dt } dt = [W(T)-W(0)]/T}$,  where the Berry phase can be defined for the lower band as $
W(t)=-i\int_{0}^{2\pi} \langle \chi_{-}^{\tilde{\mathbf{k}}} |\partial_{k} |\phi_{-}^{\tilde{\mathbf{k}}}\rangle ~dk   
$.

\begin{figure}
		\subfigure []{
	\includegraphics[width=0.15\textwidth,height=2.5cm]{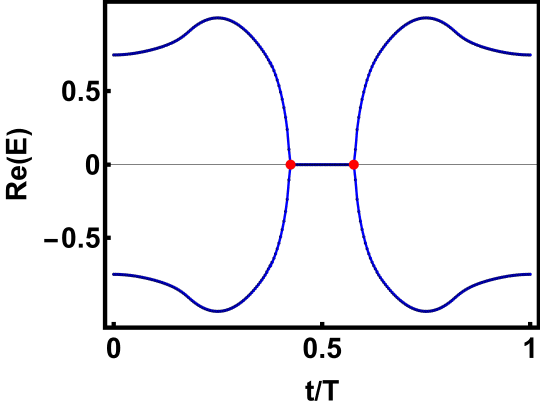}}
	\subfigure []{
	\includegraphics[width=0.15\textwidth,height=2.5cm]{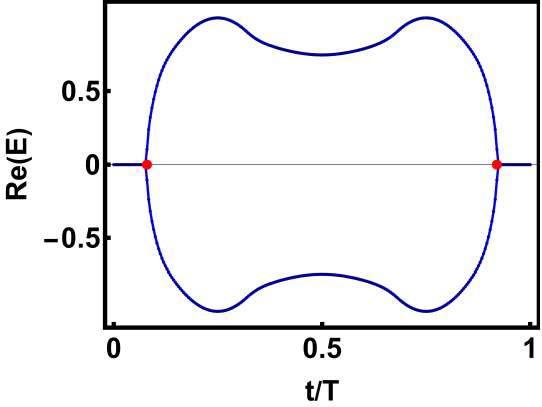}}
	\subfigure []{
	\includegraphics[width=0.15\textwidth,height=2.5cm]{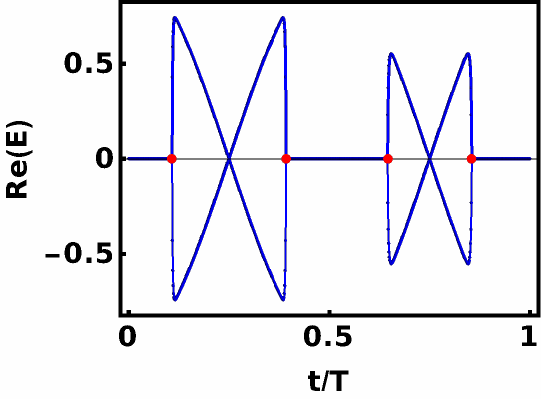}}
\\
\vspace{-1\baselineskip}
	\subfigure []{
	\includegraphics[width=0.15\textwidth,height=2.5cm]{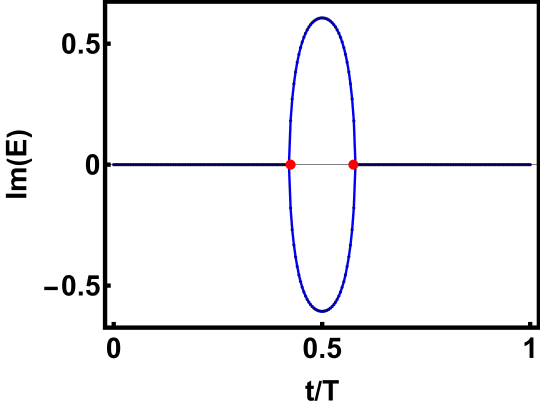}}
	\subfigure []{
	\includegraphics[width=0.15\textwidth,height=2.5cm]{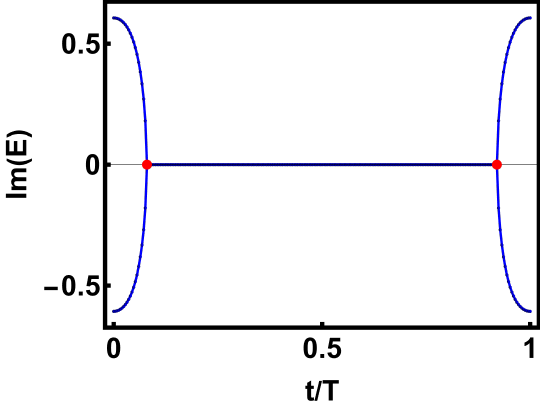}}
	\subfigure []{
	\includegraphics[width=0.15\textwidth,height=2.5cm]{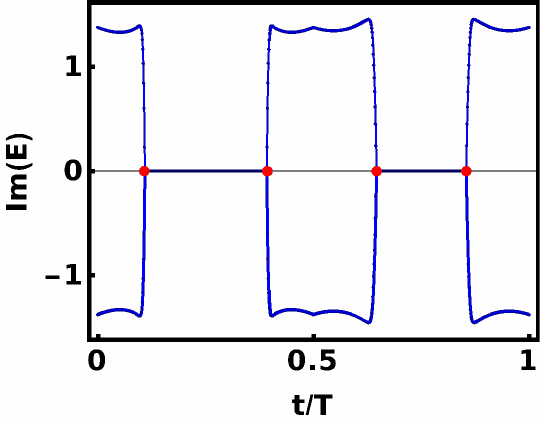}}
   
	\vspace{-1\baselineskip}
	\caption{ \textbf{Exceptional edge modes in non-Hermitian RM model: } Plot of the real and imaginary part of the Edge modes spectrum for ((a),(d)) clockwise non-trivial, ((b),(e)) anti-clockwise non-trivial protocol at $\gamma = 0.97$ and for trivial protocol ((c),(f)) at $\gamma=1.9$.  We see that there are edge modes connected by EPs (red dots) in all pumping cases. The positions of such EPs can be calculated analytically. The system size is $N=35$.}
	\label{fig:Exceptional edge_Modes}
\end{figure}

In Fig.~\ref{fig:non-bloch chern} (a-c), we show the variation of the cylindrical non-Bloch Chern number ($\mathcal{C}$) as a function of the non-Hermitian parameter ($\gamma$). Notably, we observe that for relatively lower values of non-Hermiticity ($\gamma \sim 0.0-0.8$), $\mathcal{C}$ approaches $\pm 1$, signifying the presence of pumping phenomena for both anticlockwise and clockwise non-trivial protocols, respectively. In contrast, for the trivial protocol, $\mathcal{C}=0$, implying the absence of conventional pumping. It is also worth noting that for relatively higher values of $\gamma$, specifically in the range of $(\gamma \sim 0.85-1.75)$, the gapless edge state eigenvalues develop an imaginary component, while the real part open up a gap as shown in Fig.~\ref{fig:complex_instantaneous_spectrum_gamma_0.5_clockwise_size_40}, indicative of the breakdown of pumping. Consequently, we get $\mathcal{C}=0$ for this case. Furthermore, in Section (\ref{subsec:non-hermitianpumping}), we observe an anomalous pumping phenomenon when $\gamma$ takes relatively higher values, particularly in the range of $(\gamma \sim 1.75-2.15)$. This phenomenon lacks an analogue in Hermitian systems, rendering it as a unique characteristic of the non-Hermitian RM model. In this regime, $\mathcal{C}$ attains values of $\pm 1$ for both clockwise and anticlockwise protocols. However, for the trivial protocol, despite the existence of in-gap edge modes, we get $\mathcal{C}=0$. This arises because two edge mode crossings have opposite pumping directions, thereby maintaining the topological invariant at its prior value, evident from (Fig.~\ref{fig:protocols}c). Interestingly we can achieve a non-zero Chern number in the trivial protocol by driving the protocol in half-period, as a result only one of the edge modes will be characterized, which suggests for stronger non-Hermiticity, trivial adiabatic protocol can lead to pumping.  Hence we can see that the non-Bloch Chern number perfectly captures the OBC phase of the system in the dynamic case. Important to note that, for $\gamma>2.15$, the pumping is lost.

\begin{figure}
	\centering
	\subfigure []
	{\includegraphics[width=0.15\textwidth,height=3cm]{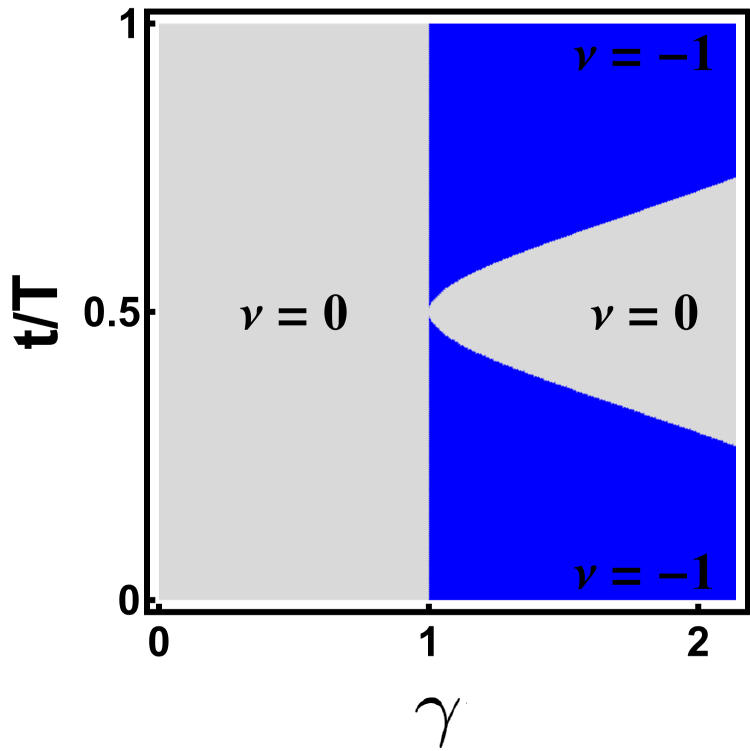}\label{gamma111}}
	\subfigure []
	{\includegraphics[width=0.15\textwidth,height=3cm]{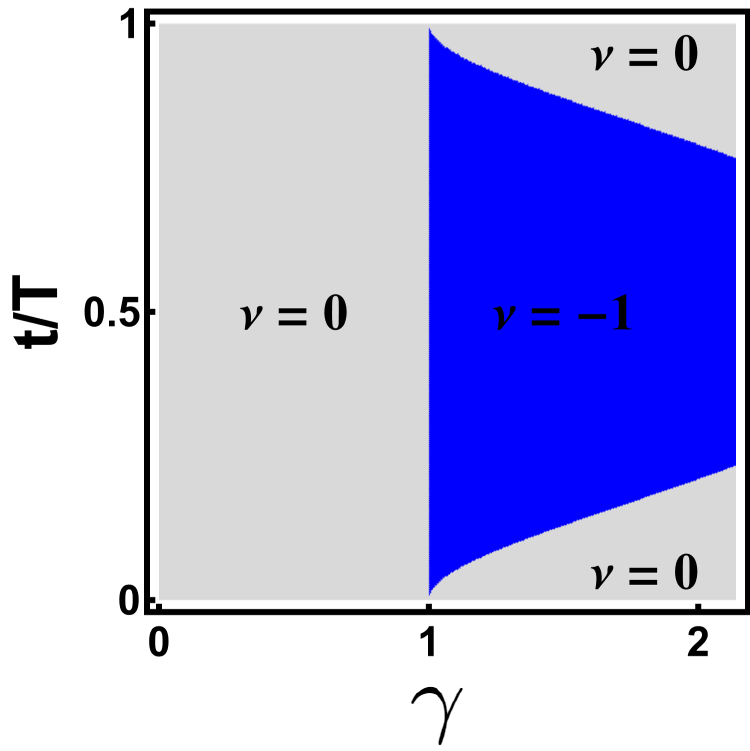}\label{gamma222}}
	\subfigure []
	{\includegraphics[width=0.15\textwidth,height=3cm]{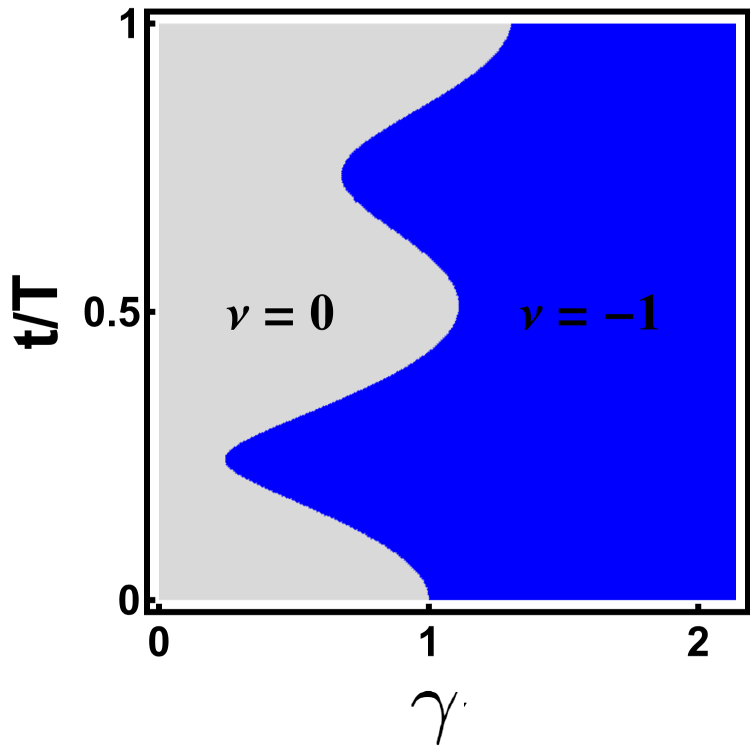}\label{gamma333}}
    \vspace{-1\baselineskip}
    \caption{\textbf{Point gap and Line gap phases in dynamic RM model: }Phase diagram in $(\gamma-t)$ space representing Point gap and Line gap phases for (a) clockwise (b) anti-clockwise and (c) Trivial protocol. In the Blue region $\nu=-1$, indicating the presence of a Point gap in the system.  While clockwise and anti-clockwise protocols exhibit both  Point gap and Line gap for higher $\gamma$, trivial protocol exhibits a Point gap consistently throughout the pumping period. }
	\label{Dynamic Winding}
\end{figure}

\subsection{Exceptional edge modes}
\label{excep edge}

In Sec. \ref{sec:Hamiltonian}(C), we have analytically demonstrated how the static non-Hermitian RM Model can host second-order EPs and their nature due to different system sizes and boundary conditions, particularly, for odd $N$. In the dynamic case, we find even richer physics, stemming from system size dependency.  In a recent study \cite{Sone_2020}, it has been shown that a tight-binding model on a ladder can host edge modes which are glued via exceptional points and are robust against disorders. Utilizing dynamical charge pumping protocols, we show that similar ``exceptional edge modes" can be realized even in our ($1+1D$) model for odd $N$. In Fig. \ref{fig:Exceptional edge_Modes}, we see that the edge modes which are flat at $E=0$, are glued by the EPs (red dots) in clockwise, anticlockwise and trivial sequences. We can analytically find out the positions of these exceptional edge modes in the dynamic case for any specific $\gamma$ and $N$ (odd). We use the same conditions that we utilized in the static case:
\begin{align}
&v_l(t)w(t)z_1^2 + z_1\bigg[v_l(t)v_r(t)+\Delta(t)^2+w(t)^2\bigg] \nonumber \\
&~~~~~~~~~~~~~~~~~~~~~~~~~~~~~~~~~~~~~~+ v_r(t)w(t) = 0, \label{eqn:EP1_dyn} \\
\nonumber \\
&v_l(t)z_1^{2N+2} + w(t)z_1^{2N+1}-w(t)r(t)^Nz_1 \nonumber\\&~~~~~~~~~~~~~~~~~~~~~~~~~~~~~~~~~~- v_l(t)r(t)^{N+1}=0 \label{eqn:EP2_dyn}
\end{align}

Solutions of Eq.~(\ref{eqn:EP1_dyn}) and Eq.~(\ref{eqn:EP2_dyn}) allow us to precisely determine the time $t$ when the exceptional edge modes occur.  For even $N$,  these equations turn out to have no solutions, implying that the exceptional edge modes here are unique characteristics of odd $N$.  It is noteworthy that the emergence of these EPs can be finely tuned by adjusting the non-Hermiticity parameter $\gamma$. We emphasize that these EPs do not arise as a consequence of finite-size effects. For odd $N$, a range of $\gamma$-values can be found for which exceptional edge modes are present. After a critical value of $\gamma$, the exceptional edge modes disappear.

\subsection{Point gap and line gap}\label{point-line}

For different pumping protocols, we now analyze the point gap and line gaps in the system by the phase diagram in $(\gamma-t)$ space similar to the static case. Fig.~\ref{Dynamic Winding}(a,b) represents how the winding invariant evolves throughout the pumping process for both clockwise and anticlockwise non-trivial protocols. It is clear that for $\gamma<1$, a line gap exists, however higher values of $\gamma$, a transition occurs for both the clockwise and anticlockwise non-trivial pumping protocols. Thus the system undergoes a transition from a line gap to a point gap. Interestingly, for the trivial protocol (Fig.~\ref{Dynamic Winding}c), the spectrum displays a point gap for higher values of $\gamma$, throughout the pumping period.

\section{Conclusions}
In conclusion, we have studied the non-Hermitian RM model, exhibiting both skin effects and exceptional points. Specifically, we show how non-Hermiticity helps or destroys pumping in both static 1D and adiabatically driven RM model. Our main results are summarised as follows- (i) An anomalous charge pumping is observed as a function of non-Hermitian parameter $\gamma$ without adiabatic and time-periodic evolution of the model parameters. Using a finite-size GBZ scheme, we find analytical solutions for both topological boundary and skin modes, which in turn helps differentiate topological boundary modes from skin modes and improve the understanding of state pumping phenomena. (ii) A system-size-dependent topological phase transition is found in the static case which also has richer implications in the dynamical scenario as the pumping phenomena can also be solely obtained by the variation of system size ($N$) in the non-Hermitian RM model. We also engineered second-order EPs in the static case due to odd system size which results in the emergence of exceptional edge modes in the dynamic case, and its existence is confirmed analytically by the finite-size GBZ scheme. (iii) During adiabatic modulation of parameters, the notion of charge pumping remains even in the presence of weak and moderate $\gamma$ for both clockwise and anticlockwise protocols. While the trivial protocol does not lead to any pumping for weak and moderate non-Hermiticity, an anomalous pumping is found for stronger non-Hermiticity, where the energy states leading to pumping remain localized at one boundary. However, the edge modes in the trivial case represent opposite pumping direction hence the topological invariant associated with the edge modes remain unaltered. For non-trivial protocols, the charge pumping is lost for moderate $\gamma$ and reemerges again with a higher value of $\gamma$. We also explain the pumping phases in the system using a non-Bloch topological invariant, representing a non-Bloch BBC. (v) Finally, we emphasize how different dynamical protocols induce a transition from line gap to point gap or vice versa in the model. We note that the results obtained here supported by analytical solutions can invite further studies on non-Hermitian topological charge pumping in various systems. Particularly, our study can be extended to spinful generalization of the Rice-Mele model\cite{PhysRevB.74.195312} and Aubry-Andr\'e model with fractional pumping\cite{PhysRevB.91.125411}. Further, it opens up possibilities for the experimental realization of non-Hermitian pumping by introducing gain and loss in dynamically controlled optical superlattices of ultracold atoms\cite{Nakajima_2016,Lohse_2015}. Also, the system-size-dependent topological phase transition achieved in this model requires further investigation for other non-Hermitian systems.

\bibliography{references}
\onecolumngrid
\appendix

\section{System-size-dependent topological transition}

\subsection{Static case}

\label{appen A}
\begin{figure}[H]
	\centering
	\subfigure [$N=20$]
	{\includegraphics[width=0.155\textwidth,height=3cm]{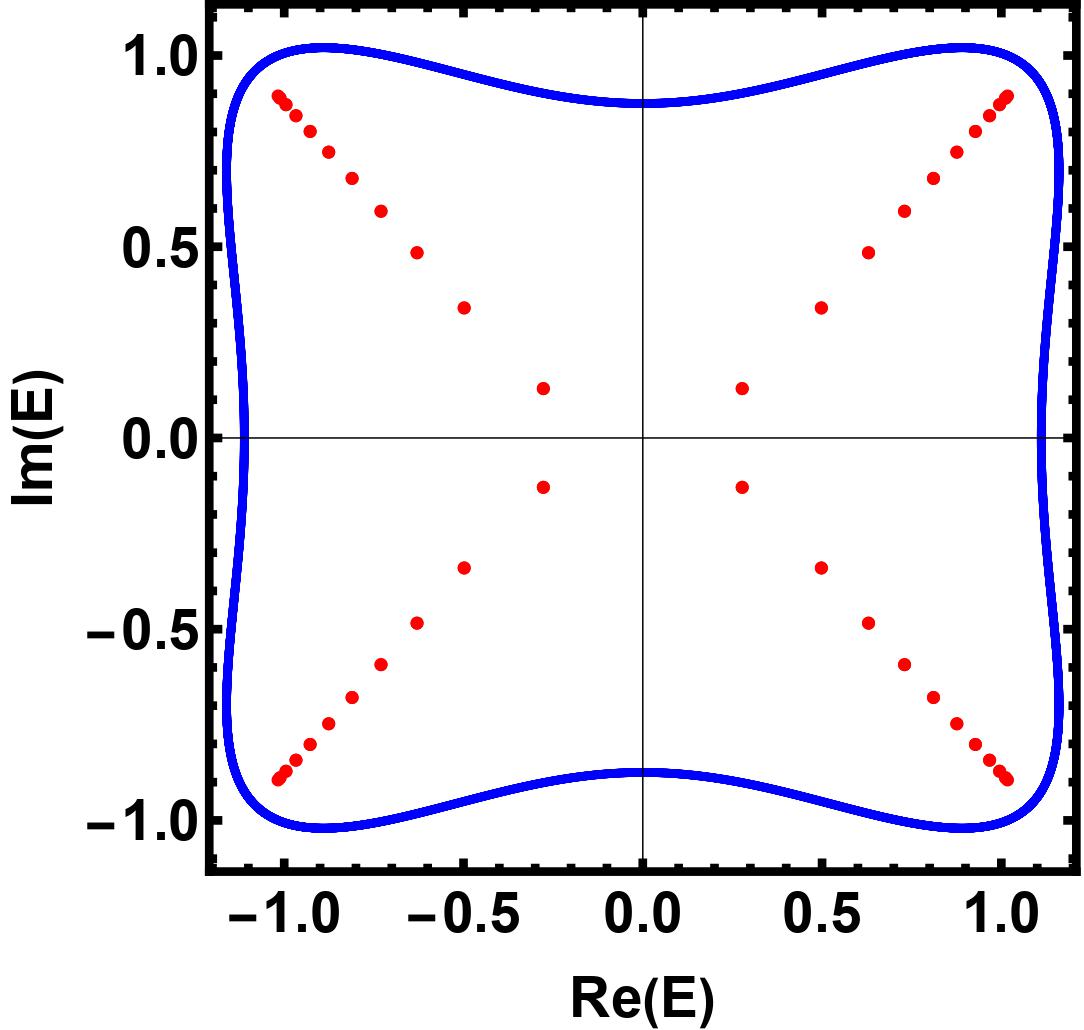}\label{fig:gbz_0.8_200}}
	\subfigure [$N=30$]
	{\includegraphics[width=0.155\textwidth,height=3cm]{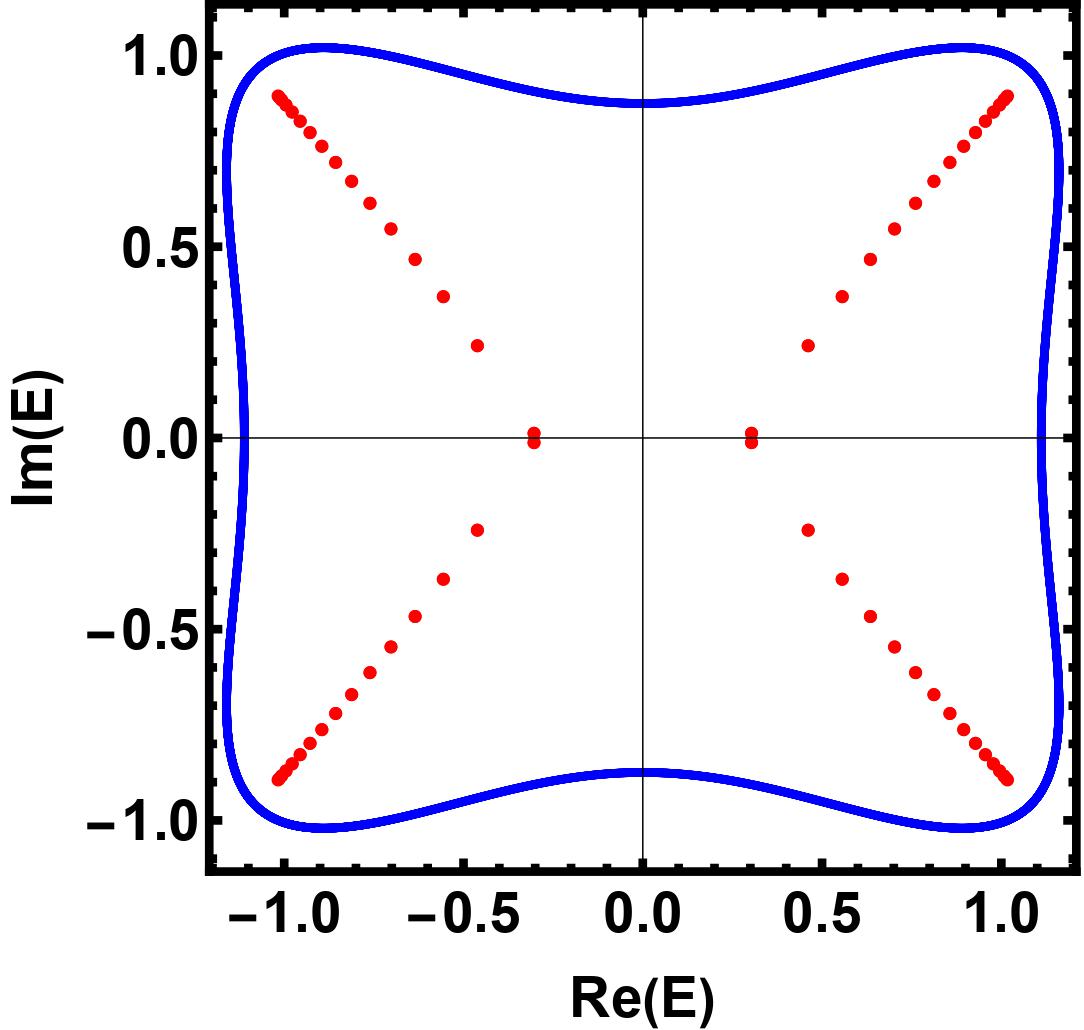}\label{fig:gbz_1.01_200}}
	\subfigure [$N=60$]
	{\includegraphics[width=0.155\textwidth,height=3cm]{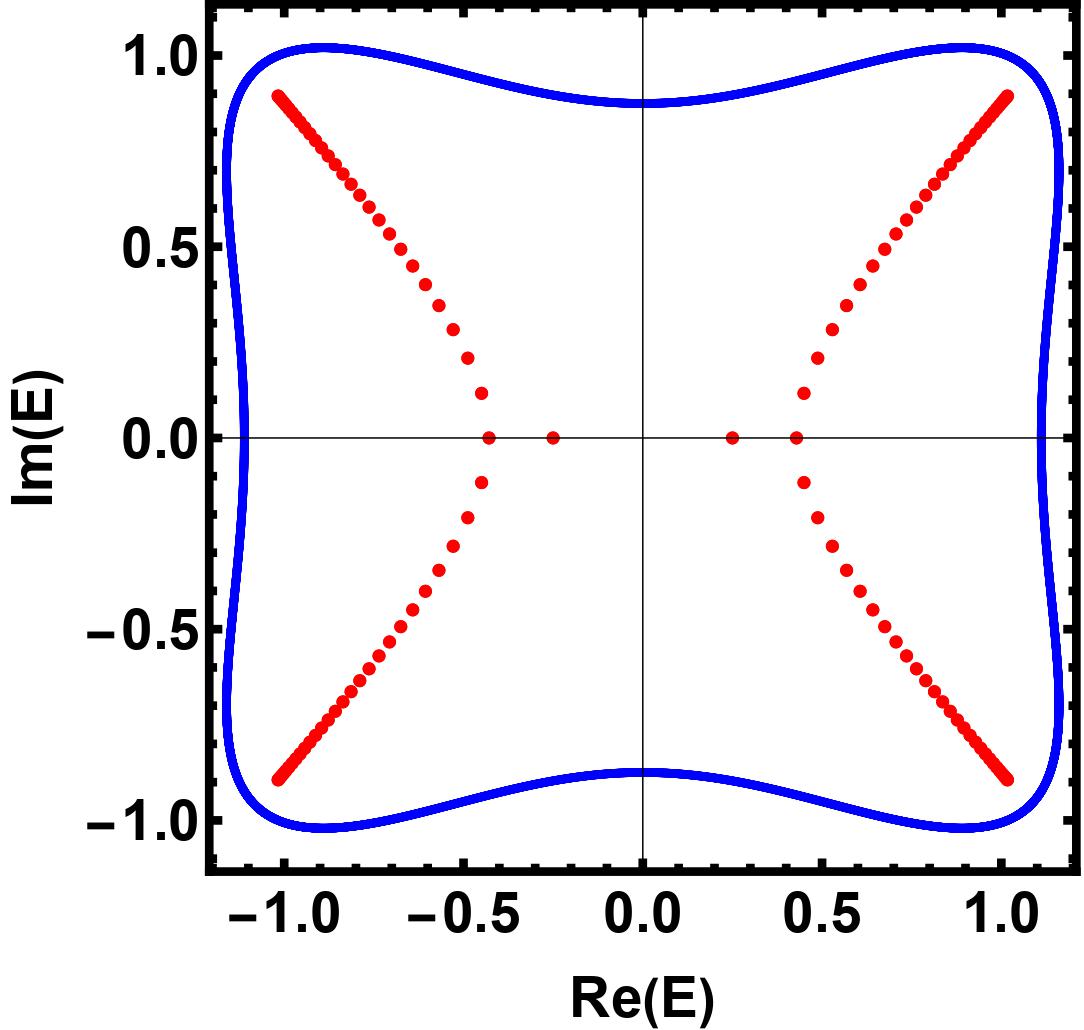}\label{fig:gbz_1.01_500}}\\
	\vspace{-1\baselineskip}
	\subfigure [$N=15$]
	{\includegraphics[width=0.155\textwidth,height=3cm]{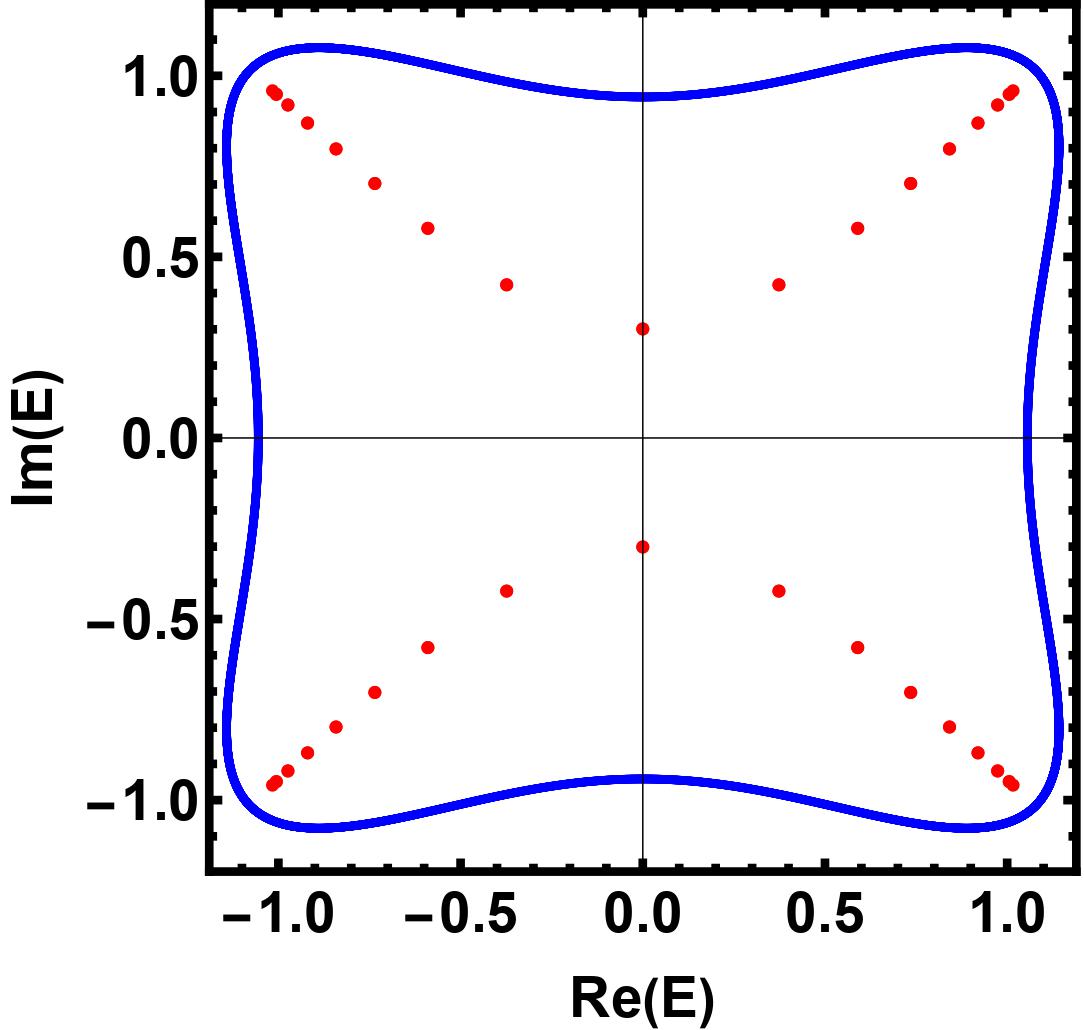}\label{fig:gbz_spectrum_0.8_200}}
	\subfigure [$N=35$]
	{\includegraphics[width=0.155\textwidth,height=3cm]{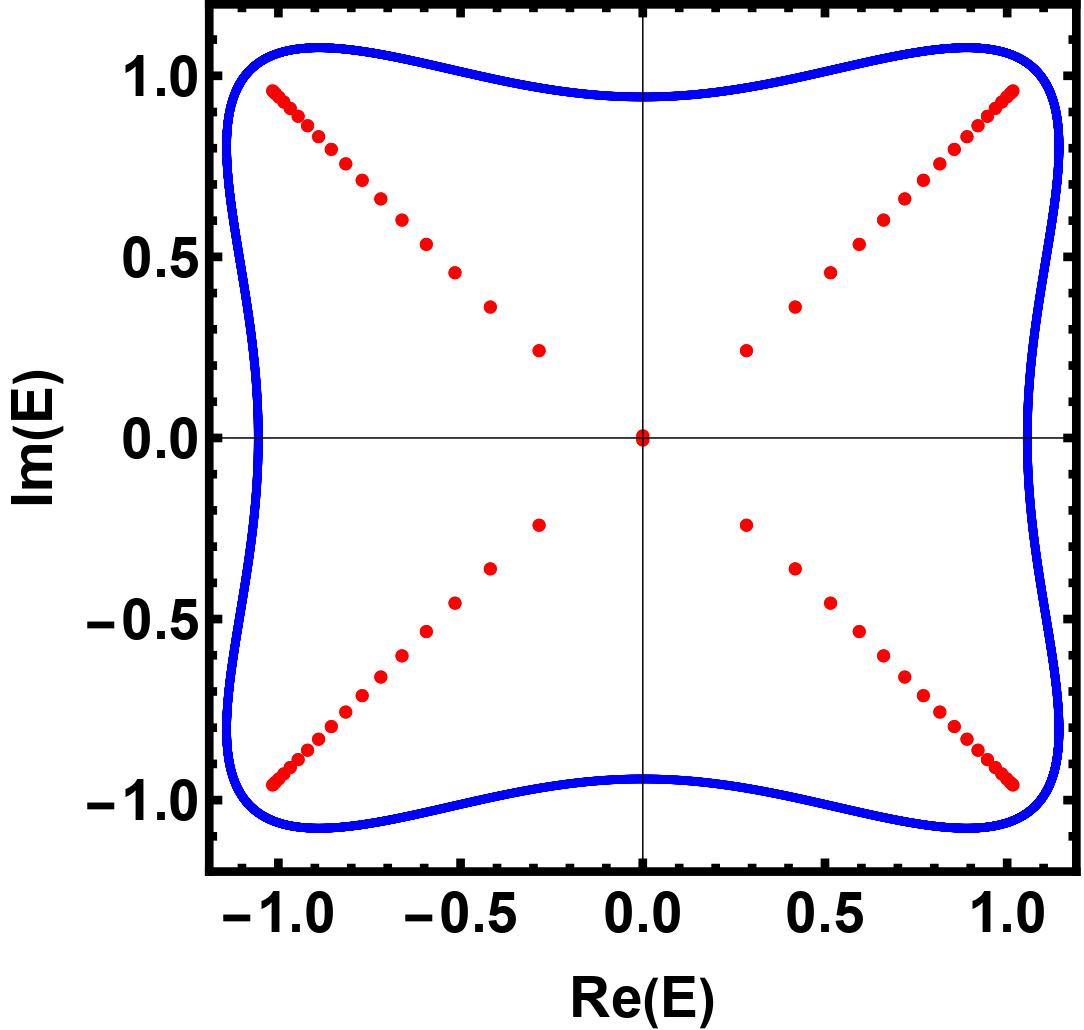}\label{fig:gbz_spectrum_1.01_200}}
	\subfigure [$N=55$]
	{\includegraphics[width=0.155\textwidth,height=3cm]{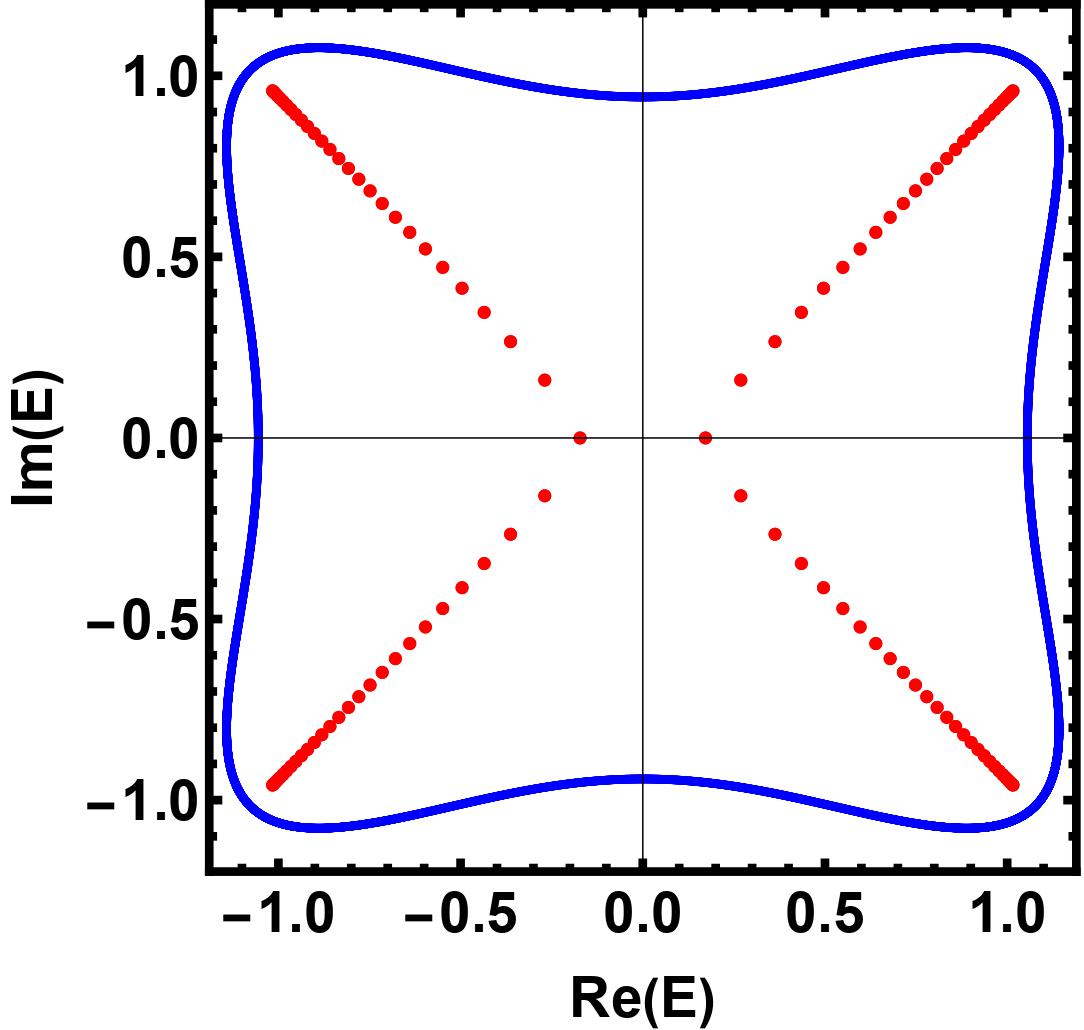}\label{fig:gbz_spectrum_1.25_200}}
    \vspace{-1\baselineskip}
    \caption{\textbf{System-size-dependent spectral evolution of non-Hermitian RM model}. Plot of the OBC spectrum
(red dots) and PBC (blue dots) spectrum. Upper panel: We fix $\gamma = 1.038$ and vary the system size as $N=20, 30 ~\text{and}~60 $. Lower panel: We fix $\gamma=1.095$ and vary system size as $N=15, 35 ~\text{and}~55 $. We see that the spectral evolution demonstrated in Fig.~\ref{fig:complex_spectrum_size_40} and Fig.~\ref{fig:wavefunction_size_odd} is achieved by varying the system size ($N$).   All other parameters are the same as in Fig.~\ref{fig:complex_spectrum_size_40}.}
	\label{fig:gbz_complex_spectrum_size_400}
\end{figure}

To explain system-size-induced topological transitions we plotted the OBC spectrum of the static RM model while fixing $\gamma$ and by varying the system size ($N$), encompassing both odd and even values. In Figure \ref{fig:gbz_complex_spectrum_size_400}, we observe the merging of edge modes with skin modes along the real energy axis, leading to the formation of an exceptional point (EP), resembling the behaviour depicted in Fig.~ \ref{fig:complex_spectrum_size_40}, \ref{fig:gbz_complex_spectrum_size_40} and \ref{fig:wavefunction_size_odd}.

\subsection{Dynamic case}
\label{appen A2}

\begin{figure}[H]
	\centering
	\subfigure [$N=10$]
	{\includegraphics[width=0.2\textwidth,height=3cm]{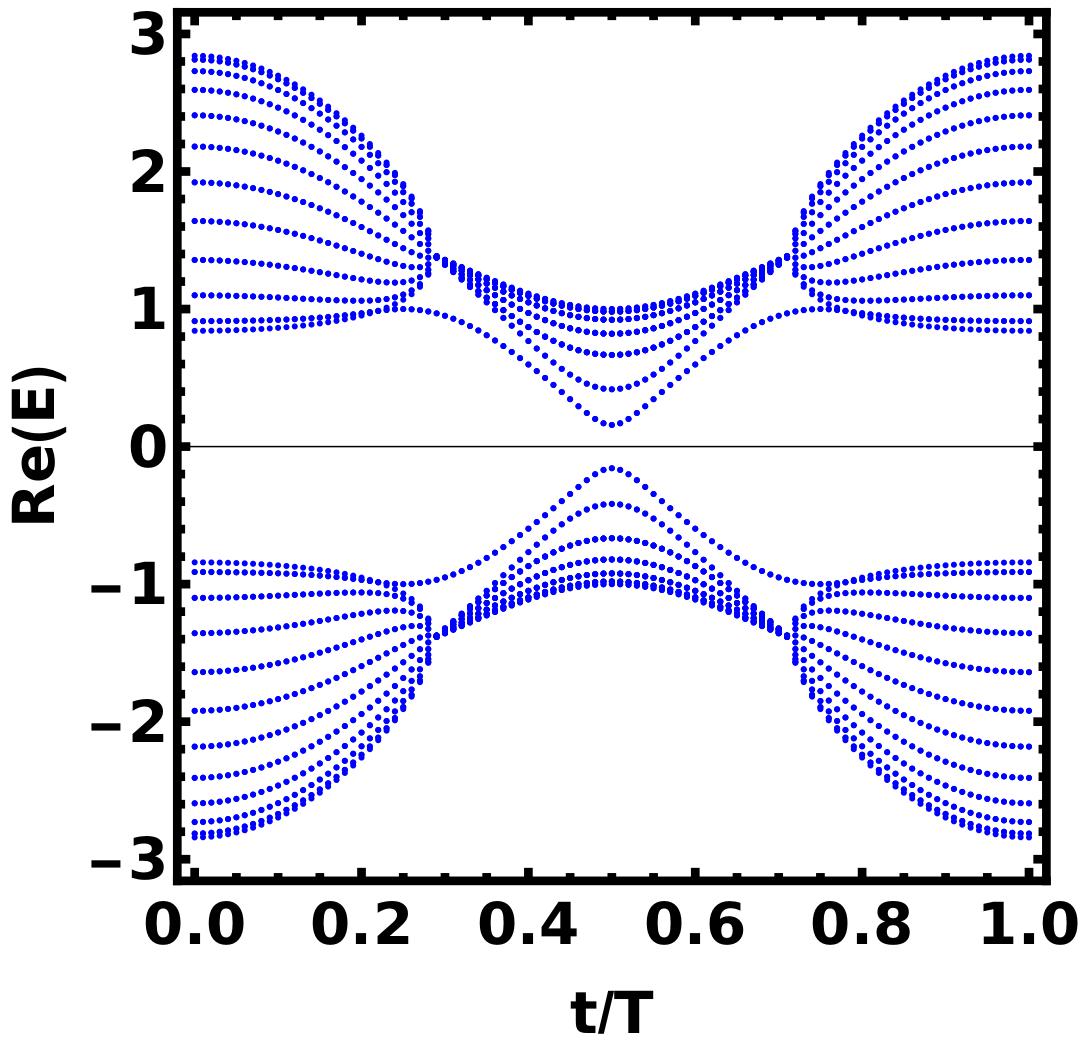}\label{fig:gbz_0.8_20001}}
	\subfigure [$N=30$]
	{\includegraphics[width=0.2\textwidth,height=3cm]{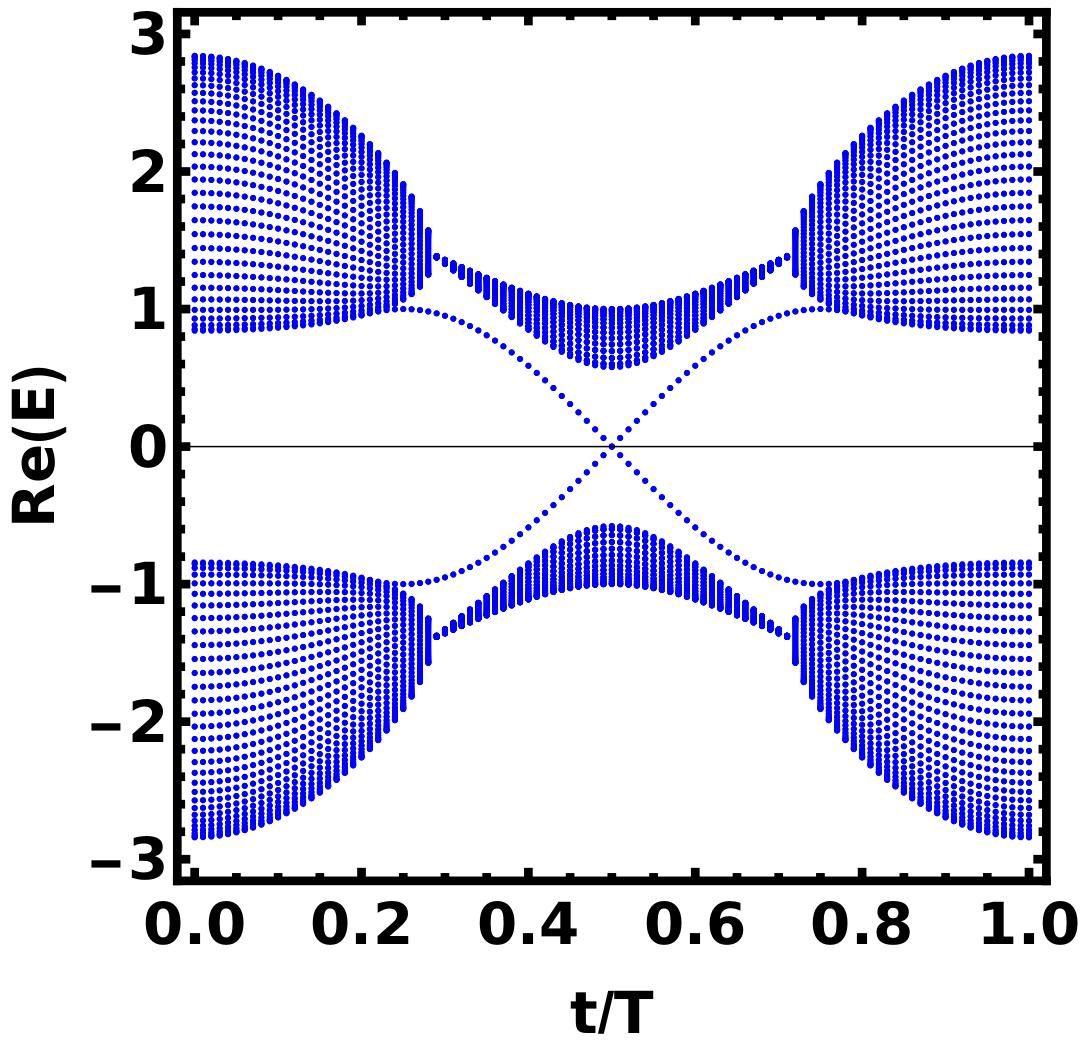}\label{fig:gbz_1.01_20001}}\\
	\subfigure [$N=15$]
	{\includegraphics[width=0.2\textwidth,height=3cm]{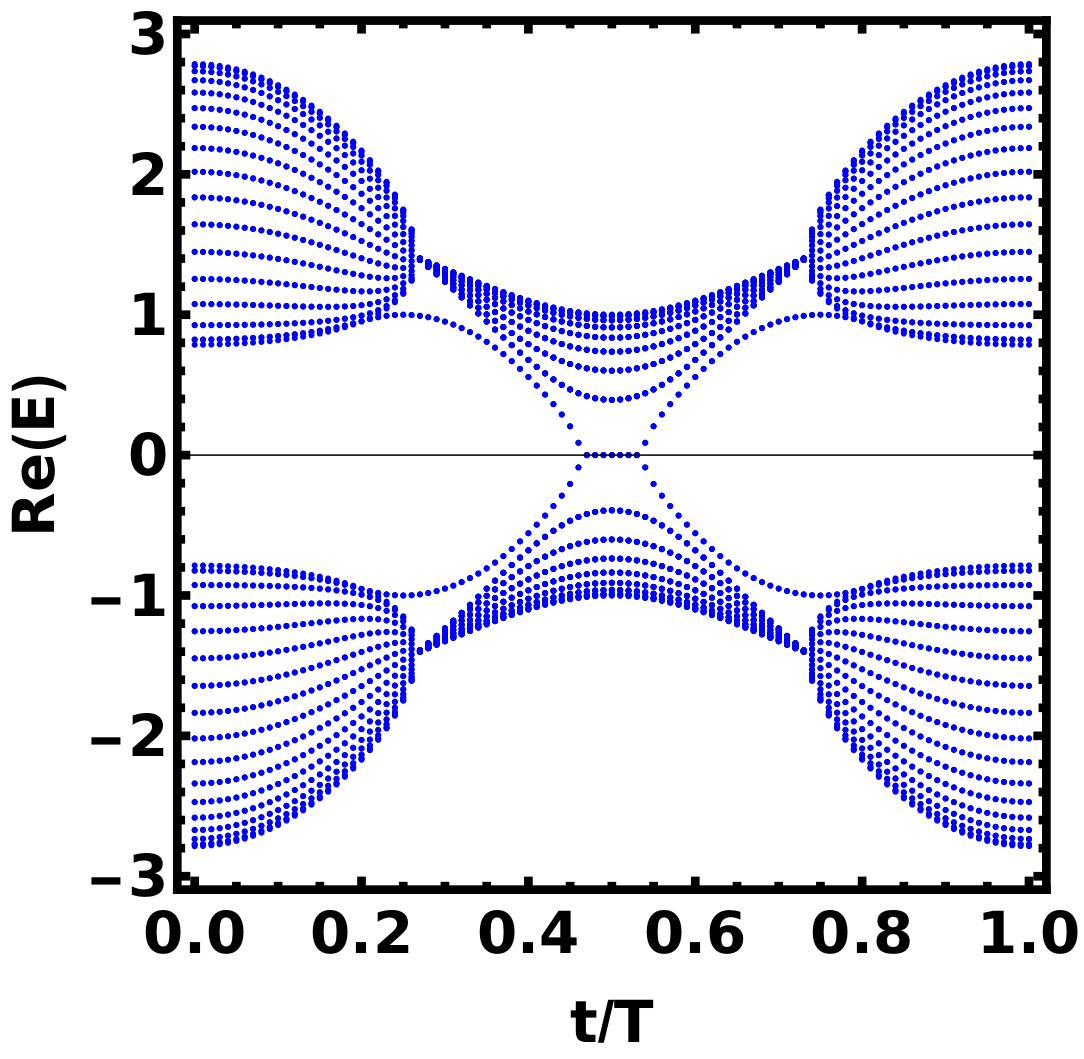}\label{fig:gbz_1.01_50001}}
	\vspace{-1\baselineskip}
	\subfigure [$N=45$]
	{\includegraphics[width=0.2\textwidth,height=3cm]{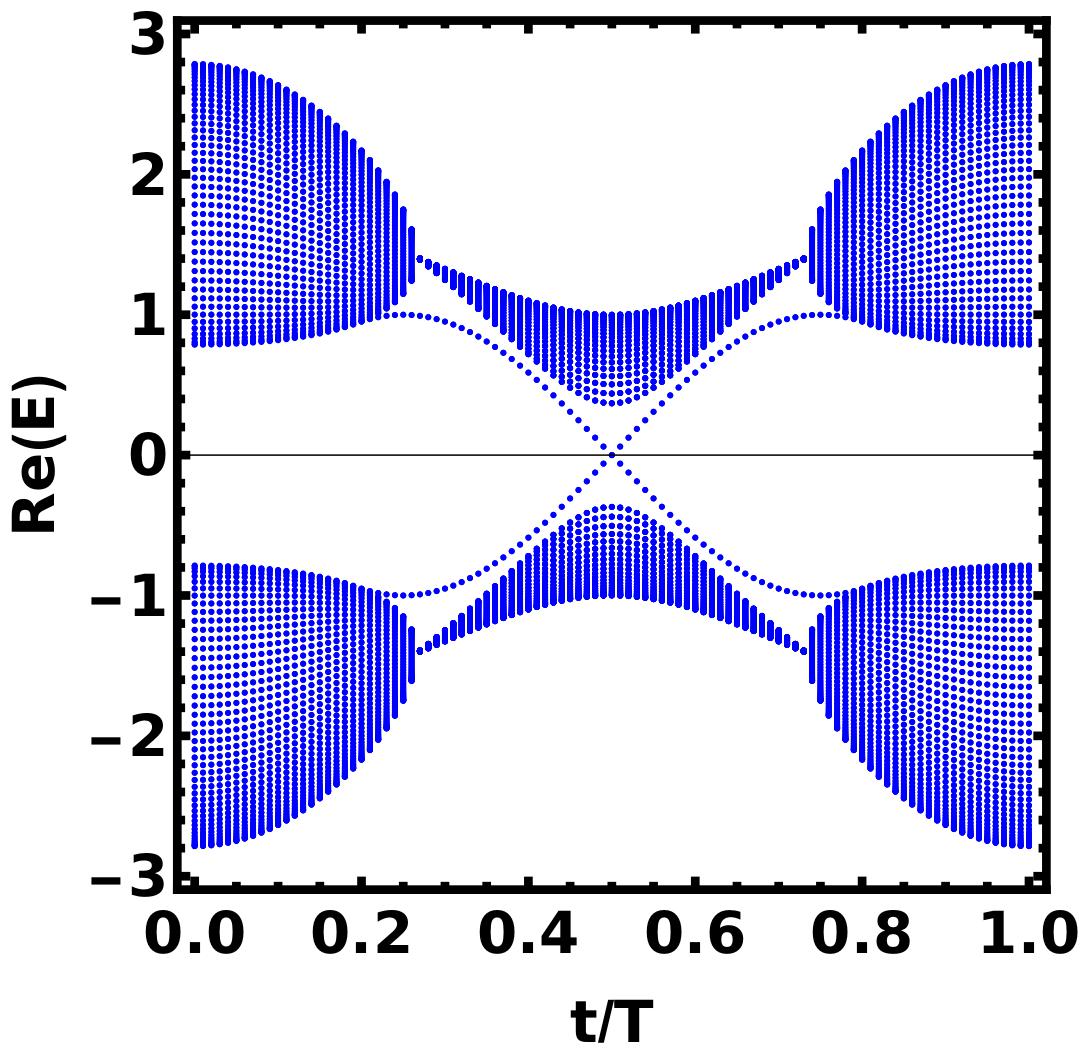}\label{fig:gbz_spectrum_0.8_20001}}

    \caption{\textbf{Real energy spectrum and system-size-induced pumping}. Upper panel (a, b): We fix $\gamma=0.78$ and vary $N$. We observe that for $N=10$, the real spectrum is gapped, while for $N=30$, there is gap-less edge mode. Lower Panel (c, d): We fix $\gamma=0.9$ and tune the system size, to obtain a transition from exceptional edge mode to a gap-less edge mode.  
	\label{fig:gbz_complex_spectrum_size_4001}}
\end{figure}

\begin{figure}[H]
	\centering
	\subfigure [$N=10$]
	{\includegraphics[width=0.2\textwidth,height=3cm]{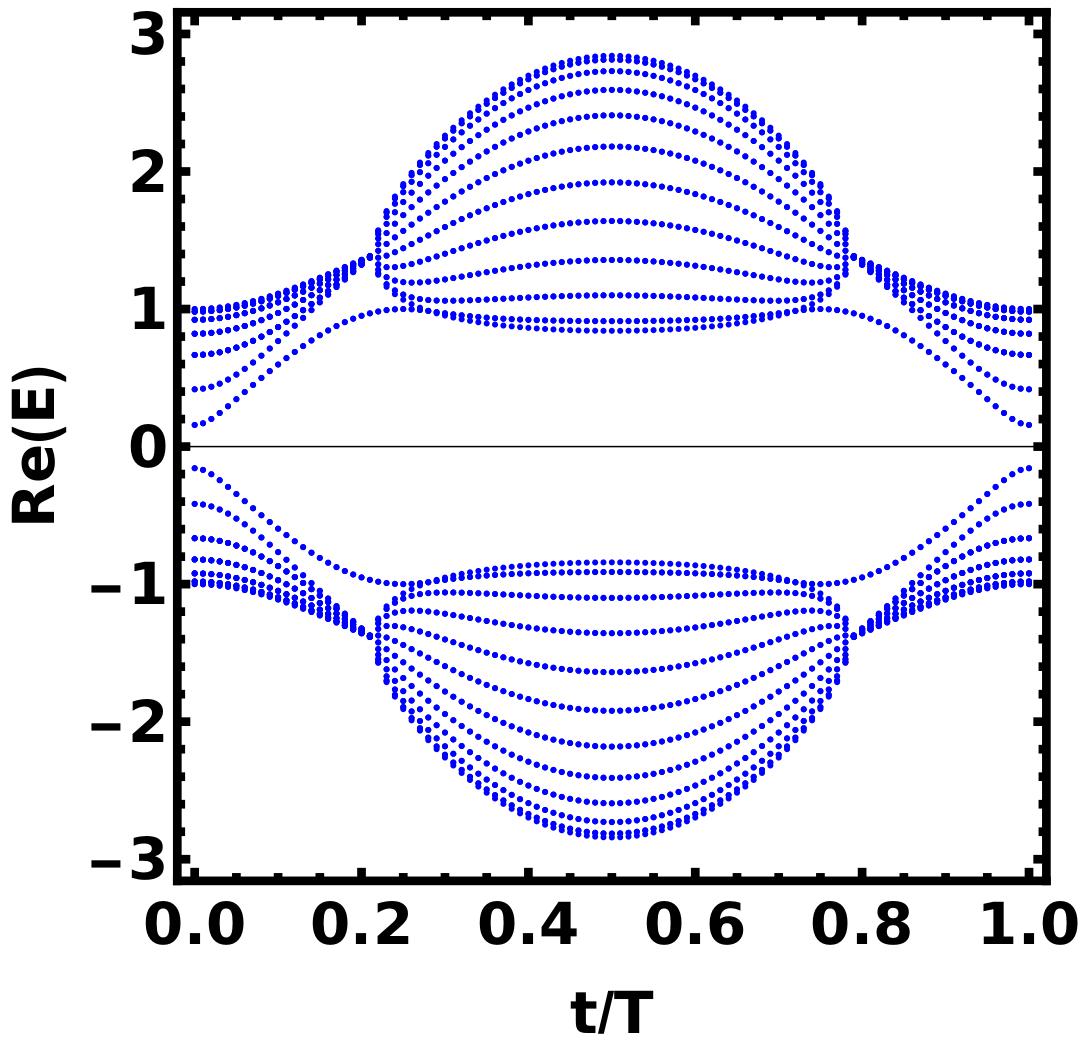}\label{fig:gbz_0.8_2000}}
	\subfigure [$N=30$]
	{\includegraphics[width=0.2\textwidth,height=3cm]{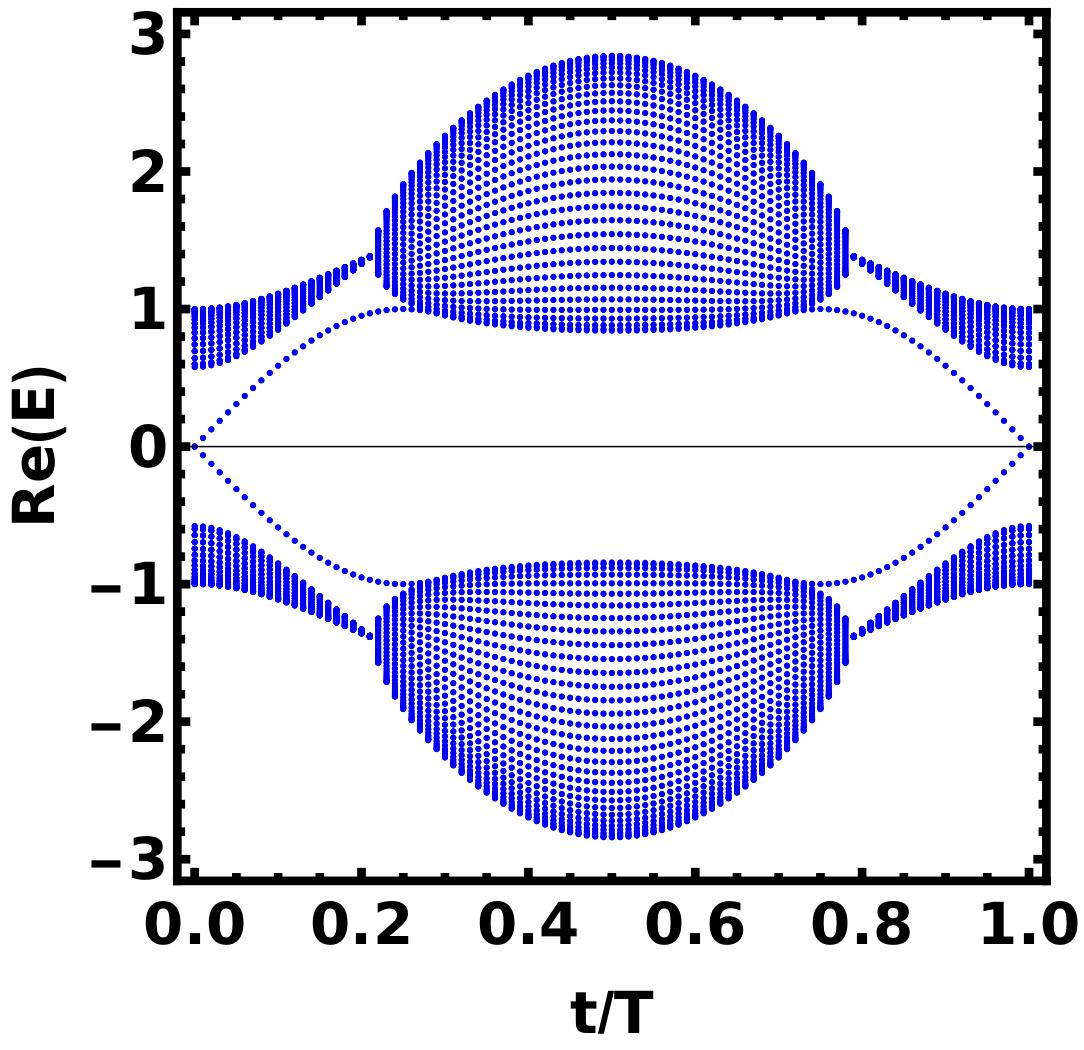}\label{fig:gbz_1.01_2000}}\\
	\subfigure [$N=15$]
	{\includegraphics[width=0.2\textwidth,height=3cm]{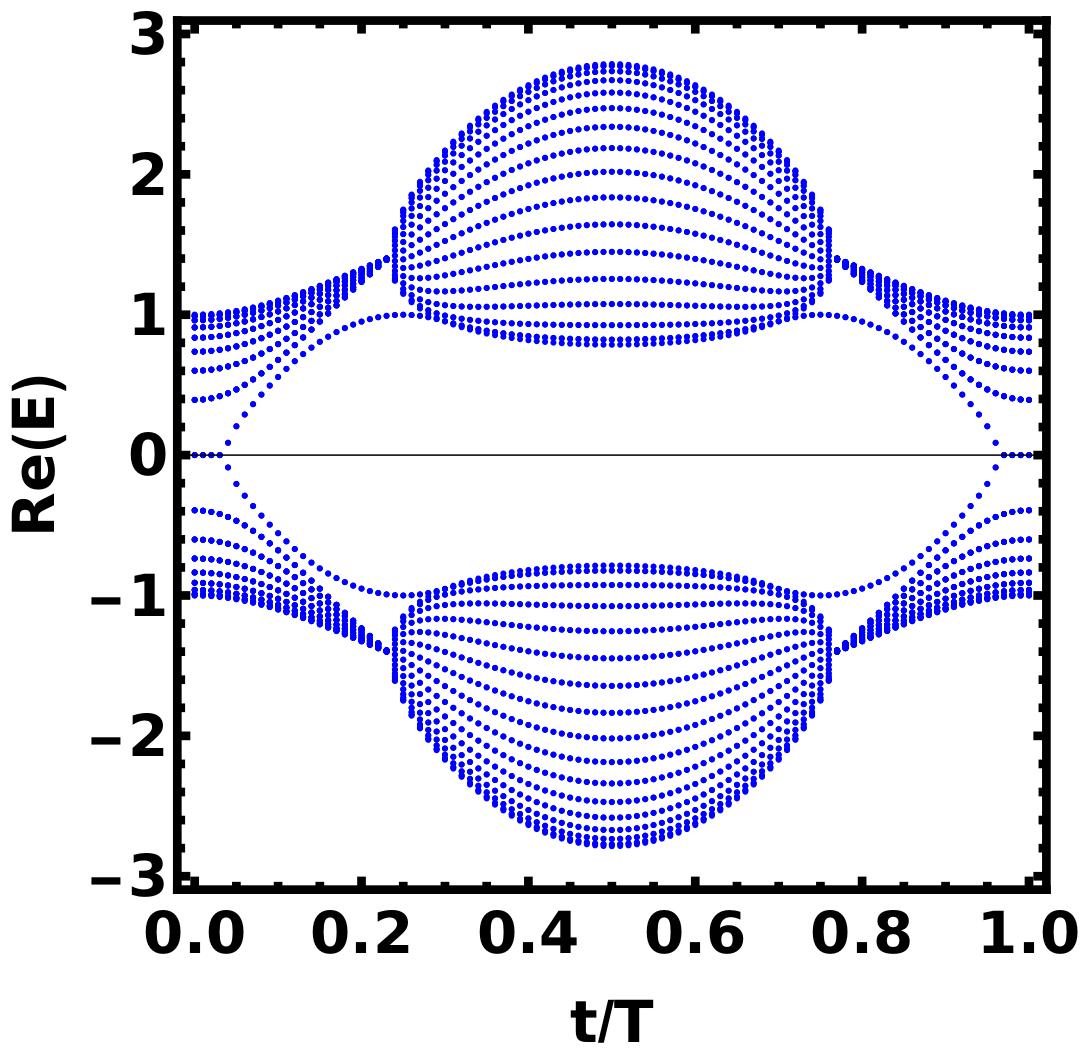}\label{fig:gbz_1.01_5000}}
	\vspace{-1\baselineskip}
	\subfigure [$N=45$]
	{\includegraphics[width=0.2\textwidth,height=3cm]{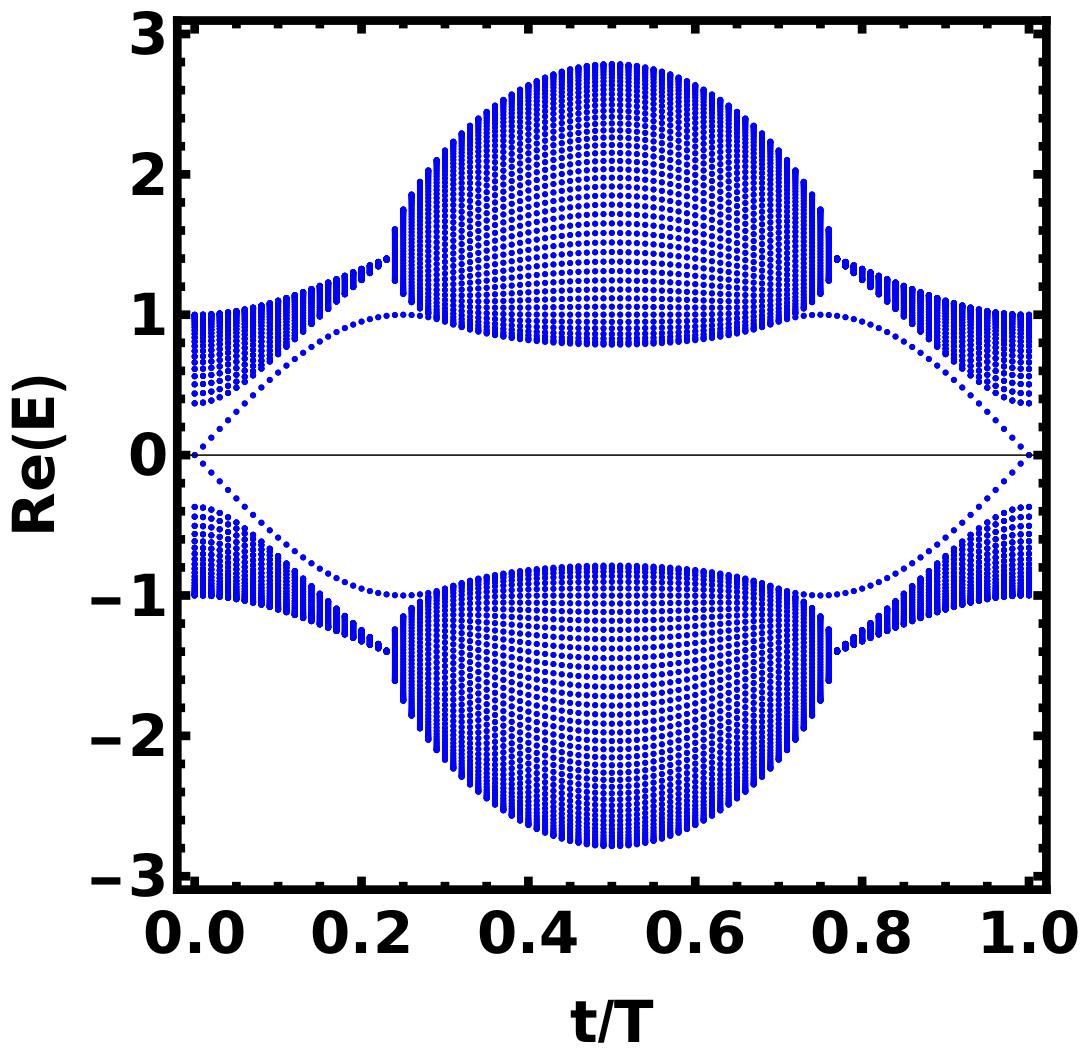}\label{fig:gbz_spectrum_0.8_2000}}

    \caption{\textbf{Real energy spectrum and system-size-induced pumping}. Upper panel (a, b): We fix $\gamma=0.78$ and vary $N$. We observe that for $N=10$, the real spectrum is gapped, while for $N=30$, there is gap-less edge mode. Lower Panel (c, d): We fix $\gamma=0.9$ and tune the system size, to obtain a transition from exceptional edge mode to a gap-less edge mode.  
	\label{fig:gbz_complex_spectrum_size_4000}}
\end{figure}

\begin{figure}[H]
	\centering
	\subfigure [$N=20$]
	{\includegraphics[width=0.2\textwidth,height=3cm]{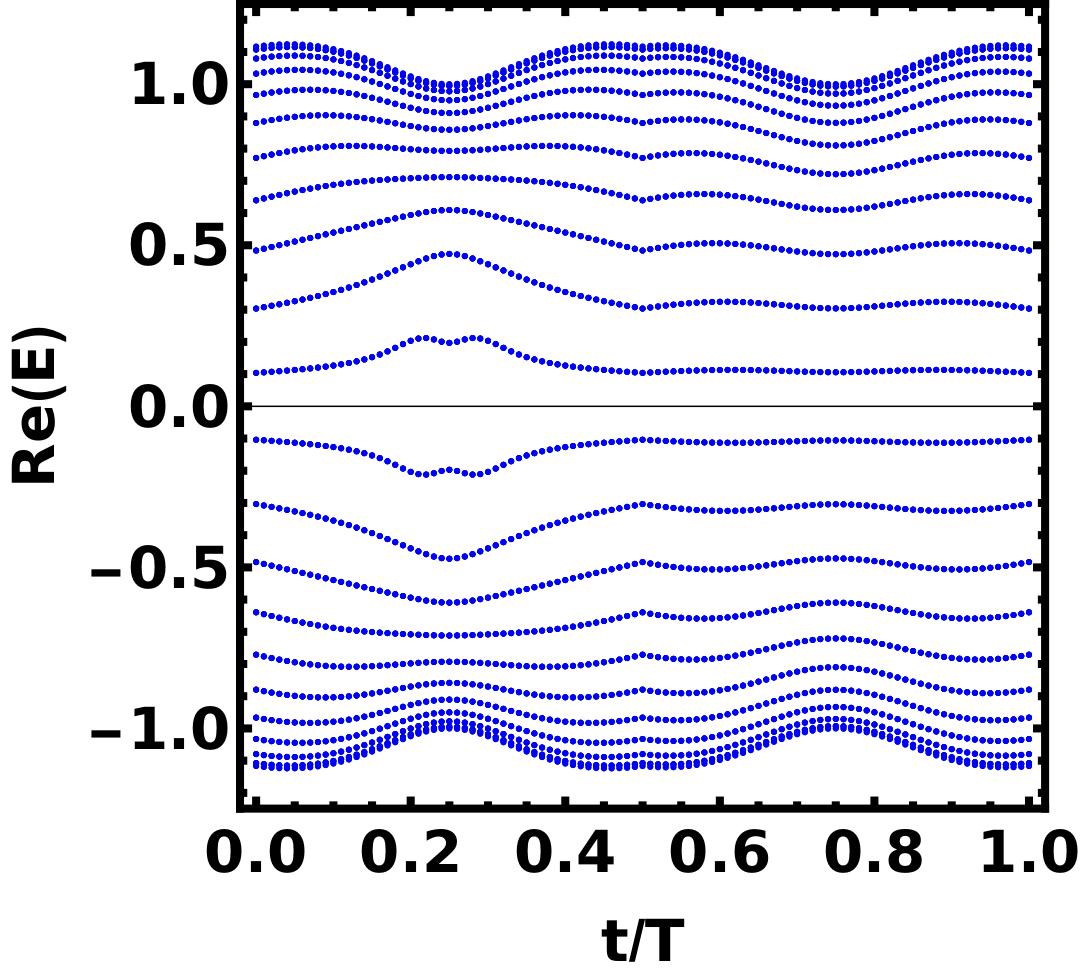}\label{fig:gbz_0.8_200101}}
	\subfigure [$N=50$]
	{\includegraphics[width=0.2\textwidth,height=3cm]{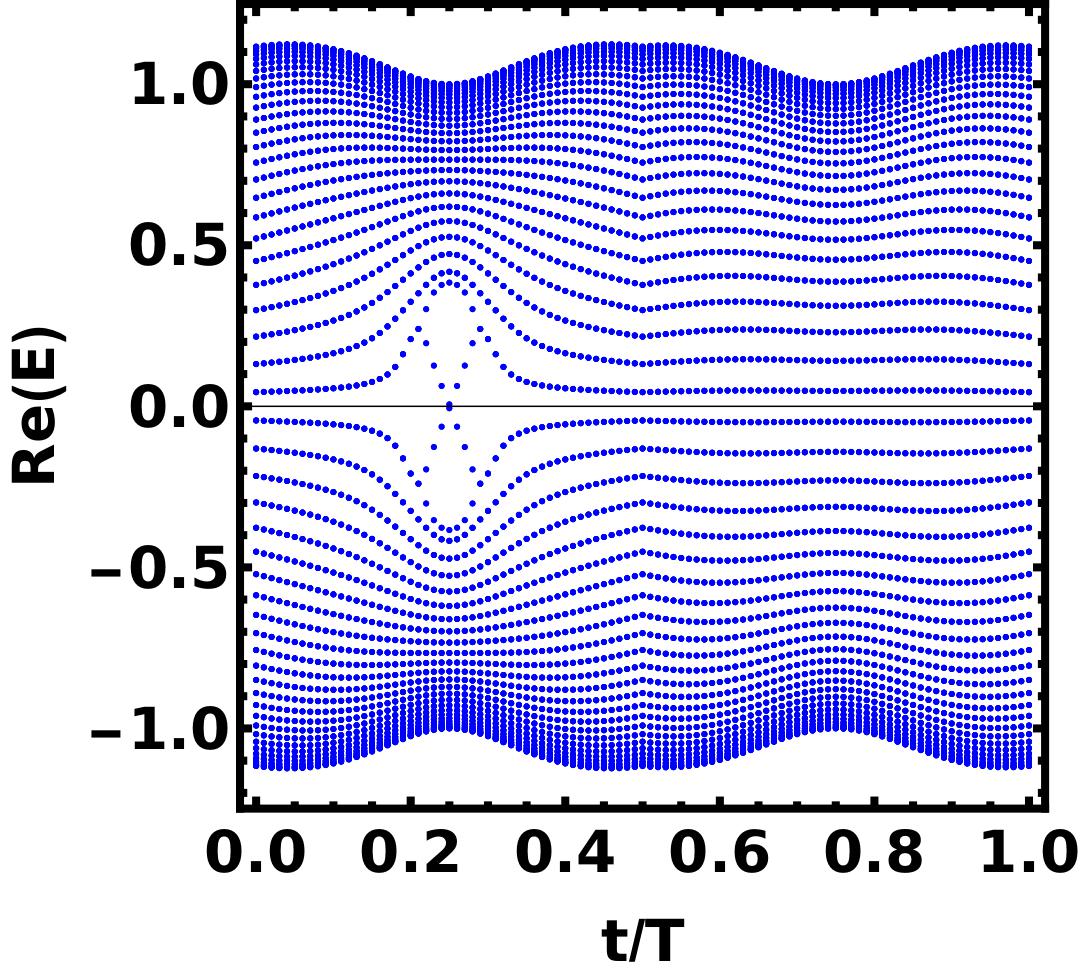}\label{fig:gbz_1.01_201001}}\\
	\subfigure [$N=19$]
	{\includegraphics[width=0.2\textwidth,height=3cm]{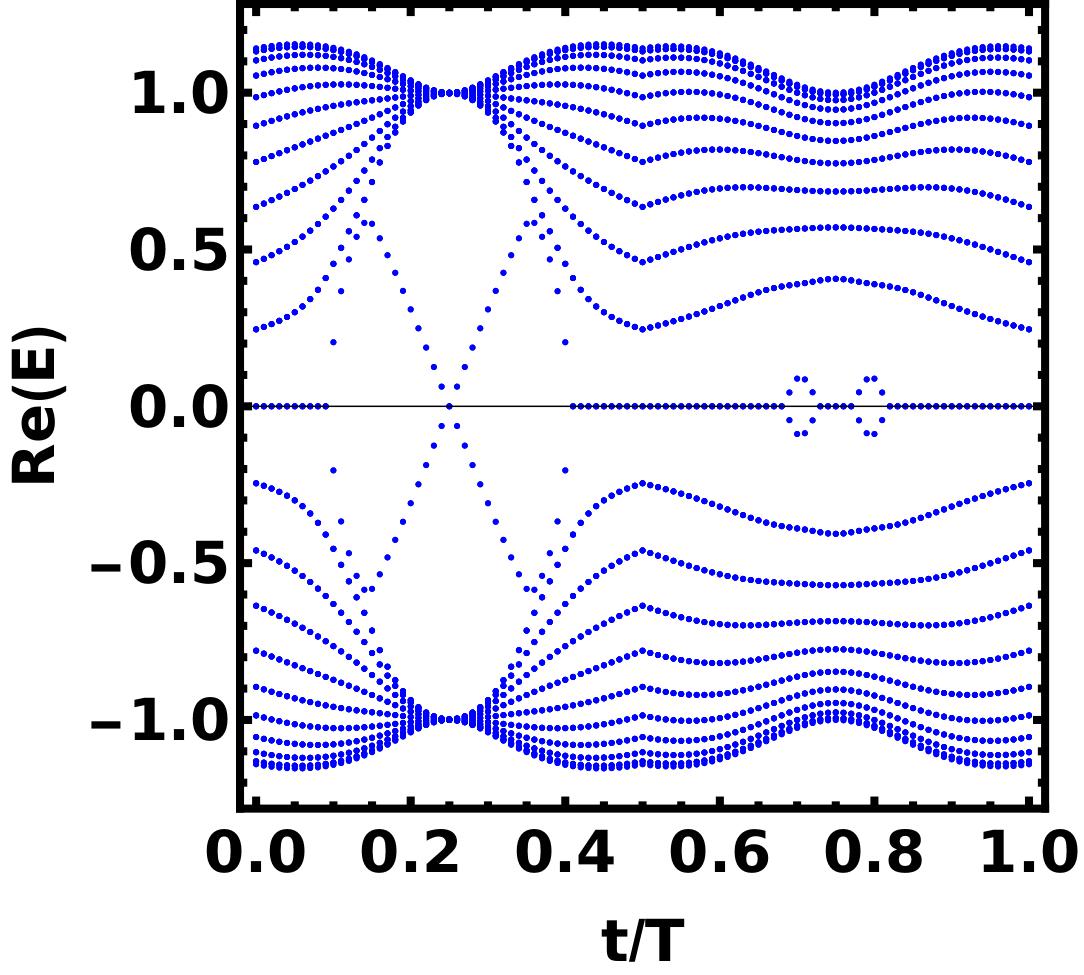}\label{fig:gbz_1.01_510001}}
	\vspace{-1\baselineskip}
	\subfigure [$N=59$]
	{\includegraphics[width=0.2\textwidth,height=3cm]{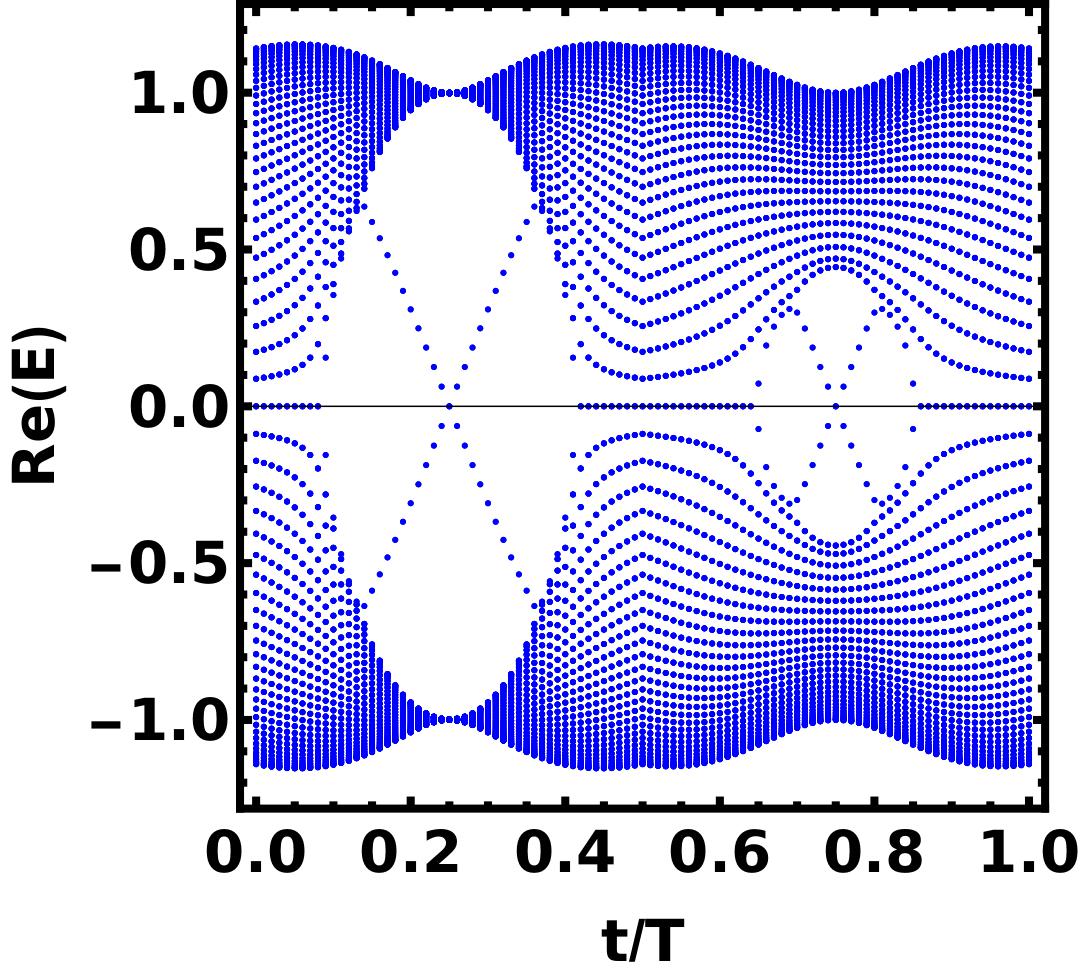}\label{fig:gbz_spectrum_0.8_201001}}

    \caption{\textbf{Real energy spectrum and system-size-induced pumping}. Upper panel (a, b): We fix $\gamma=2.19$ and vary $N$. We observe that for $N=20$, the real spectrum does not have any gap-less edge modes, while for $N=50$, there is a gap-less edge mode. Lower Panel (c, d): We fix $\gamma=2.001$ and tune the system size, and obtain a transition from exceptional edge mode to a gap-less edge mode in the second half of the pumping period.  	\label{fig:gbz_complex_spectrum_size_40001}}
\end{figure}

In the dynamic scenario, we observe that changes in the system size lead to topological transitions. We plotted the instantaneous real energy spectrum including both odd and even system sizes for a fixed $\gamma$, in case of different pumping protocols. In Fig.~\ref{fig:gbz_complex_spectrum_size_4001} (a), \ref{fig:gbz_complex_spectrum_size_4000} (a) and \ref{fig:gbz_complex_spectrum_size_40001}(a) clearly illustrate that the real part of the spectrum exhibits a gap, and in-gap edge states emerge as a consequence of alterations in system size. Consequently for odd system sizes, in Fig.~\ref{fig:gbz_complex_spectrum_size_4001} (b), \ref{fig:gbz_complex_spectrum_size_4000} (b) and \ref{fig:gbz_complex_spectrum_size_40001} (b), reveal the presence of exceptional edge modes that disappear when the system size is modified. As mentioned previously, these EPs do not arise as a consequence of finite-size effects. For odd $N$, a range of $\gamma$-values can be found for which exceptional edge modes are present in the system. Evidently, for the hermitian counterpart ($\gamma=0$), the spectra remain unchanged due to system size variations.

\subsubsection{Clockwise non-trivial protocol}

\subsubsection{Anti-clockwise non-trivial protocol}

\subsubsection{Trivial protocol}

\end{document}